\documentclass[11pt]{./class/elsarticle}

\usepackage{amssymb}
\usepackage{amsmath}
\usepackage{color}
\usepackage{multicol,multicap,graphicx}
\usepackage[mathlines]{lineno}
\usepackage{ulem}
\usepackage{multirow}
\def\beq{\begin{equation}}
\def\eeq{\end{equation}}
\def\bea{\begin{eqnarray}}
\def\eea{\end{eqnarray}}

\def\Re{\mbox{Re}}

\def\pl{\partial}

\definecolor{MyDarkBlue}{rgb}{0,0.08,0.45}
\definecolor{MyDarkRed}{rgb}{0.65,0.00,0.05}
\definecolor{Mygrey}{rgb}{0.8,0.8,0.8}
\definecolor{Mygrey1}{rgb}{0.92,0.92,0.92}
\definecolor{Mygrey2}{rgb}{0.96,0.96,0.92}

\begin{document}
%\pagewiselinenumbers
%\linenumbers

\begin{frontmatter}

%% Title, authors and addresses

%% use the tnoteref command within \title for footnotes;
%% use the tnotetext command for theassociated footnote;
%% use the fnref command within \author or \address for footnotes;
%% use the fntext command for theassociated footnote;
%% use the corref command within \author for corresponding author footnotes;
%% use the cortext command for theassociated footnote;
%% use the ead command for the email address,
%% and the form \ead[url] for the home page:
%% \title{Title\tnoteref{label1}}
%% \tnotetext[label1]{}
%% \author{Name\corref{cor1}\fnref{label2}}
%% \ead{email address}
%% \ead[url]{home page}
%% \fntext[label2]{}
%% \cortext[cor1]{}
%% \address{Address\fnref{label3}}
%% \fntext[label3]{}

\title{An immersed boundary method for the fluid--structure--thermal interaction in rarefied gas flow}

\author[a]{Li Wang}

\author[a]{Fang-Bao Tian \footnote{Correspondence author. Email addresses: f.tian@adfa.edu.au}}

\author[a]{John Young}

\address[a]{School of Engineering and Information Technology, University of New South Wales, Canberra ACT, 2600, Australia}

\begin{abstract}
An immersed boundary method for the fluid--structure--thermal interaction in rarefied gas flow is presented. In this method, the slip model is incorporated with the penalty immersed boundary method to address the velocity and temperature jump conditions at the fluid--structure interface in rarefied gas flow within slip regime. In this method, the compressible flow governed by Navier-Stokes equations are solved by using high-order finite difference method; the elastic solid is solved by using finite element method; the fluid and solid are solved independently and the fluid--structure--thermal interaction are achieved by using a penalty method in a partitioned way. Several validations are conducted including Poiseuille flow in a 2D pipe, flow around a 2D NACA airfoil, moving square cylinder in a 2D pipe, flow around a sphere and moving sphere in quiescent flow. The numerical results from present method show good agreement with the previous published data obtained by other methods, and it confirms the the good ability of the proposed method in handling fluid--structure--thermal interaction for both weakly compressible and highly compressible rarefied gas flow. To overcome the incapability of Navier-Stokes equations at high local Knudsen numbers in supersonic flow, an artificial viscosity is introduced to ease the sharp transition at the shock wave front. Inspired by Martian exploration, the application of proposed method to study the aerodynamics of flapping wing in rarefied gas flow is conducted in both 2D and 3D domains, to obtain some insights for the flapping-wing aerial vehicles operating in Martian environment.

\end{abstract}

\begin{keyword}

Rarefied gas; immersed boundary method; fluid--structure-acoustics interaction; large deformations

\end{keyword}

\end{frontmatter}

%% \linenumbers

\section{Introduction}
Fluid--structure interaction (FSI) is not only ubiquitous in the nature, e.g., flapping flags~\cite{wang2019numerical}, flapping-wing insects~\cite{tian2010interaction} and flow around tall buildings~\cite{toja2018review}, it is also important in many engineering areas such as aeronautics engineering~\cite{kamakoti2004fluid} and biological flows~\cite{huang2021transition}. For this reason, it has drawn considerable research interests in the past hundred years~\cite{dowell2001modeling}. Recently, with the development of hypersonic vehicles, microelectromechanical systems (MEMS) and vacuum technologies, the accurate modelling FSI in rarefied gas flow is of importance to uncover the associated flow physics~\cite{yang2019improved}.

The fluid flow can be categorized into four regimes based on Knudsen number ($Kn$) defined as the ratio of mean-free-path to the characteristic length, i.e, continuum regime ($Kn< 0.001$), slip flow regime ($0.001 \leq Kn \leq 0.1$), transition regime ($0.1 \leq Kn \leq 10.0$) and free molecular regime ($Kn \geq 10.0$)~\cite{tang2005lattice,ahangar2019simulation,bhagat2019implementation}. The traditional numerical method which solves the Navier-Stokes equations (NSE) is not applicable for the flow out of continuum regime for two main reasons, i.e., the rarefied gas could slide over a surface and the inequalities of temperature could rise a force driving the gas slide over a surface from colder to hotter regions which was also known as ``thermal creep''~\cite{lockerby2004velocity}. To overcome this drawback, the first-order slip model was derived according to the theoretical analysis of Maxwell~\cite{maxwell1879vii}. In such models, the slippery velocity and temperature are calculated based on the local velocity and temperature gradients. To achieve accurate modelling of the slip boundary conditions induced by rarefied gas effects, significant efforts have been involved to improve the slip model. For example, the 1.5-order~\cite{mitsuya1993modified} and second order slip model~\cite{loyalka1975some,maccormack1989nonequilibrium,wu2008slip} were proposed to improve the accuracy of the original 1st-order model~\cite{kennard1938kinetic,lockerby2004velocity}; Wu~\cite{wu2008slip} derived a slip model for wall bounded rarefied gas flows from kinetic theory to extend the slip model to whole Knudsen number range. These models are typically categorized into first-order and second-order model based on the derivative terms used in the calculation~\cite{wu2008slip}. The factors of the derivatives depending on gas--solid interactions are obtained from analytical models or experimental data~\cite{tucny2020comparison}. Thanks to the development of these slip models, a lot of successful applications of incorporating the such models into computational fluid dynamics (CFD) solvers have confirmed their capability of solving flow in slip flow regime and the early transition regime. For example, Fan et al.~\cite{fan2001computation} investigated flow around a 2D NACA0012 airfoil in various conditions from subsonic to supersonic by using both slip model based NSE solver and particle approach at a Knudsen number up to 0.026, and found that good agreements can be achieved with these solvers. It is also noted that the excessive statistical fluctuations of Direct simulation Monte Carlo (DSMC) makes it failure for the low subsonic flow simulation, and an information preservation technique was used in this work.  Similarly, Le et al.~\cite{le2015rarefied} implemented a second order slip model into the OpenFoam to simulate flow around a 2D stationary NACA airfoil in the supersonic and rarefied gas flow up to a Knudsen number of 0.05, and concluded that such an implementation achieves good agreements compared with the DSMC. Lofthouse et al.~\cite{lofthouse2008velocity} systematically examined flow around a stationary cylinder at Knudsen numbers up to 0.25 and Mach numbers up to 25, and concluded that the slip model (proposed by Maccormack~\cite{maccormack1989nonequilibrium}) based solver can achieve comparable results with the DSMC with the maximum deficits of surface properties (i.e., pressure, fraction and heating coefficients) around 5\%. The successful application of slip model provides a simple way to adapt the conventional computational fluid dynamics (CFD) solvers for the rarefied gas flow in the slip regime. Although the inherent continuum assumptions of NSE have limited such solvers at slip and early transition regimes (e.g., $Kn < 1.0$), they are still attractive as the excellent efficiency compared with particle method such as DSMC and discrete velocity method (DVM) where excessive memory and computational capacity are required. In addition, the shock wave-induced high local Knudsen numbers leads to the failure of continuum consumption as well, which has been discussed in Refs.~\cite{boyd1995predicting,lofthouse2008velocity}. It is noted such local violations do not influence the overall aerodynamic force and surface properties, but the discontinuities of shock wave from NSE solvers tend to be much sharper than those from DSMC computation~\cite{boyd1995predicting,lofthouse2008velocity}. Interestingly, this point has long been noticed but we see no attempts have been made to resolve it. Here, we will present a simple model with artificial viscosity into the NSE solvers to ease the failure at the shock induced discontinuities.

Although the incorporation of slip model into NSE solvers have been implemented and validated in previous studies. It is noted that most of the previous studies focused on the conformed-mesh based solvers for simple and stationary geometries, leaving the complex geometries with large displacements and deformations not explored. The immersed boundary method (IBM) as an non-conformed mesh method has been extensively been studied for many physical problems such as turbulent flow and viscoelastic flow involving complex moving geometries and two-way FSI. For example, as a versatile technique to handle the arbitrarily complex boundary conditions at the fluid--structure interface, the IBM based wall models have been developed for turbulence modelling~\cite{shi2019wall,capizzano2011turbulent}; the heat transfer in a complex system with FSI is developed~\cite{wang2018heat}. These successful implementations and applications inspired us to develop a slip model based IBM for the numerical modelling of rarefied gas flow in slip regime. The IBM can be typically categorized into diffusive and sharp interface method, where the former one is generally achieved by applying body force and the later one is achieved by applying the velocity boundary conditions reconstructed by interpolation~\cite{huang2019recent}. The sharp interface IBM provides a second-order accuracy, but it suffers spurious fluctuations and is less efficient for complex moving objects~\cite{ghias2007sharp,mittal2008versatile}. Although such fluctuations can be relived by using such as flow reconstruction~\cite{luo2009hybrid} and cut-cell approach~\cite{seo2011sharp}, but it involves complex computation and decreases the computational efficiency. Alternatively, the diffusive IBM has been extensively employed in a range of problems such as FSI in compressible and multi-phase flow~\cite{wang2017immersed} and turbulence wall modelled large eddy simulation (LES)~\cite{shi2019wall}, for the smooth force, simple implementation and good computational efficiency compared with its sharp interface counterpart. 

A diffusive IBM for the fluid--structure--thermal interaction in rarefied gas flow is presented. In this method, the slip model is incorporated with the penalty immersed boundary method to address the velocity slip and temperature jump conditions at the fluid--structure interface. It should be noted that the proposed IBM can be easily incorporated into other fluid solvers such as high order LBM~\cite{meng2011accuracy} to extend it to a wide range of Knudsen numbers. The arrangements of the paper are as follows: the numerical method is presented in Section 2. Several validations including Poiseuille flow in a pipe at very low Mach numbers, supersonic flow around a stationary NACA airfoil, a moving square cylinder in a pipe, flow around a sphere and moving sphere in quiescent flow are presented in Section 3. Inspired by Martian exploration, the application of proposed method to study the aerodynamics of flapping wing hovering in rarefied gas flow are presented in Section 4. A final conclusion is given in Section 5.

%%%%%%%%%%%%%%%%%%%%%%%%%%%%%%
\section{Numerical method}
%%%%%%%%%%%%%%%%%%%%%%%%%%%%%%

\subsection{Fluid solver}
The fluid dynamics considered here are governed by the 3D compressible viscous Navier--Stokes equations
\begin{eqnarray}
&&\frac{\pl Q}{\pl t} + \frac{\pl F}{\pl x} + \frac{\pl G}{\pl y}+ \frac{\pl H}{\pl z}-
\frac{1}{\Re}(\frac{\pl F_u}{\pl x}+\frac{\pl G_v}{\pl y}+\frac{\pl H_v}{\pl z})= S,\\
&&Q=[\rho,\rho u,\rho v,\rho w, E]^T,\quad F=[\rho u,\rho u^2+P,\rho u v,\rho u w, (E+P) u]^T, \nonumber \\
&&G=[\rho v,\rho u v, \rho v^2+ P,\rho v w, (E+P)v]^T,\quad H=[\rho w,\rho u w,\rho v w, \rho w^2+ P,(E+P)w]^T,\nonumber \\
&&F_u=[0,\tau_{xx},\tau_{xy},\tau_{xz},b_x]^T,\quad G_v=[0,\tau_{xy},\tau_{yy},\tau_{yz},b_y]^T,\quad H_v=[0,\tau_{xz},\tau_{yz},\tau_{zz},b_z]^T,\nonumber \\
&&b_x=u \tau_{xx}+ v \tau_{xy}+w \tau_{xz} - q_x,\quad b_y=u \tau_{xy}+ v \tau_{yy}+ w \tau_{yz} - q_y,\quad b_z=u \tau_{xz}+ v \tau_{yz}+ w \tau_{zz} - q_z,\nonumber 
\end{eqnarray}
where $\rho$ is the density of the fluid, $u$, $v$ and $w$ are the velocity components, $P$ is the pressure, $E$ is the total energy, $S$ is a general source term including the IB-imposed Eulerian force and other body forces, Re is the Reynolds number, and $\tau_{ij}$ is the shear stress. The thermal flux $q_x$, $q_y$ and $q_z$ are expressed according to Fourier’s law as
\begin{equation}
q_x = -\frac{\mu}{Pr (\gamma - 1)}\frac{\pl T}{\pl x}, \quad q_y = -\frac{\mu}{Pr (\gamma - 1)}\frac{\pl T}{\pl y}, \quad q_z = -\frac{\mu}{Pr (\gamma - 1)}\frac{\pl T}{\pl z},
\end{equation}
where $Pr$ is Prandtl number, $\gamma$ is the adiabatic coefficient, and $\mu$ is the viscosity of the fluid. The temperature of the compressible fluid is given by
\begin{equation}
T = \frac{\gamma P}{\rho (\gamma - 1) c_p},
\end{equation}
where $c_p$ is the specific heat coefficient. Without loss of generality, the ideal gas equation of state is used here for the fluid, and thus the total energy is given by
\begin{equation}
E = \frac{P}{\gamma - 1} + \frac{\rho(u^2 + v^2 + w^2)}{2}.
\end{equation}

In the fluid solver, the convective term and viscous term are respectively discretized by the fifth-order accuracy WENO scheme~\citep{liu1994weighted} and a fourth-order central difference scheme. The third-order Runge-Kutta scheme is adopted for temporal discretization for all unsteady equations in flow solver. 

\subsection{Solid solver}
In this work, the dynamics of the solid is governed by~\citep{zienkiewicz2000finite}
\begin{equation}
\rho_s \dot{\boldsymbol{v}}  = \nabla \cdot \boldsymbol{\sigma} + \rho_s \boldsymbol{b},
\label{eq:solid}
\end{equation}
where $\rho_s$ is the density of the solid, $\boldsymbol{v}$ is the velocity vector with the dot representing the temporal derivative, $\boldsymbol{\sigma}$ is stress tensor, and $\boldsymbol{b}$ is the body force exerting on the solid. By using the principle of virtual work, the governing equation can be transformed into its weak form, and the discrete form is expressed as~\cite{wang2017projection}
\begin{equation}
\int_V N_I \rho_s N_J \ddot{v}_{iJ} {\rm d}V + \int_V \frac{\pl N_I}{\pl x_j} \sigma_{j,i} {\rm d}V= \int_V N_I \rho_s b_i {\rm d}V + \int_{A_t} N_I \bar{t_i} {\rm d}A,
\label{eq:solidweakform}
\end{equation}
where,$b_i$ is the external body force, $\bar{t_i}$ is the external force exerting on the surface boundary $A_t$, and $N_I$ is the shape function of the $I$-th node. The subscripts meet Einstein summation convention. The discrete form of Eq.~\eqref{eq:solidweakform} can be written in a simplified form as
\begin{equation}
\boldsymbol{M} \ddot{\boldsymbol{v}} + \boldsymbol{f}^{\rm int} = \boldsymbol{f}^{\rm ext},
\label{eq:soliddiscrete}
\end{equation}
where, 
\begin{eqnarray}
\boldsymbol{M}_{IJ} =\int_V \rho_s \boldsymbol{N}_I^T \boldsymbol{N}_J {\rm d} V, \quad
\boldsymbol{f}_I^{\rm int} = \int_V \boldsymbol{B}_I^T \boldsymbol{\sigma} {\rm d}V, \quad
\boldsymbol{f}_I^{\rm ext} = \int_V \boldsymbol{N}_I^T \rho_s \boldsymbol{b} {\rm d}V + \int_{A_t} \boldsymbol{N}_I^T \boldsymbol{\bar{t}} {\rm d} A,
\label{eq:solidmatrix}
\end{eqnarray} 
where, $\boldsymbol{N}_I=N_I \boldsymbol{I}$ and 
\begin{equation}
\boldsymbol{B}_I = \left[ 
\begin{array}{cccccc}
\frac{N_I}{x_1} & 0 & 0 & 0 & \frac{N_I}{x_3} & \frac{N_I}{x_2}\\
0 & \frac{N_I}{x_2} & 0 & \frac{N_I}{x_3} &0 & \frac{N_I}{x_1}\\
0 & 0 & \frac{N_I}{x_3} & \frac{N_I}{x_2} & \frac{N_I}{x_1} & 0
\end{array} 
\right ]^T.
\end{equation}
Here, eight-nodes solid element is employed and its shape function is
\begin{equation}
N_I(\xi,\eta,\zeta) = \frac{1}{8} (1+\xi_I \xi)(1+\eta_I \eta)(1 + \zeta_I \zeta),
\label{eq:shapfunc}
\end{equation}
where $\xi_I$, $\eta_I$ and $\zeta_I$ are natural coordinates of the interpolation point,  $\xi$, $\eta$ and $\zeta$ are natural coordinates of the node of the element. In this work, two-points Gauss-Legendre numerical integral is used for Eq.~\eqref{eq:solidmatrix}. Lumped mass matrix is used to reduce the computation cost and the solid is assumed to be ideally elastic. More details of the finite element method can be found in Refs.~\cite{zienkiewicz2000finite,tian2014fluid}.

\subsection{Fluid--structure--thermal coupling method}
In this work, the dynamics of the fluid and solid are solved independently and the fluid-structure coupling is achieved by using a feedback law~\citep{goldstein1993modeling} based on the penalty immersed boundary (pIB) method. The interaction force is calculated explicitly by
\begin{equation}
\boldsymbol{F}_f = \alpha \int_0^t (\boldsymbol{U}_{ib} - \boldsymbol{U}) dt + \beta (\boldsymbol{U}_{ib} - \boldsymbol{U}),
\label{eq:penatly}
\end{equation}
where $\boldsymbol{U}_{ib}$ is the velocity of solid node integrated from the flows, $\boldsymbol{U}$ is the real velocity of the solid node, and $\alpha$ and $\beta$ are large positive constants. Eq.~\eqref{eq:penatly} gives the Lagrangian force acting on the solid and the force exerting on the fluid by the immersed boundary is -$\boldsymbol{F}_f$, which is spread onto the fluid nodes adjacent to the solid nodes to achieve the desired boundary condition. The interpolation of the velocity and the spreading of the Lagrange force are respectively expressed as
\begin{equation}
\boldsymbol{U}_{ib} (s, t) = \int_{V} \boldsymbol{u} (x, t) \delta_h (\boldsymbol{X}(s,t) - \boldsymbol{x}) d \boldsymbol{x},
\label{eq:ibvelocity}
\end{equation}
\begin{equation}
\boldsymbol{f} (\boldsymbol{x}, t) = -\int_{\Gamma} \boldsymbol{F}_f (\boldsymbol{X}, t) \delta_h (\boldsymbol{X}(s,t) - \boldsymbol{x}) d\boldsymbol{X},
\label{eq:fluidforce}
\end{equation}
where $\boldsymbol{u}$ is the fluid velocity, $\boldsymbol{X}$ is the coordinates of solid nodes, $\boldsymbol{x}$ is the coordinates of fluid, $V$ and $\Gamma$ are respectively the is the fluid and solid domain, and $\delta_h$ is the smoothed Dirac delta function~\citep{peskin2002immersed}, which is expressed as
\begin{equation}
\delta_h(x,y,z) = \frac{1}{h^2} \phi(\frac{x}{h}) \phi(\frac{y}{h}) \phi(\frac{z}{h}).
\label{eq:phifun}
\end{equation}

Here, the four-point delta function introduced by Peskin~\citep{peskin2002immersed} is used
\begin{equation}
\phi (r) =
\begin{cases}
\frac{1}{8} (3 - 2 |r| + \sqrt{1 + 4|r| - 4|r|^2}), & 0\leq|r|<1 \\
\frac{1}{8} (5 - 2 |r| - \sqrt{-7 + 12|r| - 4|r|^2}), & 1\leq|r|<2 \\
0, & 2\leq|r|.
\end{cases}
\label{eq:deltfun}
\end{equation}

In addition to the fluid--structure coupling, the rarefied gas flow sometimes also involves heat transfer, such as the thermal effects of hypersonic flight. To address the thermal coupling between the fluid and the solid, a similar pIB method is adopted. Specifically, the heat transferred from the immersed boundary to the fluid can be expressed as~\cite{wang2018heat}
\begin{equation}
q=\int_{\Gamma} Q (s, t) \delta_h (\boldsymbol{X}(s,t) - \boldsymbol{x}) d\boldsymbol{X},
\label{eq:heatim}
\end{equation}
where $Q$ is the heat flux transferred from the immersed boundary to fluid. When the temperature boundary condition is used, then $Q$ can be calculated by using penalty immersed boundary method, which can be expressed as
\begin{equation}
Q=\alpha_T (T_{ib} - T_w), 
\label{eq:ibq}
\end{equation}
where $\alpha_T$ is a large factor, $T_w$ is the real temperature of the solid, and $T_{ib}$ is the temperature of interpolated from the fluid domain, which can be expressed as
\begin{equation}
T_{ib} (s, t) = \int_{V} T (x, t) \delta_h (\boldsymbol{X}(s,t) - \boldsymbol{x}) d \boldsymbol{x}.
\label{eq:ibtempr}
\end{equation}

In the rarefied gas flow, the gas could slip over a surface and the inequalities of temperature also gives rise to a force driving the gas sliding over a surface from colder to hotter regions. Therefore, the no-slip boundary condition is not valid at the fluid--structure interface, the velocity and temperature jump boundary conditions are desired. The slip velocity in the kinetic theory is given by~\cite{hadjiconstantinou2003comment}
\begin{equation}
u_s = c_1 \lambda \frac{\pl u}{\pl n}|_w + c_2 \lambda^2 \frac{\pl^2 u}{\pl^2 n}|_w + u_w,
\label{eq:slipvel}
\end{equation}
where $\lambda$ is the mean free path of the molecule, $n$ is the normal direction pointing from the wall into the fluid, the subscript $w$ indicates the variables at the wall, $u_w$ is the velocity of the wall, $c_1$ and $c_2$ are parameters of the slip model. The different choice of $c_1$ and $c_2$ leads to various slip models. Here, the slip model proposed by Wu~\cite{wu2008slip} is adopted in this study. This is a second-order slip model, which is suitable for gas flow at all Knudsen numbers and it is given by 
\begin{equation}
u_s = \frac{2}{3} [\frac{3 - \sigma_s s_f^3}{\sigma_s} - \frac{3}{2}\frac{1 - s_f^2}{Kn}] \lambda \frac{\pl u}{\pl n}|_w - \frac{1}{4}[s_f^4 + \frac{2}{Kn^2} (1 - s_f^2)] \lambda^2 \frac{\pl^2 u}{\pl n^2}|_w,
\label{eq:slipwu}
\end{equation}
where $\sigma_s$ is the accommodation coefficient which represents the portion of total wall-colliding molecules that are diffusively reflected back by wall and have a bulk velocity equal to the wall velocity after collision, and $s_f = min(1/Kn, 1.0)$. 
%As the continuum N-S equations are not valid for $Kn > 1.0$, only $Kn < 1.0$ is considered here and Eq.~\eqref{eq:slipwu} reduces to
%\begin{equation}
%u_s = \frac{6 - 2 \sigma_s}{3 \sigma_s} \lambda \frac{\pl u}{\pl n} - \frac{1}{4} \lambda^2 \frac{\pl u}{\pl n}.
%\end{equation}
It should be noted that the slip velocity is calculated on the local coordinate system in the tangential direction. Similarly, the temperature at the wall with the slippery modification is given by
\begin{equation}
T_w^s = \frac{2 - \sigma_t}{\sigma_t} \frac{2 \gamma}{\gamma + 1} \frac{1}{Pr} \lambda \frac{\pl T}{\pl n}|_w + T_w,
\label{eq:sliptemp}
\end{equation}
where $s_t$ is the energy accommodation coefficient~\cite{fan2001computation,le2015rarefied}. $\sigma_s = 1.0$ and $\sigma_t = 1.0$ are used in all simulations as in Refs.~\cite{le2015rarefied}.

\begin{figure}
  \begin{center}
  \includegraphics[width=3in]{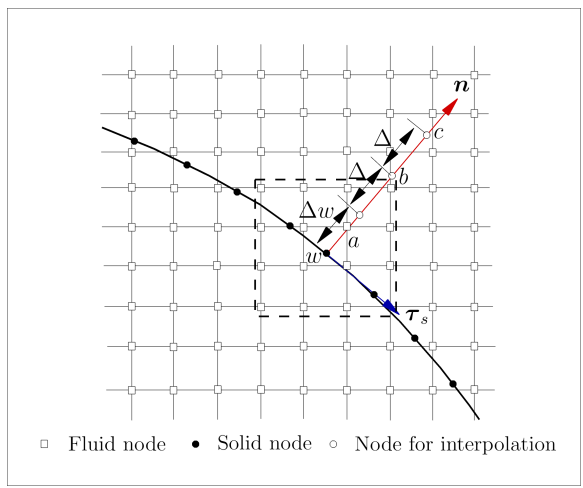}
  \end{center}
\caption{Schematic of the velocity interpolation in the slip model.}
\label{Fig:slpsch}
\end{figure}

To incorporate the velocity slip boundary conditions into the IBM framework, interpolation is necessary for the calculation of velocity derivatives and thus the slip velocity. A schematic of the interpolation is shown in Fig.~\ref{Fig:slpsch}, where 2 points ($a$ and $b$) are required to calculate the first-order normal derivatives and 3 points ($a$, $b$ and $c$) are required to calculate the second-order derivatives. The velocities at these points, as shown in Fig~\ref{Fig:slpsch}, can be interpolated either by using inverse distance function or weighted least square interpolation~\cite{capizzano2011turbulent}, here the former one is unitized. Once the velocities at these interpolation points are obtained, a project procedure is required to obtain the normal components, i.e., $u^n = \boldsymbol{u} \cdot \boldsymbol{n}$ with $\boldsymbol{n}$ being the normal vector at the wall, then the normal derivatives can be calculated as 
\begin{equation}
\frac{\pl u}{\pl n} = \frac{u^n_b - u^n_a}{\Delta}, \quad \frac{\pl^2 u}{\pl^2 n} = \frac{u^n_c - 2 u^n_b + u^n_a}{\Delta^2}.
\label{eq:normalderivat}
\end{equation}
It is noted that the derivatives at the solid point $w$ is evaluated at a point $a$ with a normal distance of $\Delta w$. Such treatment is employed to enhance the numerical stability. As the slip-boundary conditions at the fluid--structure interface is only approximately fulfilled in the IBM, the interpolation of velocity at point $w$ involving non-fluid nodes presents numerical fluctuations at high Mach numbers. Here, $\Delta w = 0.5 \Delta x \sim \Delta x$ is used, with $\Delta x$ being the fluid mesh spacing, which shows excellent robustness according to the validations considered.

%There are two reasons to implement such treatment, i.e., maintain the exact boundary condition and enhance the numerical stability. A schematic of the velocity profile at the boundary layer is shown in Fig.~\ref{Fig:knvprofsch}~\cite{guo2008lattice}. As the NSE results are not able to reproduce the exact velocity profiles at the Knudsen layer, using the velocities of fluid nodes located in the Knudsen layer may lead to errors in the calculation of the derivatives. On the other hand, the slip-boundary conditions at the fluid--structure interface is only approximately fulfilled in the IBM, therefore the interpolation of velocity at point $w$ involving non-fluid nodes shows numerical fluctuations at high Mach numbers. Here, $\Delta w = 0.5 \Delta x \sim \Delta x$ is used, with $\Delta x$ being the fluid mesh spacing, which shows excellent robustness according to the validations considered.

%\begin{figure}
%  \begin{center}
%  \includegraphics[width=3in]{./Figs/knvelprof.eps}
%  \end{center}
%\caption{Schematic of the velocity profiles at the boundary layer.}
%\label{Fig:knvprofsch}
%\end{figure}

Substitute Eq.~\eqref{eq:normalderivat} into Eq.~\eqref{eq:slipvel}, the slip velocities in the two tangential directions ($\boldsymbol{\tau}_s^1$ and $\boldsymbol{\tau}_s^2$) denoted by $\boldsymbol{U}_s^1$ and $\boldsymbol{U}_s^2$ can be obtained. Then, the slip velocities calculated on the tangential directions need to be projected back onto the overall Cartesian coordinate system. Due to the slip boundary conditions at the fluid--solid interface, the structure velocity $\boldsymbol{U}$ in Eq.~\eqref{eq:penatly} need to be reconstructed to include the slippery effects as 
\begin{equation}
\boldsymbol{U}^{s} = \boldsymbol{U} + \boldsymbol{U}_{s}^1 \cdot \boldsymbol{\tau}_s^1 + \boldsymbol{U}_{s}^2 \cdot \boldsymbol{\tau}_s^2.
\label{eq:uibslip}
\end{equation} 
This equation indicates that the fluid in the ambient of the structure is not moving with the velocity of structure, i.e., $\boldsymbol{U}$ (which is the commonly used no-slip boundary condition), the fluid is moving with the modified velocity of structure, i.e., $\boldsymbol{U}^{s}$. By replacing $\boldsymbol{U}$ in Eq.~\eqref{eq:penatly} with $\boldsymbol{U}^{s}$ calculated by using Eq.~\eqref{eq:uibslip}, the boundary conditions with slippery effects can be fulfilled. The temperature jump at the fluid--structure interface can be handled in the same way by replacing $T_{w}$ in Eq.~\eqref{eq:ibtempr} with $T_w^s$ calculated by using Eq.~\eqref{eq:sliptemp}. It should be noted that the proposed amendment of the original pIB can be easily incorporated into other fluid solvers such as FVM and LBM to take the rarefied gas effects into consideration. 

The pIB approximates the boundary condition by applying body force which introduces errors at the fluid-structure interfaces, and this error is pronounced when slip velocity is introduced. Such errors can be minimized by increasing the values of $\alpha$, $\beta$ and $\alpha_T$ in Eqs.~\eqref{eq:penatly} and \eqref{eq:ibtempr}, but it will inevitably introduce instabilities to the numerical simulation, especially for moving geometries in highly compressible flow with slip-velocity boundary conditions. Alternatively, an explicit iterative procedure is adopted for moving geometries in this work instead of the original pIB in Ref.~\cite{wang2017immersed}. In the iterative IBM, the IB force from the displacement is not valid, i.e., $\alpha = 0$, Eq.~\eqref{eq:penatly} is iteratively used to obtain the IB force which is distributed to the fluid until the expected error at the fluid-structure interface is achieved or maximum iterations reached. Here, the converge criteria is set as $10^{-3} U_{ref}$ for ($\boldsymbol{U}_{ib}$ - $\boldsymbol{U}$) and $10^{-3} T_{ref}$ for $T - T_{ib}$ with $U_{ref}$ being the reference velocity and $T_{ref}$ being the reference temperature, and 10 iterations maximum is used. The accumulated force is then applied to the solid when solving the solid dynamics.

%%%%%%%%%%%%%%%%%%%%%
\section{Validations}
%%%%%%%%%%%%%%%%%%%%%
\subsection{Poiseuille flow in a pipe}
In this section, the Poiseuille flow in a 2D pipe is considered. The pipe has a size of $2h \times 1.2h$, with $h$ being the height of the pipe. The extra nodes in the transversal direction are used for the implementation of IB method. The wall of the pipe is represented by solid nodes, and the slip boundary conditions considering the rarefied gas effects are achieved by the aforementioned IB method and slip model. The mesh spacings of the fluid and solid domain are respectively $h/100$ and $h/200$. A velocity boundary condition is applied at the left inlet with a parabolic velocity profile, i.e., $u = u_0 [1-(y/h)]^2$, where $u_0$ is the maximum velocity and $y$ is the vertical coordinate. The zero velocity gradient and pressure boundary condition is used for the right outlet. The governing parameter of this physical problem is $Kn = \lambda / h$ and $M = u_0/c$ with $c$ being the sound speed. Here, two Knudsen numbers, i.e., 0.04 and 0.2 are considered, where $Kn=0.04$ is in the slip regime and $Kn=0.2$ is in the transitional regime. A small Mach number of 0.05 is used, thus the fluid is nearly incompressible and the thermal effects are not considered.

\begin{figure}
  \begin{center}
  \includegraphics[width=3in]{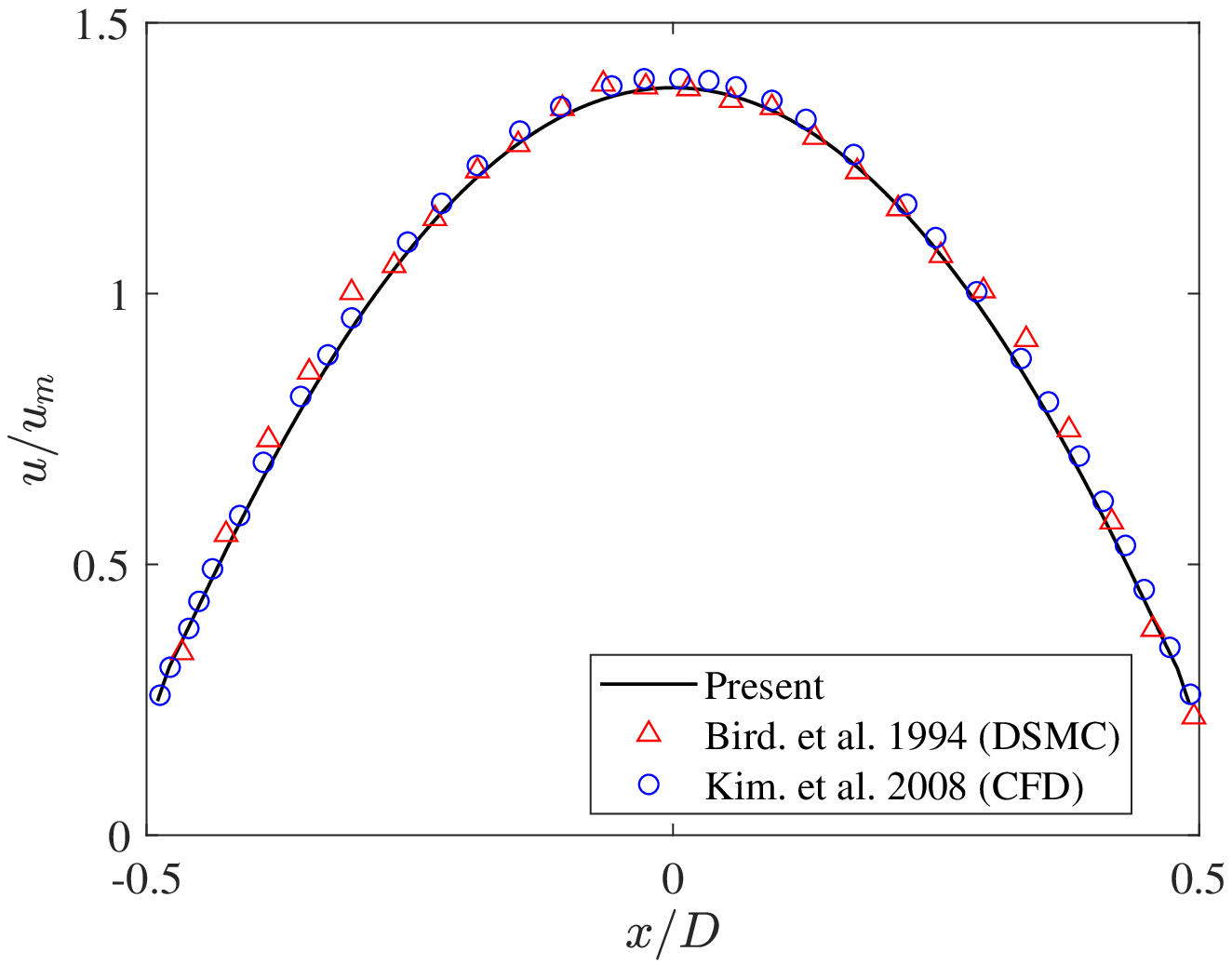}
  \includegraphics[width=3in]{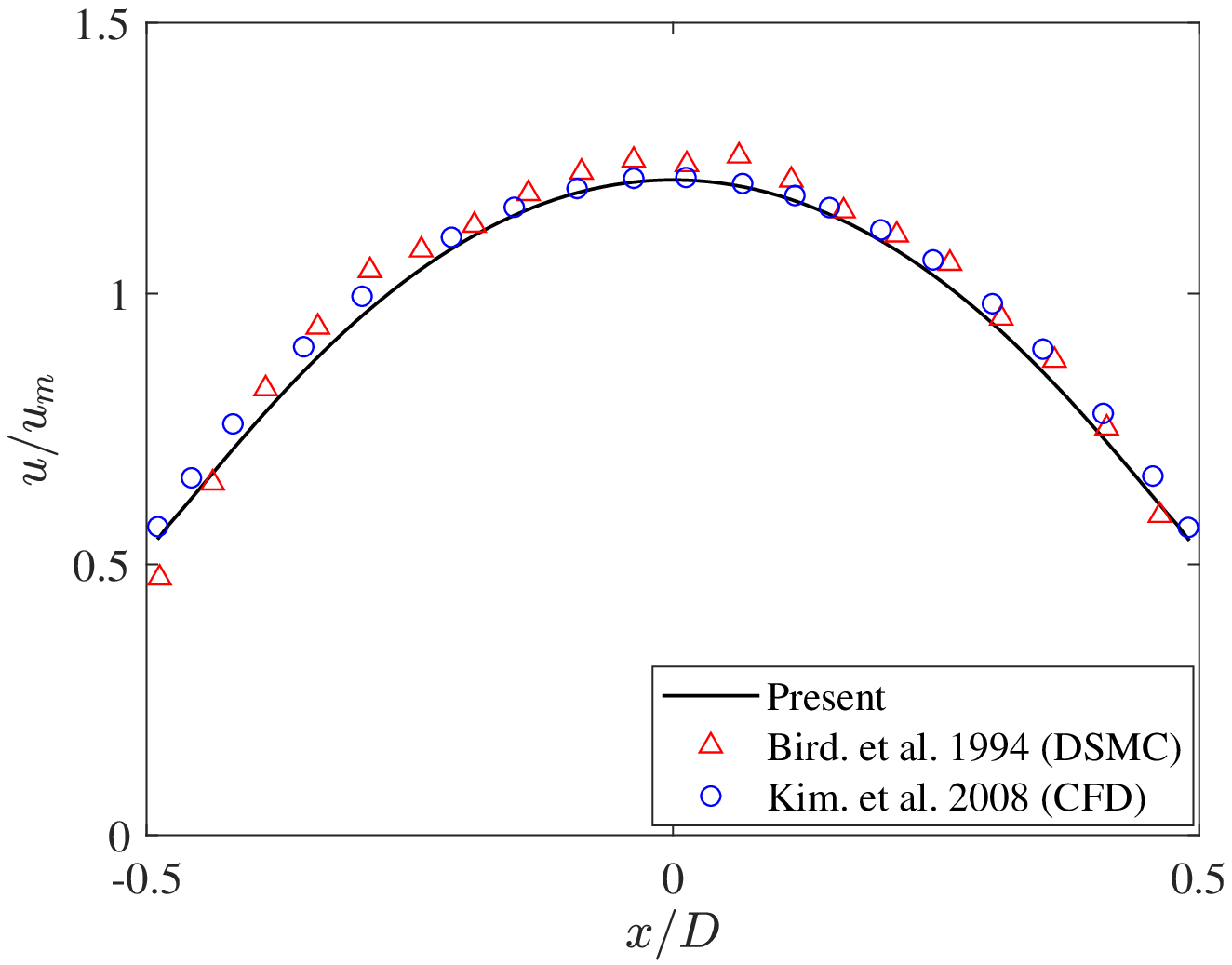}
  \end{center}
\caption{Stream-wise velocity profile of Poiseuille flow in a 2D pipe at $kn=0.04$ (left) and 0.2 (right).}
\label{Fig:pfvelprof}
\end{figure}

%\begin{figure}
%  \begin{center}
%  \includegraphics[width=3in]{./Figs/pipe2dconvg.eps}
%  \end{center}
%\caption{Mass ratio convergence of Poiseuille flow in a 2D pipe at $Kn=0.04$.}
%\label{Fig:pfmc}
%\end{figure}

The velocity profiles computed by using the present method and those by using DSMC and solving NS equations~\cite{kim2008slip} are shown in Fig.~\ref{Fig:pfvelprof}, where the velocity is normalized by the mean velocity at the inlet, i.e., $u_m = \frac{1}{h}\int_{-0.5h}^{0.5h} u dy$. It shows that the present results agree well with the published data, which confirms the good accuracy of the present method. 

%To further quantify the performance of the present method, a mesh convergence study is conducted considering $kn = 0.04$. Another four fluid mesh spacings, i.e., $h/200$,  $h/80$, $h/50$ and $h/40$, are examined with the results obtained from the finest mesh $h/200$ are used as the reference solution. It should be noted that the mesh spacing of the solid is kept as half of the fluid mesh spacing in this work unless otherwise stated. The convergence of normalized mass ratio ($Q_m = \frac{1}{u_m k_n}\int_{-0.5h}^{0.5h} u dy$) is shown in Fig.~\ref{Fig:pfmc}, it is found that the convergence performance of the present method is first-order, which is consistent with the previous observations of penalty IB method. In addition, a higher convergence is observed when the mesh is refined from $h/40$ to $h/50$, which is attributed to the poor mesh resolution to resolve the height of the pipe.

\subsection{Flow around a NACA airfoil}
After the validation of physical problems involving internal flow, we further consider external flow problems by examining flow around a NACA0012 airfoil in 2D domain. This problem involves fluid--structure--thermal interaction and has been considered in previous studies by using both DSMC and slip velocity model, which is a good validation for the present method. A non-uniform mesh is used to discretize the fluid domain which has a size of $30L \times 30L$, with $L$ being the airfoil chord length. The finest fluid mesh spacing is $L/400$, and it has a size of $2L \times 0.3L$ which is large enough to contain the airfoil. The governing parameters of this problem are Knudsen number and Mach number, which are defined as
\begin{equation}
Kn = \frac{\lambda}{L}, \quad M = \frac{u_0}{c},
\end{equation}
where $u_0$ is the velocity of incoming flow. The Reynolds number can be calculate as $Re = \frac{M}{Kn}\sqrt{0.5\gamma \pi}$. The fluid domain is initialized with a uniform flow, velocity inlet boundary condition is applied on the left and zero-gradient boundary conditions are used for the lateral and outlet. The airfoil is fixed in the fluid and its surface has a constant temperature $T_w$. Both the velocity and the temperature boundary conditions at the airfoil surface are achieved by the IB method. The simulation is conducted at $Kn = 0.014$ and $M=0.8$, and the details of the fluid properties are shown in Tab.~\ref{table:nacamat}. 

\begin{table}
 \caption{Fluid properties and flow conditions of flow around a NACA airfoil.}
 \label{table:nacamat}
\begin{center}
\renewcommand{\arraystretch}{1.5}
\setlength\tabcolsep{10pt}
\begin{tabular}{ccccccc}
  \hline
    $p_0 (Pa)$ &$\rho (kg/m^3)$ &$T_0 (K)$  & $T_w (K)$ & $u_0 (m/s)$ & $M$ & $Kn$\\
  \hline
   	%2.78443 & $6.026\times10^{-5}$ & 161 &290 &509 &2.0 & 0.026\\
    8.23150 & $1.116\times10^{-4}$ & 257 &290 &257 &0.8 & 0.014\\
  \hline
\end{tabular}
\end{center}
\end{table}

\begin{figure}
  \begin{center}
  \includegraphics[width=3in]{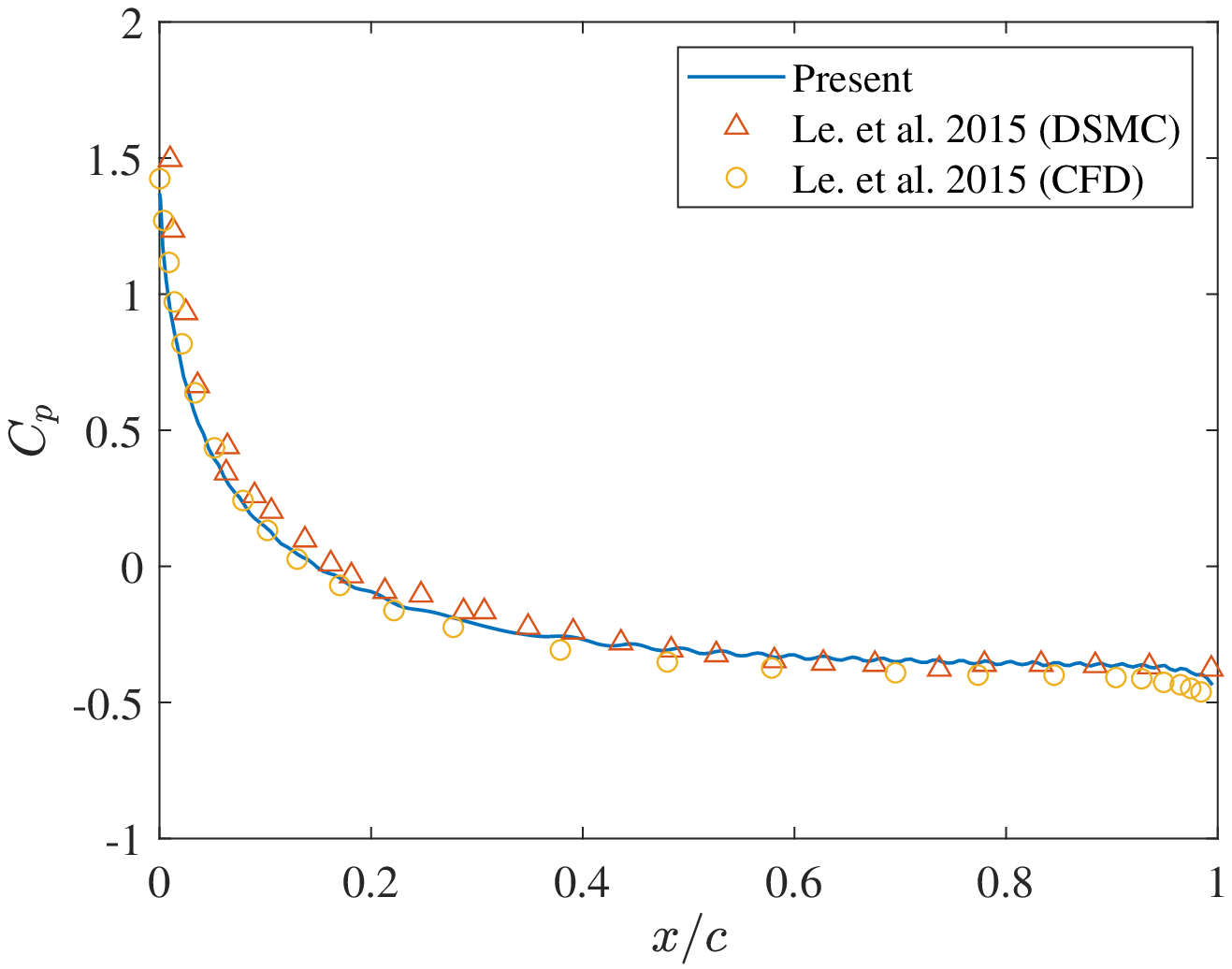}
  \includegraphics[width=3in]{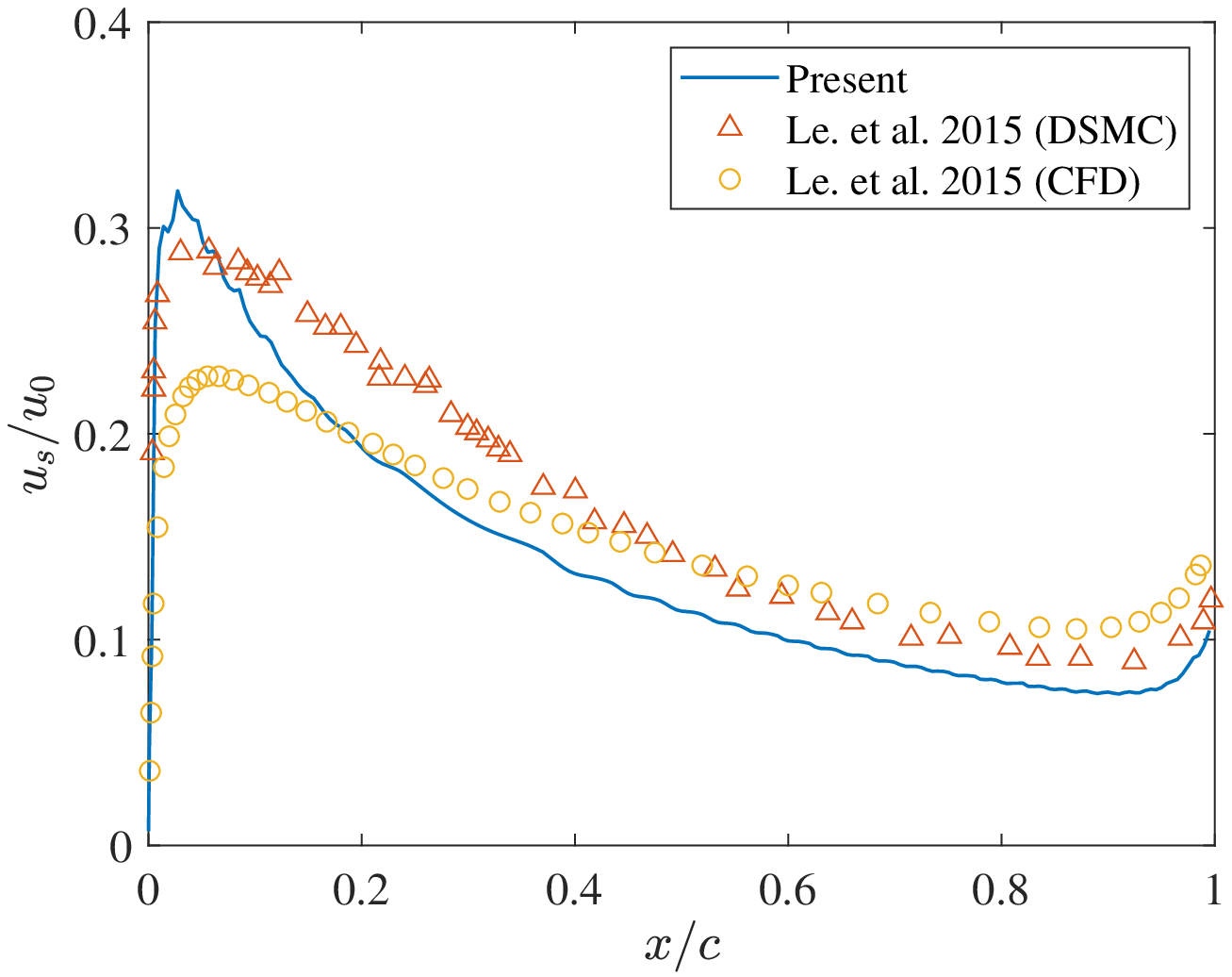}\\
  \includegraphics[width=3in]{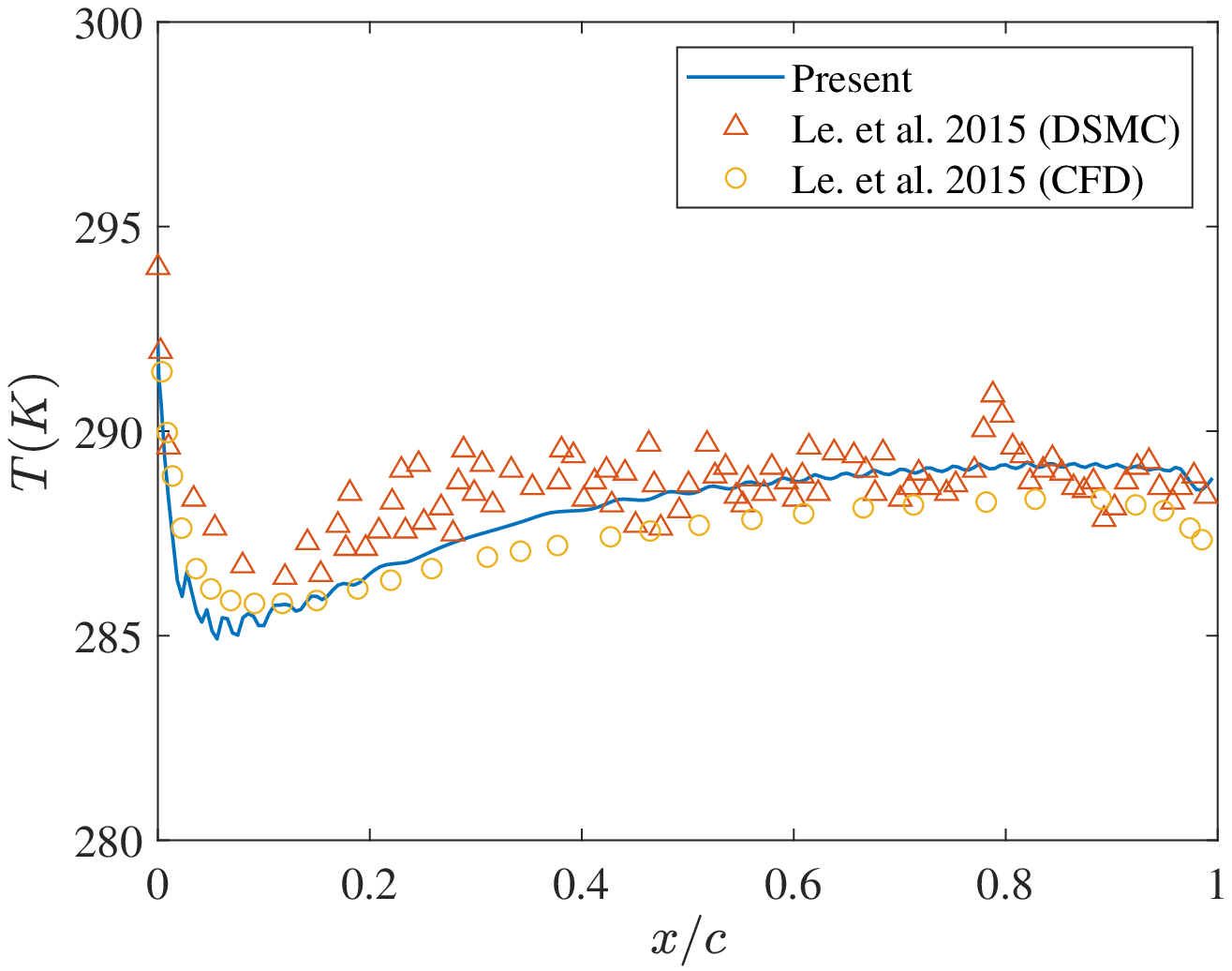}
  \includegraphics[width=3in]{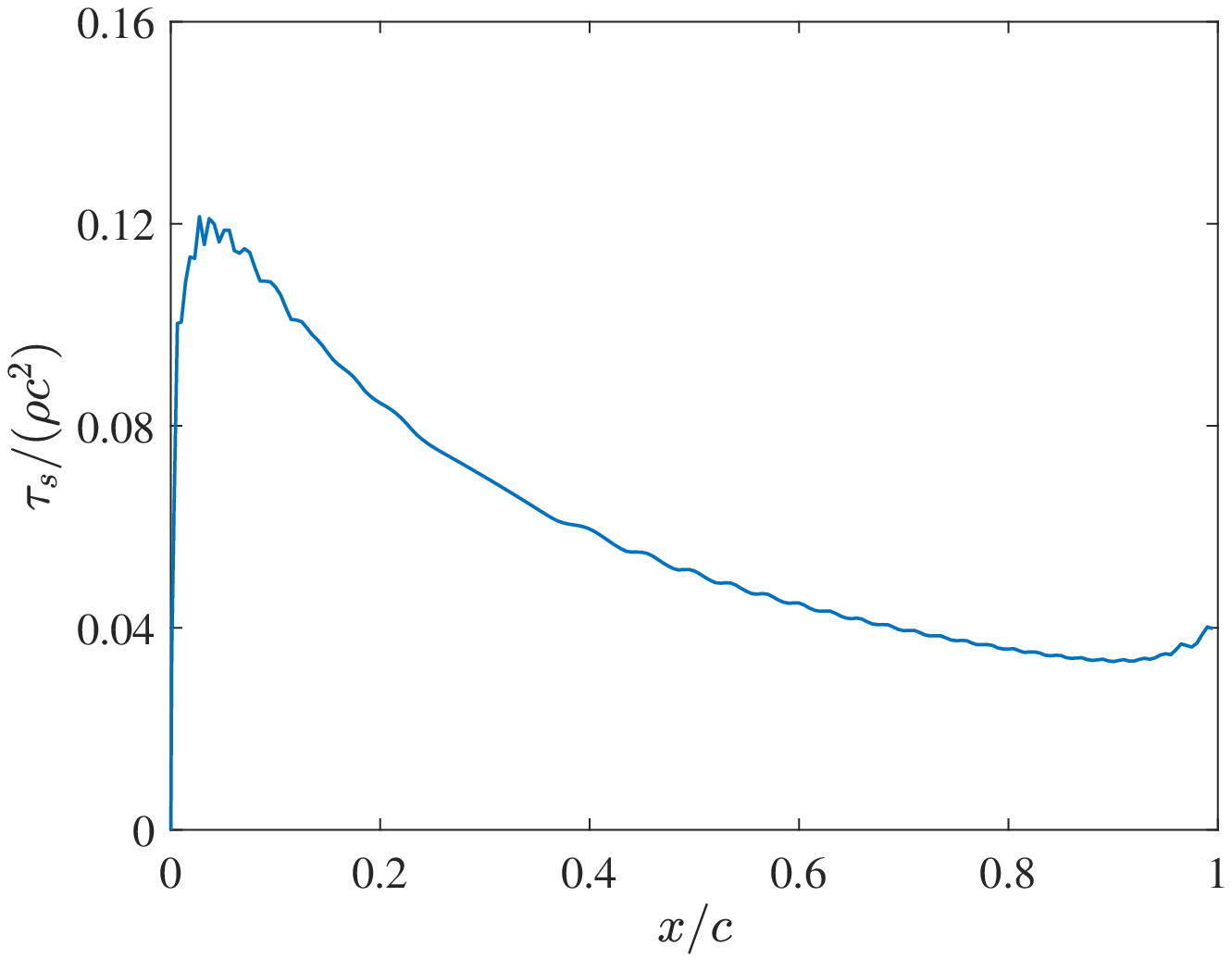}
  \end{center}
\caption{Pressure coefficient, slip velocity and temperature along the airfoil surface at $M = 0.8$ and $Kn = 0.014$.}
\label{Fig:naca2dfr}
\end{figure}

\begin{figure}
  \begin{center}
  \includegraphics[width=3in]{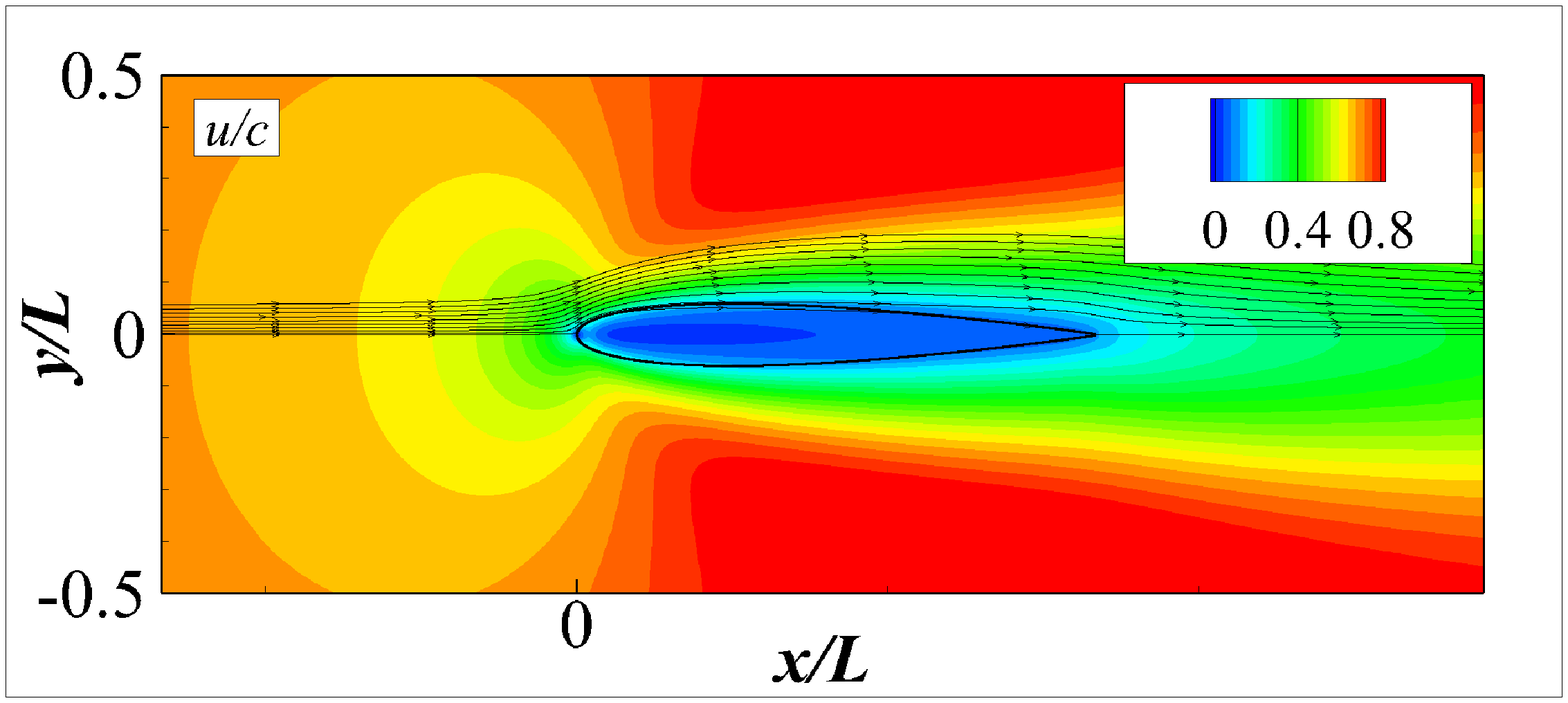}
  \includegraphics[width=3in]{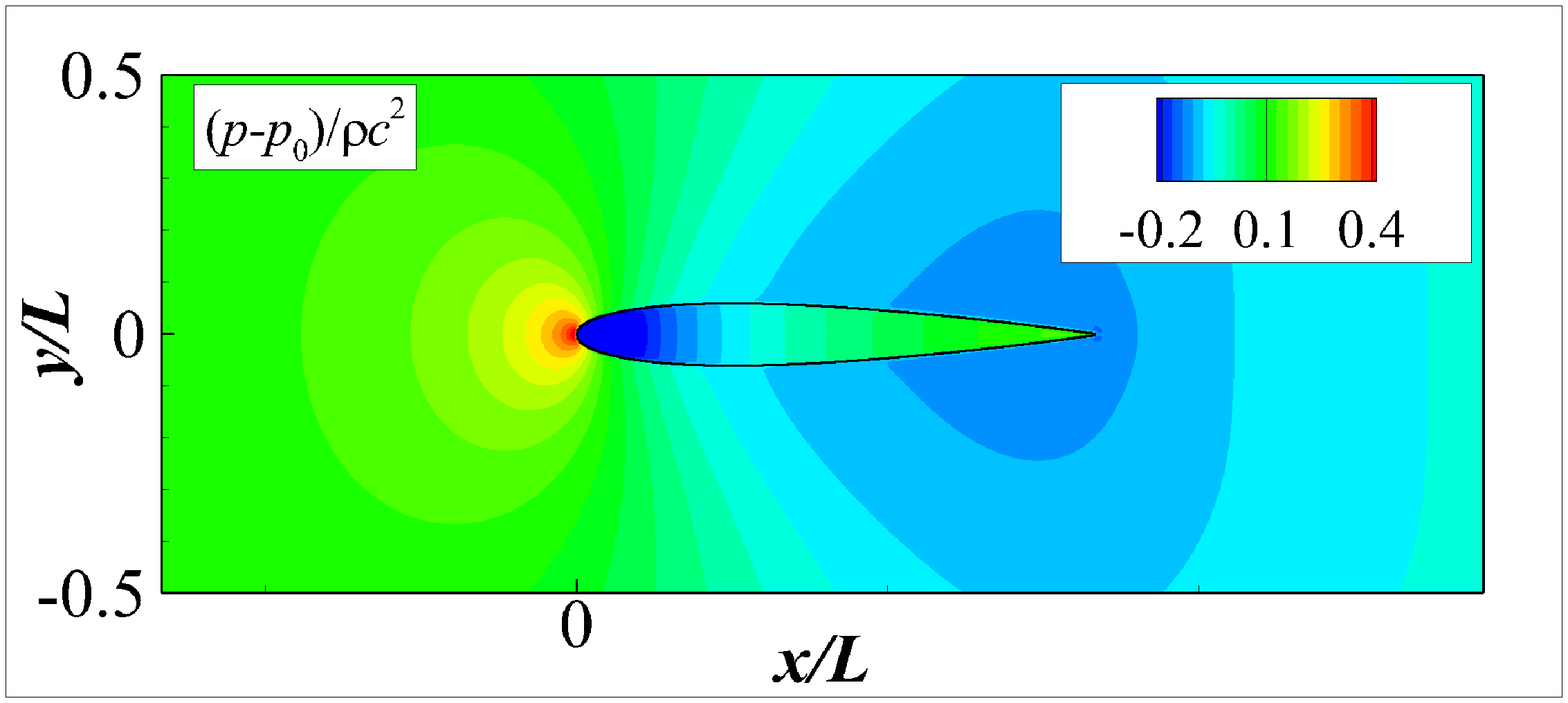}\\
  \includegraphics[width=3in]{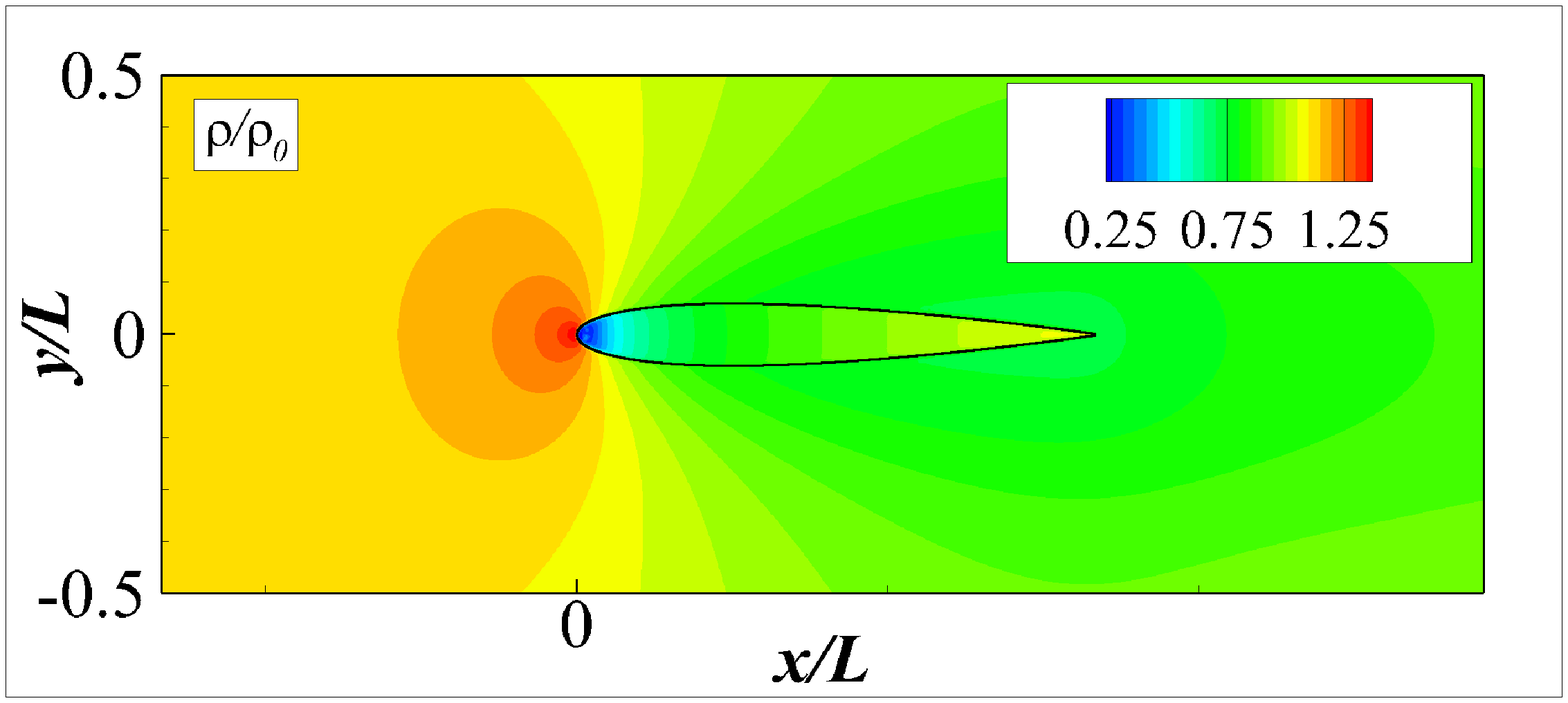}
  \includegraphics[width=3in]{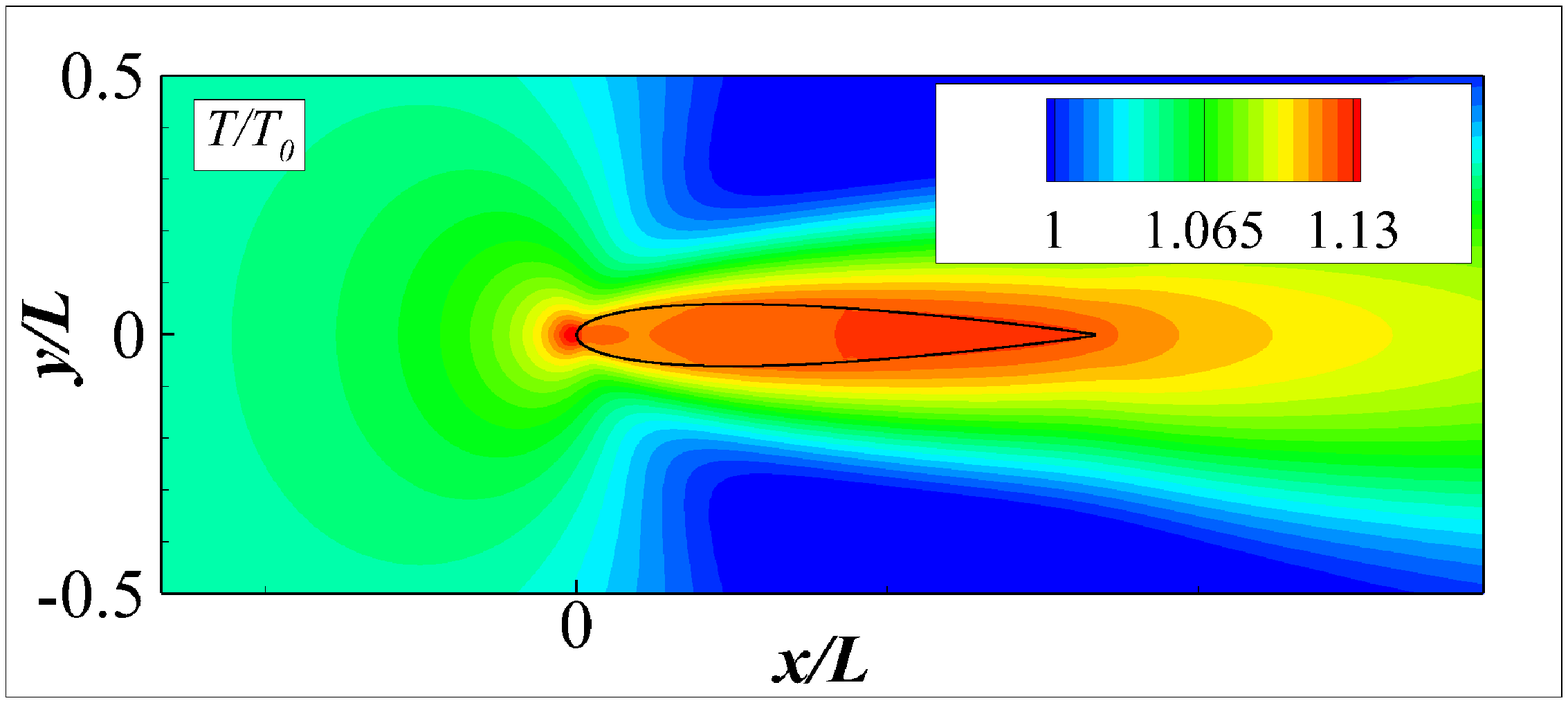}\\
  \end{center}
\caption{Stream-wise velocity, pressure, temperature and density contours of flow around a NACA airfoil with $M = 0.8$ and $Kn = 0.014$ at $tc/L = 5.0$.}
\label{Fig:nacacontour}
\end{figure}

Fig.~\ref{Fig:naca2dfr} shows the pressure coefficient ($C_p$), slip velocity ($u_s$), temperature and shear stress along the surface of the airfoil. The available data from Ref.~ obtained by DSMC and CFD is also included for comparison. It is found that the present results agree well with those published data, especially the $C_p$ and temperature. The slip velocity shows discrepancies, but the overall trend agree well with the DSMC and CFD simulation, the present results achieves a higher prediction of the slip velocity at the leading edge which is in better agreement with the DSMC results. These agreements indicate that the present method has good ability in handling fluid--structure--thermal interaction in rarefied gas flow. The instantaneous flow fields at $tc/L = 5.0$ are also shown in Fig.~\ref{Fig:nacacontour} for the future comparison.

\subsection{Moving square cylinder in a pipe}
\begin{figure}
  \begin{center}
  \includegraphics[width=3in]{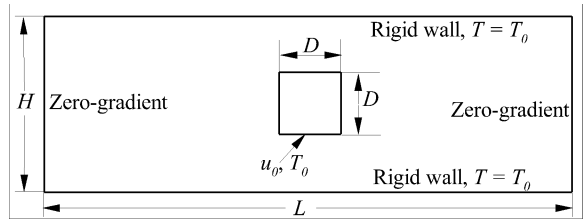}
  \end{center}
\caption{Schematic of a moving square cylinder in a pipe.}
\label{Fig:mssch}
\end{figure}

Here, a moving square cylinder in 2D domain is considered to demonstrate the superiority of the present method in handling fluid--structure interactions in rarefied gas flow. In this problem, the fluid in the pipe is at rest initially, a square cylinder is moving in the pipe with a fixed velocity $u_0 / c= 3.2$, as show in Fig.~\ref{Fig:mssch}. The computational domain has a size of $20D \times 3.2D$, with $D = 2.49\times 10^{-6} m$ being the width of the square cylinder. At $t=0$, the cylinder is located at a distance of $D$ from the left boundary. Both the rigid pipe and the cylinder have a constant temperature $T_w = T_0$ with $T_0$ being the initial temperature of the fluid field. The velocity and temperature boundary conditions of the pipe and cylinder are achieved by the IB method, other boundary conditions are shown in Fig.~\ref{Fig:mssch}.

\begin{table}
 \caption{Fluid properties of a moving square cylinder in a pipe.}
 \label{table:sqcymat}
\begin{center}
\renewcommand{\arraystretch}{1.5}
\setlength\tabcolsep{10pt}
\begin{tabular}{ccccccc}
  \hline
    $p_0 (Pa)$ &$\rho (kg/m^3)$ &$T_0 (K)$  & $T_w (K)$ & $u_0 (m/s)$ & $M$ & $Kn$\\
  \hline
    $5.34132\times10^{4}$ & 0.64687 & 288 &288 &1187.5 &3.2 & 0.05\\
  \hline
\end{tabular}
\end{center}
\end{table}

\begin{figure}
  \begin{center}
  \includegraphics[width=3in]{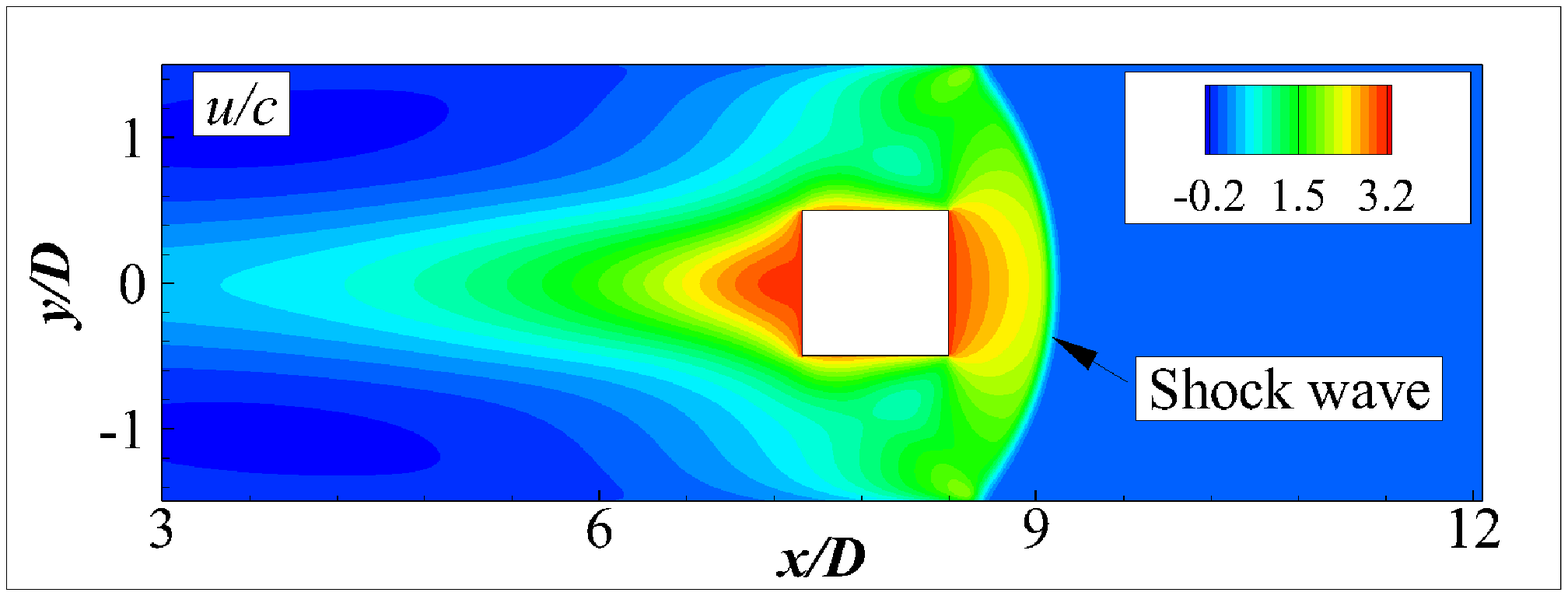}
  \includegraphics[width=3in]{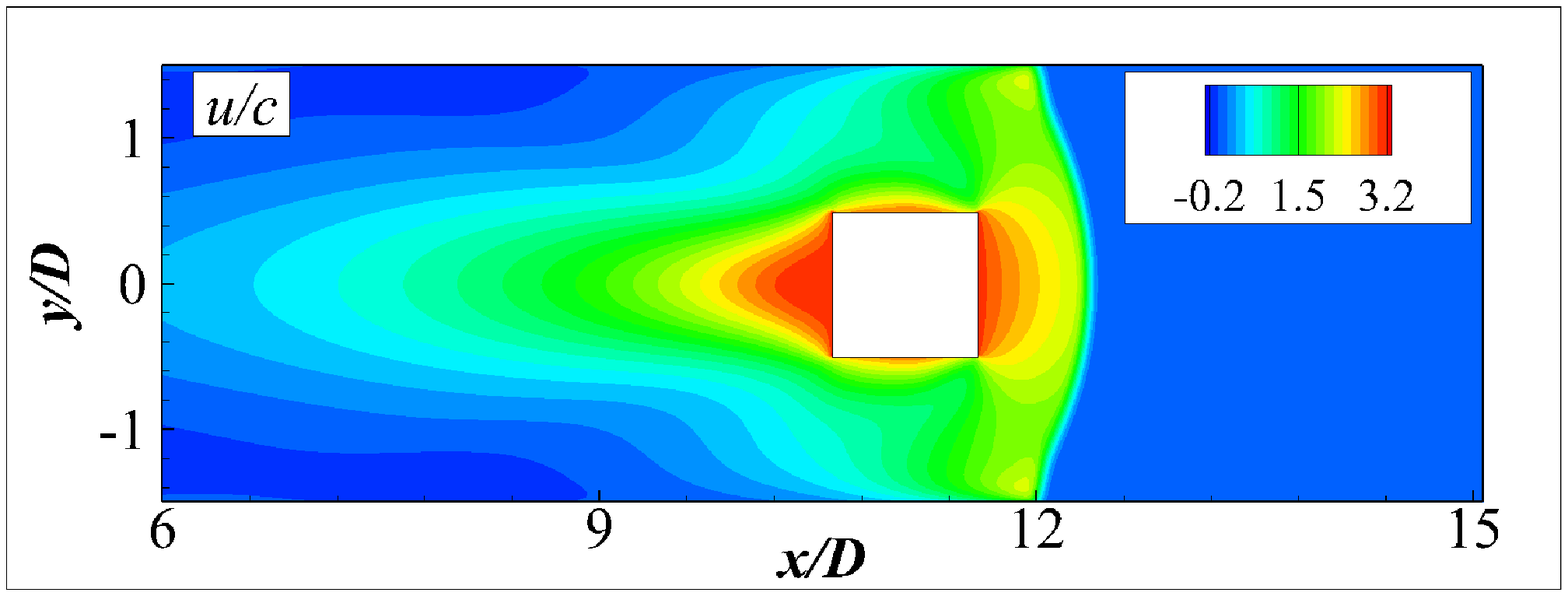}\\
  \includegraphics[width=3in]{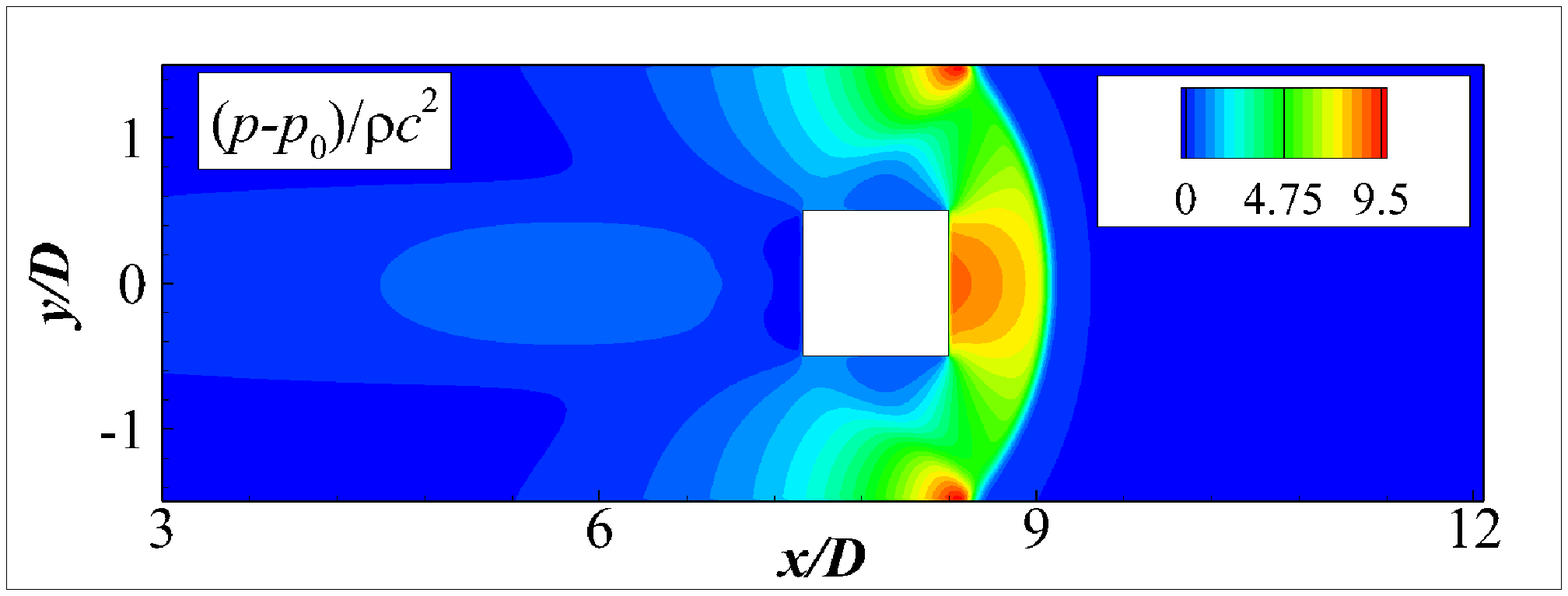}
  \includegraphics[width=3in]{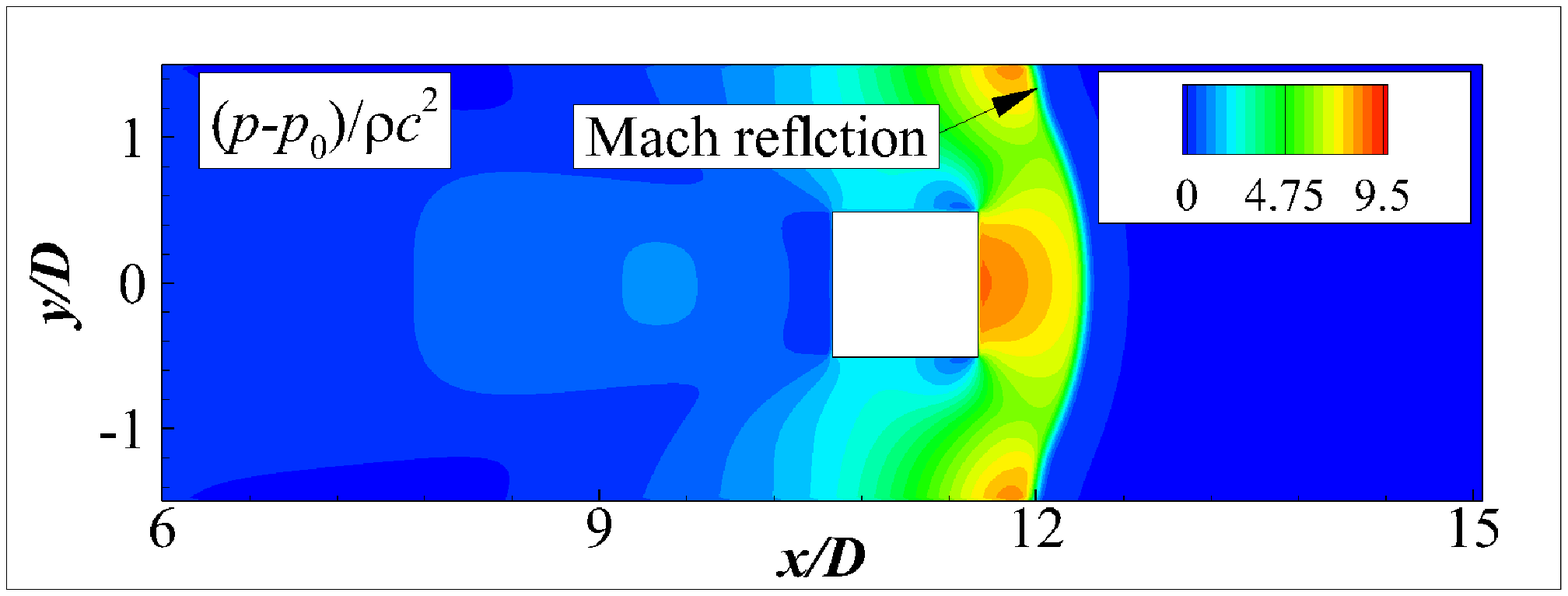}\\
  \includegraphics[width=3in]{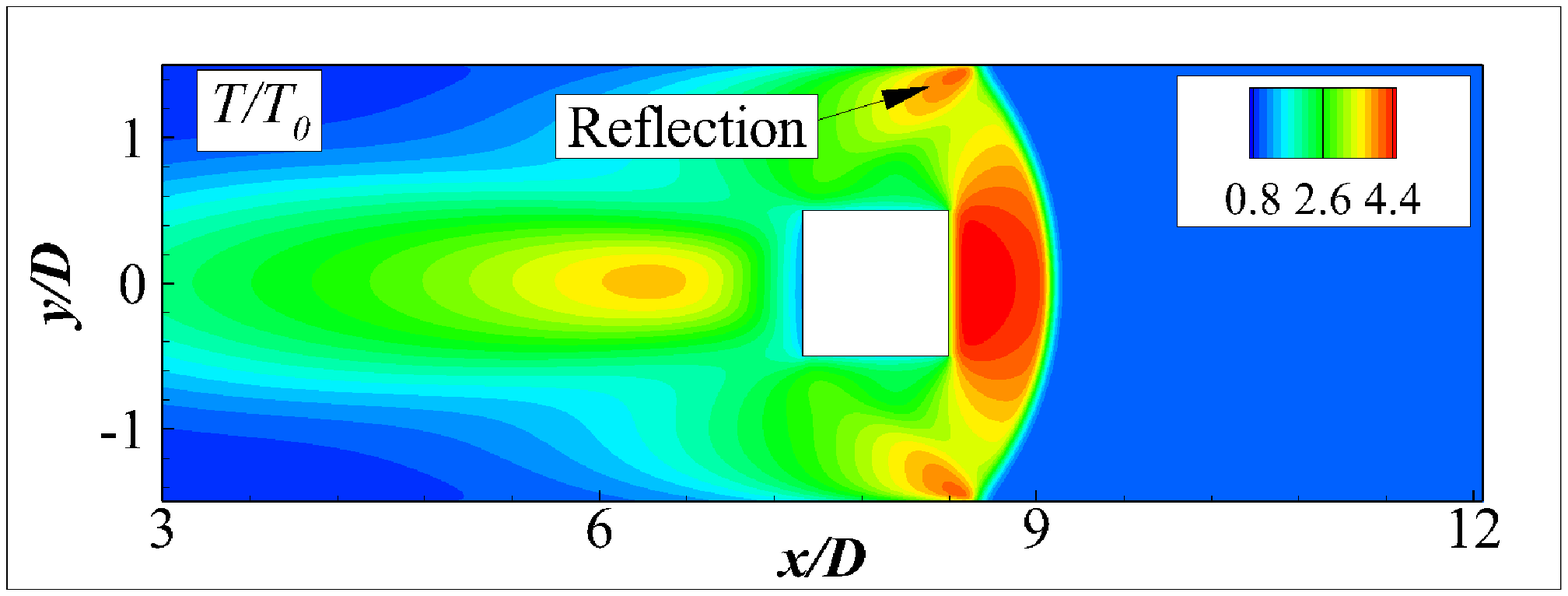}
  \includegraphics[width=3in]{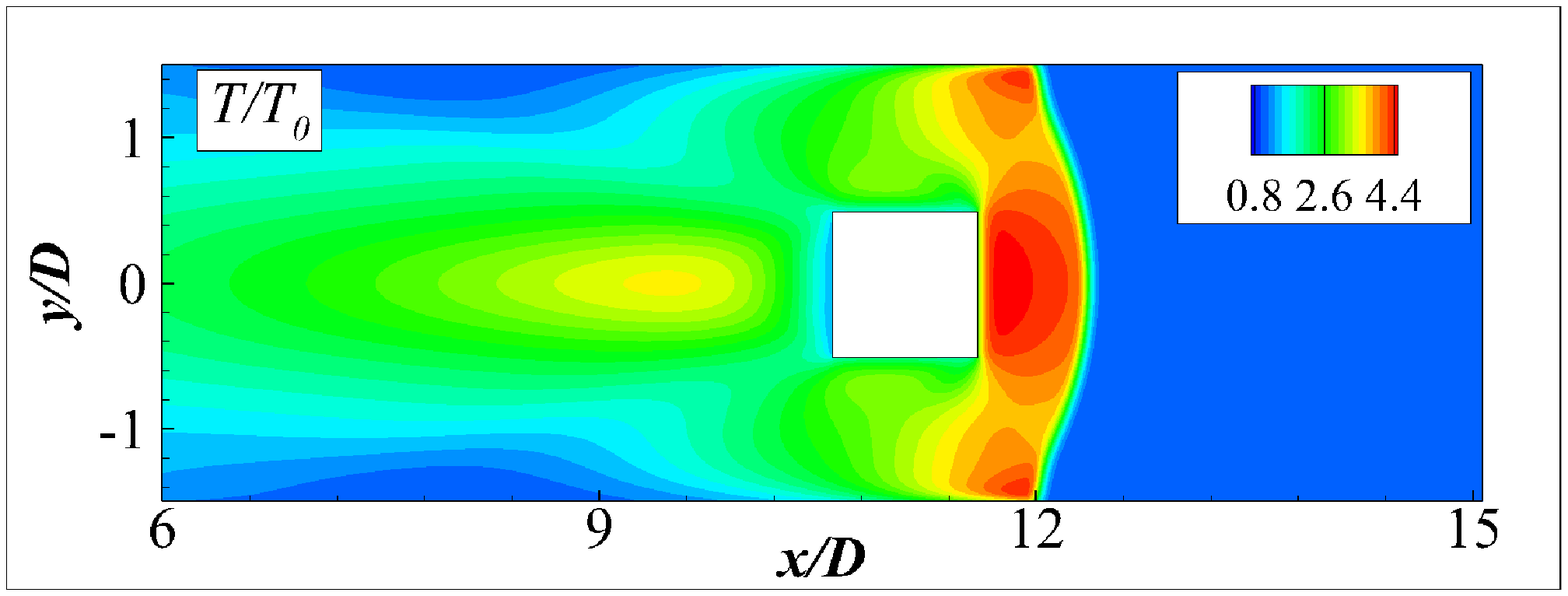}\\
  \includegraphics[width=3in]{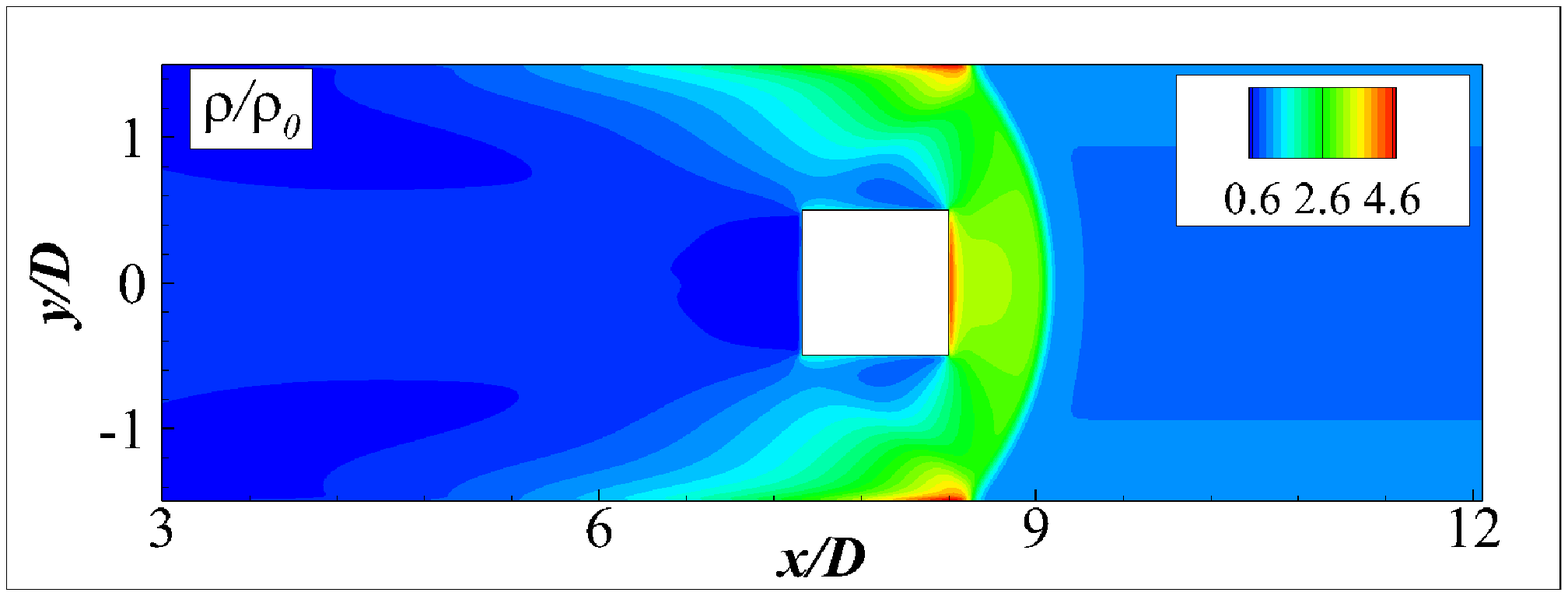}
  \includegraphics[width=3in]{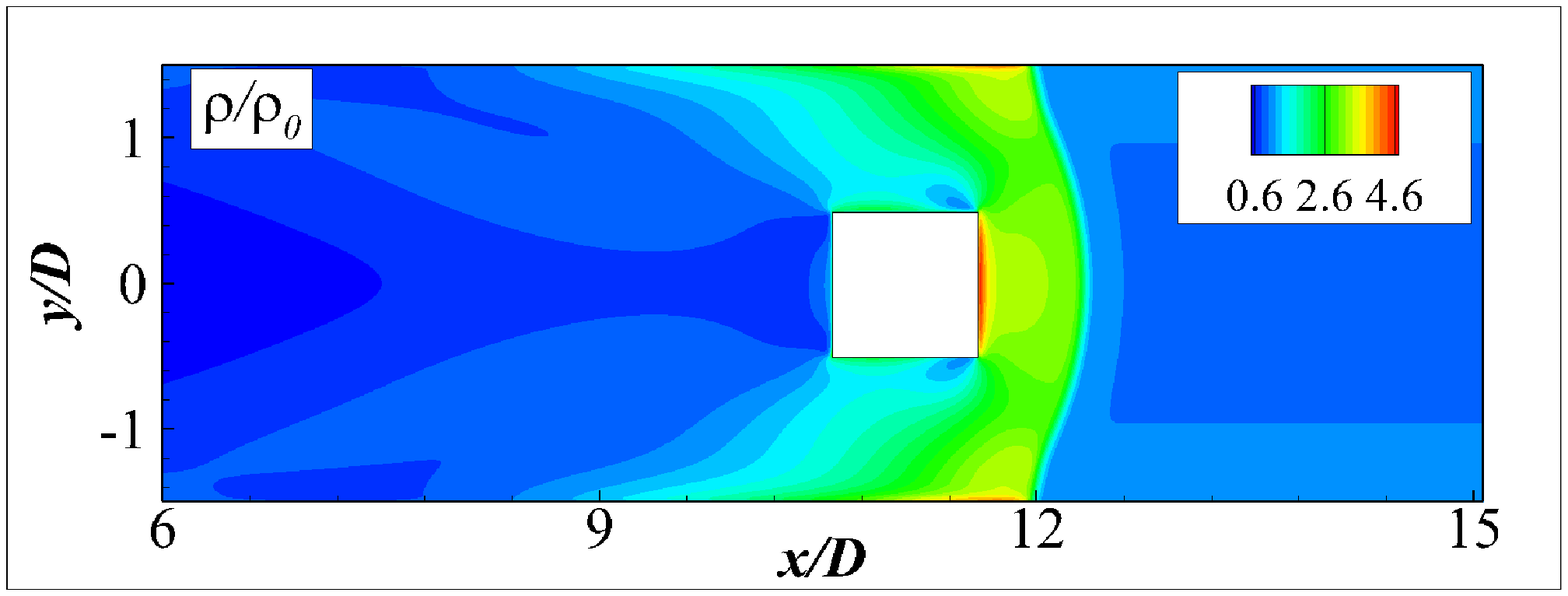}\\
  \end{center}
\caption{Stream-wise velocity, pressure, temperature and density contours of moving square cylinder with $M = 3.2$ and $kn = 0.05$ at $tc/D = 2.0$ (left column) and 3.0 (right column).}
\label{Fig:mscontour}
\end{figure}

\begin{figure}
  \begin{center}
  \includegraphics[width=3.5in]{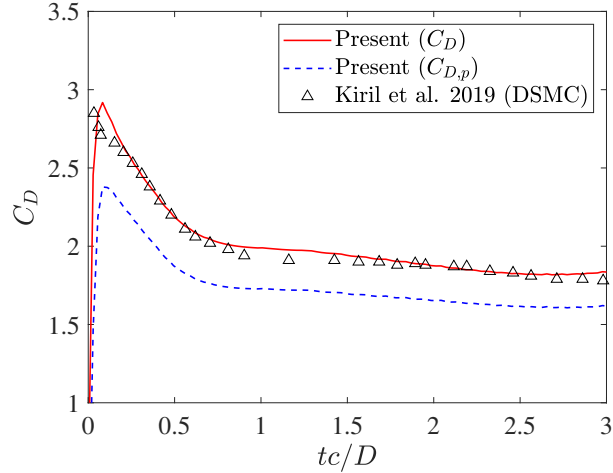}
  \end{center}
\caption{Time histories of the drag coefficient of a moving square cylinder at $M = 3.2$ and $kn = 0.05$.}
\label{Fig:mscd}
\end{figure}

Fig.~\ref{Fig:mscontour} shows the contours of stream-wise velocity, pressure, temperature and density at two instantaneous time. It shows that the detached shock wave in front of the cylinder and the reflections of the front shock wave on the rigid wall are well captured by the present simulation. With the forward movement of the cylinder, the the incident angles of the front shock waves on the rigid walls increase, and Mach reflections are thus observed as shown in the pressure contours at $tc/D = 3.0$. The time histories of the drag coefficient $$C_D = \frac{F_x}{0.5\rho u_0^2}$$ with $F_x$ being the force exerting on the cylinder) are shown in Fig.~\ref{Fig:mscd}, along with those data available in Ref.~\cite{shterev2019hybrid} obtained by DSMC. It is found that the present results agree well with that of DSMC, except the discrepancies at the early stage which is attributed to the IB method. The contributions of the pressure to the drag coefficients ($C_{D,p}$) are also shown in Fig.~\ref{Fig:mscd}, which can be used for the future validation for others. It is found that the pressure and shear stress contribute about 90\% and 10\% of the total drag.

\subsection{Supersonic flow around a sphere}
Flow around blunt body is a classical problem which has been extensively studied. The re-entry object is a typical example of this problem in the rarefied gas flow. Here, supersonic flow around a sphere is examined at $M = u_0/c = 3.834$ with $u_0$ and $c$ respectively being the free stream velocity and sound speed. This problem has been previously studied by using several different methods including discrete velocity method (DVM)~\cite{yang2019improved}, DSMC and experiments~\cite{vogenitz1968theoretical}. Knudsen number is defined based on the sphere diameter as $kn = D/\lambda = 0.03$. The variable hard sphere model is used to estimate the dynamic viscosity, i.e., $\mu = \mu_0 (T/T_0)^{\chi}$ with $\chi = 0.75$, where the reference viscosity is defined as
\begin{equation}
\mu_0 = \frac{5\rho_0 \lambda (2 \pi R_g T_0)^{0.5}}{16},
\label{eq:hsvis}
\end{equation} 
with $R_g$ being ideal gas constant. Prandtl number and specific heat ratio are respectively chosen as $Pr = 2/3$ and $\gamma = 5/3$. The temperature of the sphere is fixed at its stagnation value, i.e., $T_w = [1 + (\gamma - 1) M^2 / 2] T_0$. A mesh spacing of $0.02D$ is adopted to discretize the fluid domain ($16D \times 16D \times 16D$), and the sphere surface is discretized by triangles with a mean side length of $0.02D$. 

\begin{figure}
  \begin{center}
  \includegraphics[width=2in]{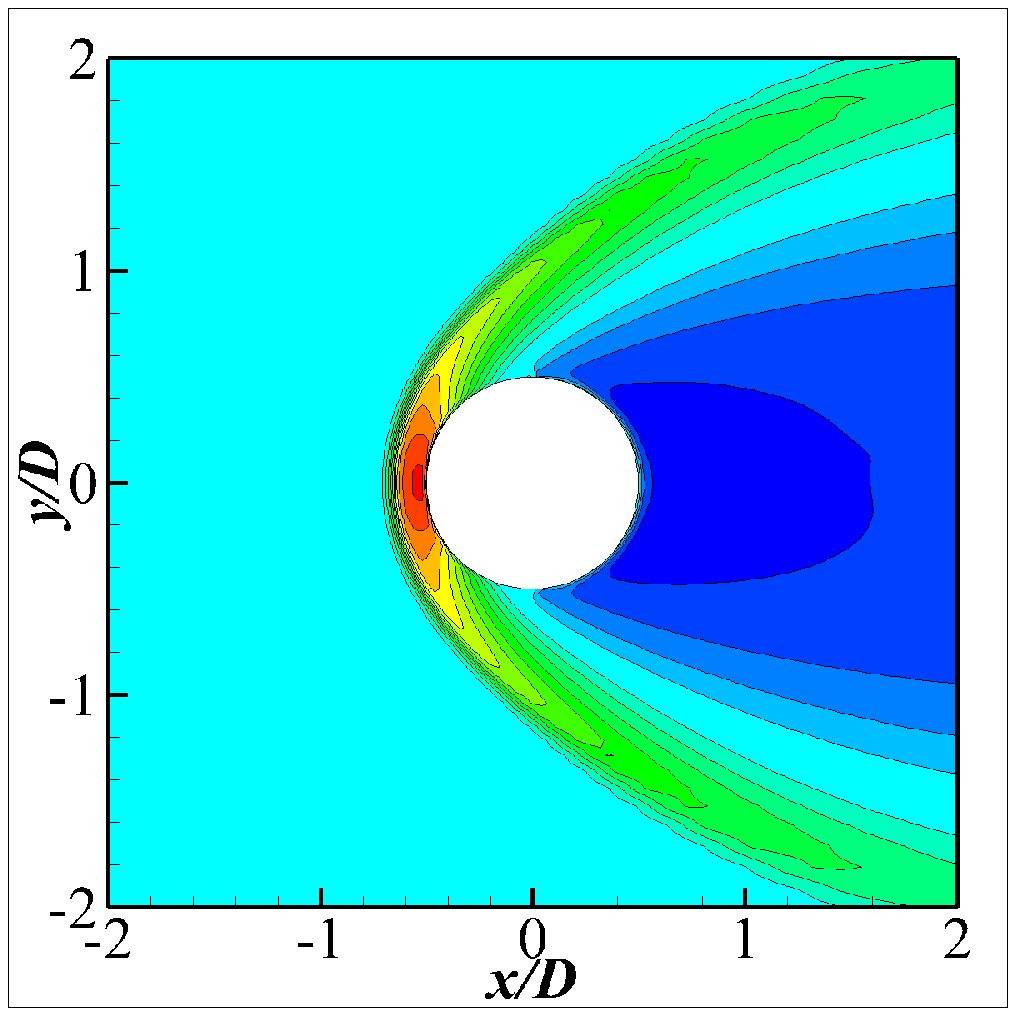}
  \includegraphics[width=2in]{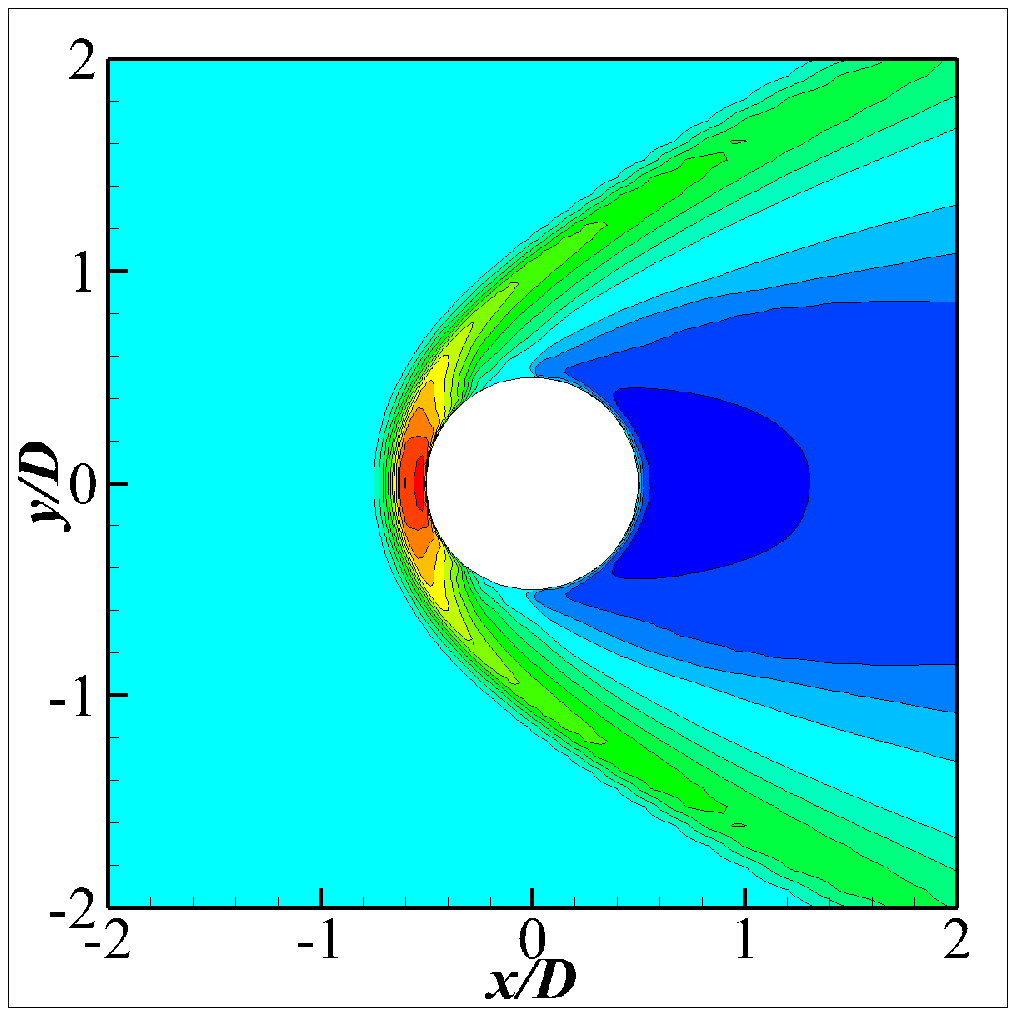}
  \includegraphics[width=2in]{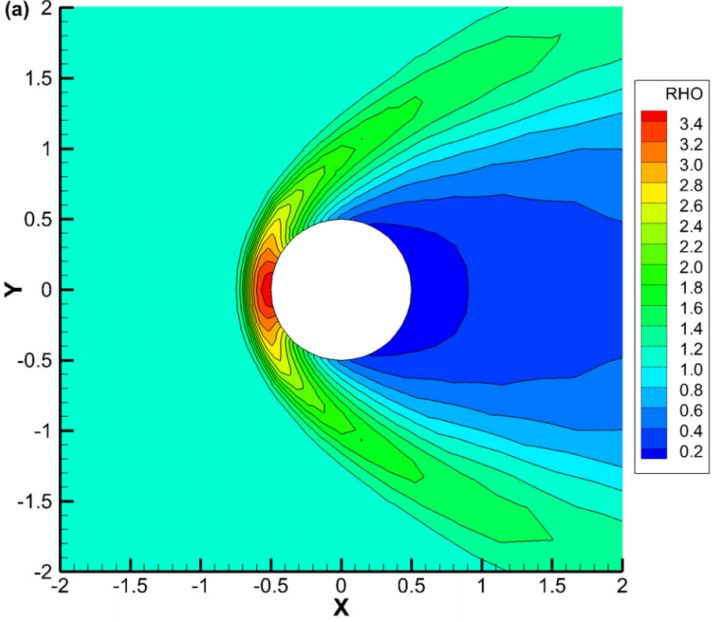}\\
  \includegraphics[width=2in]{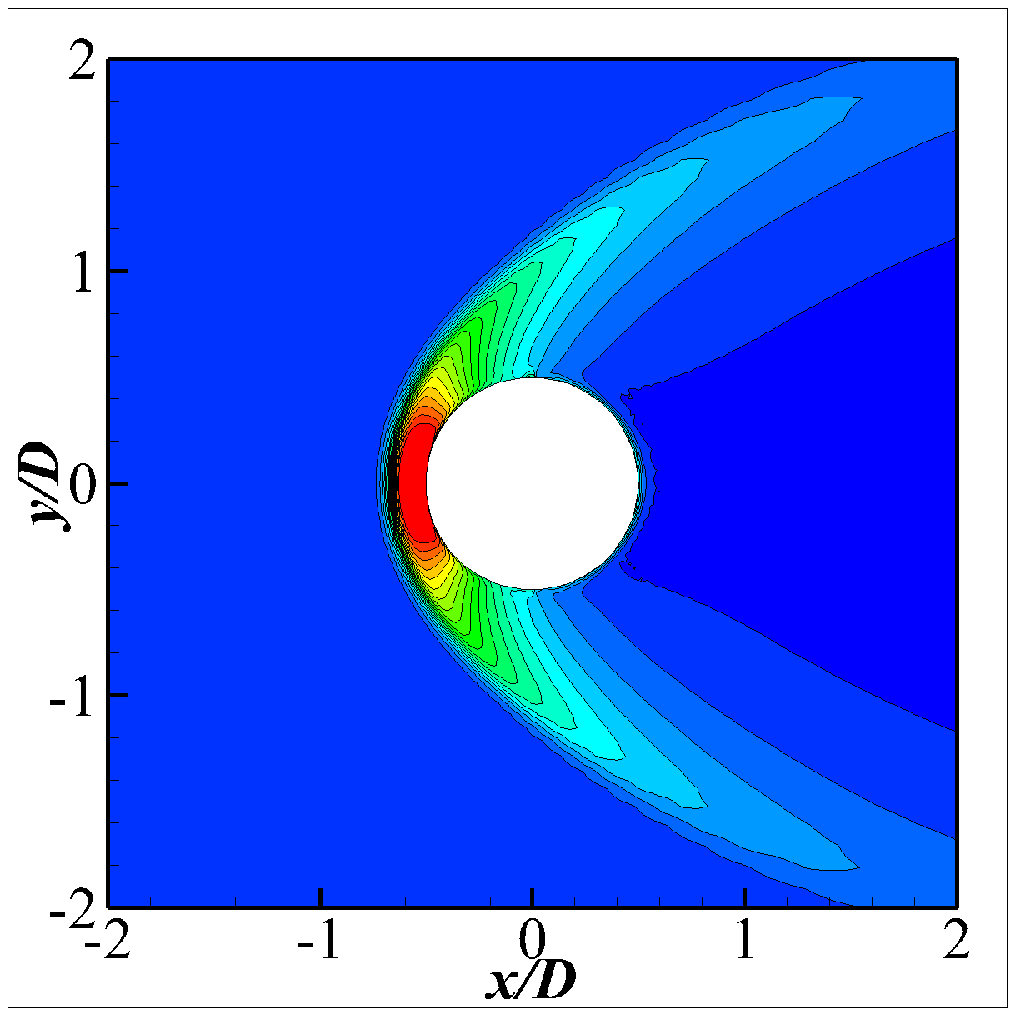}  
  \includegraphics[width=2in]{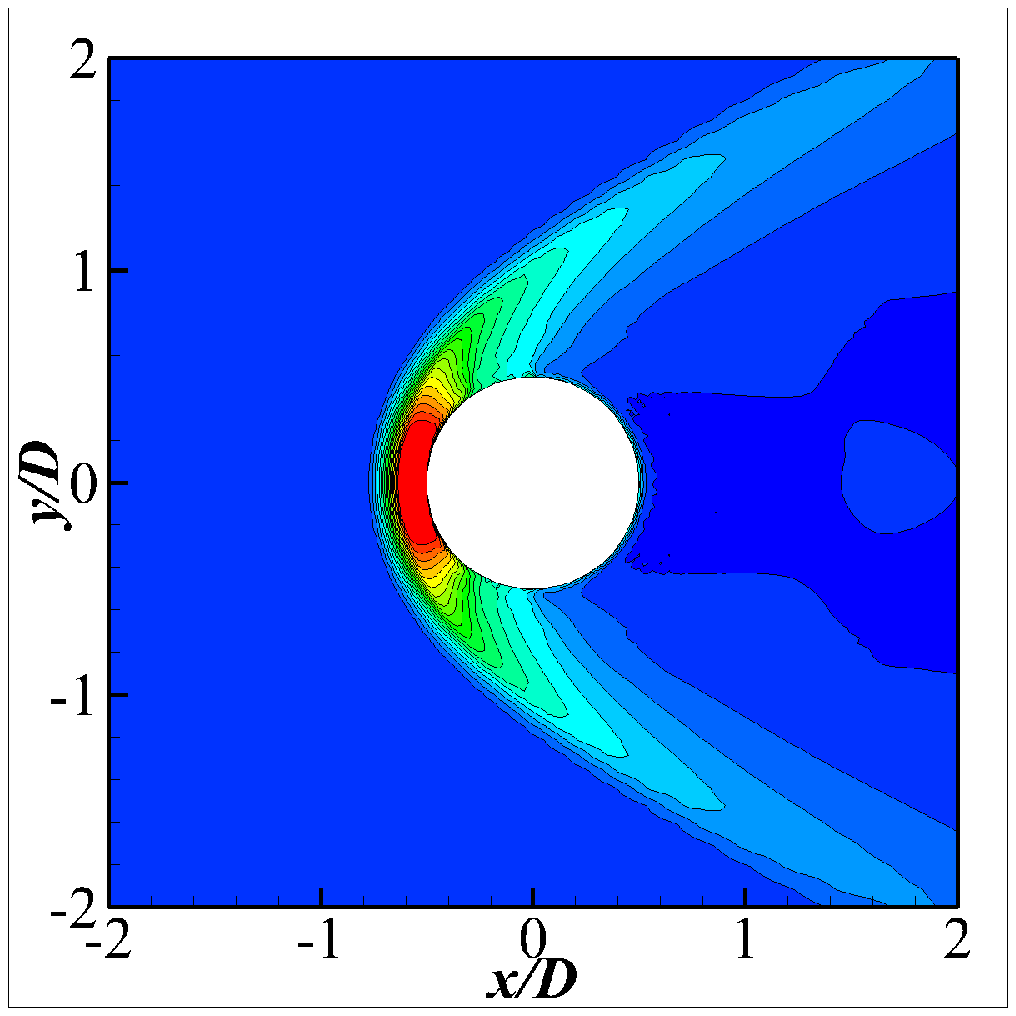}
  \includegraphics[width=2in]{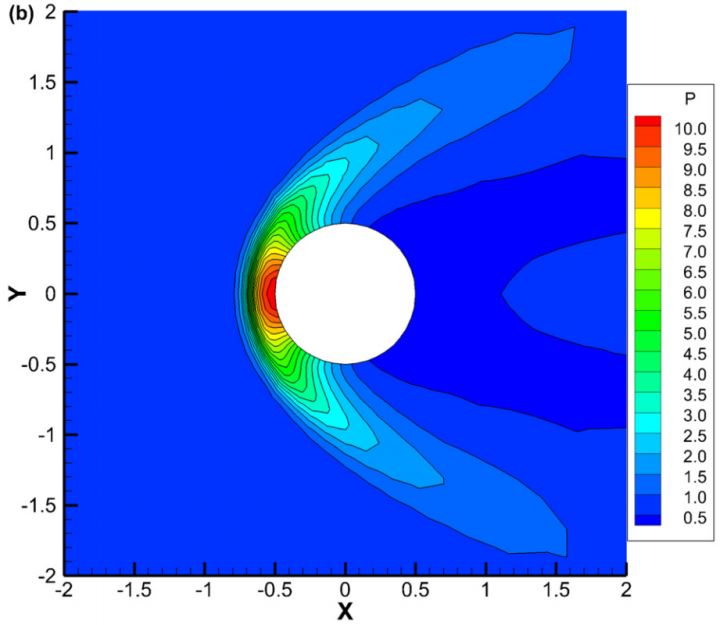}\\
  \includegraphics[width=2in]{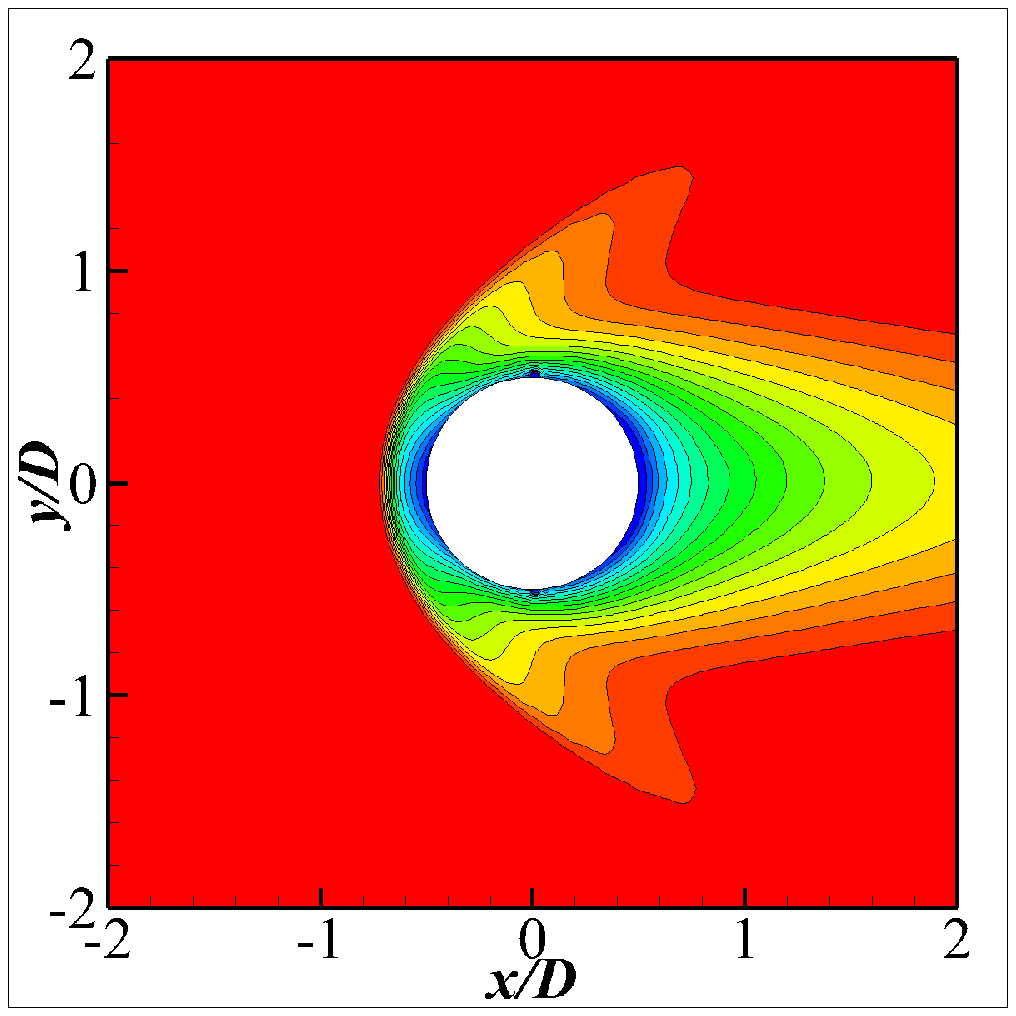}
  \includegraphics[width=2in]{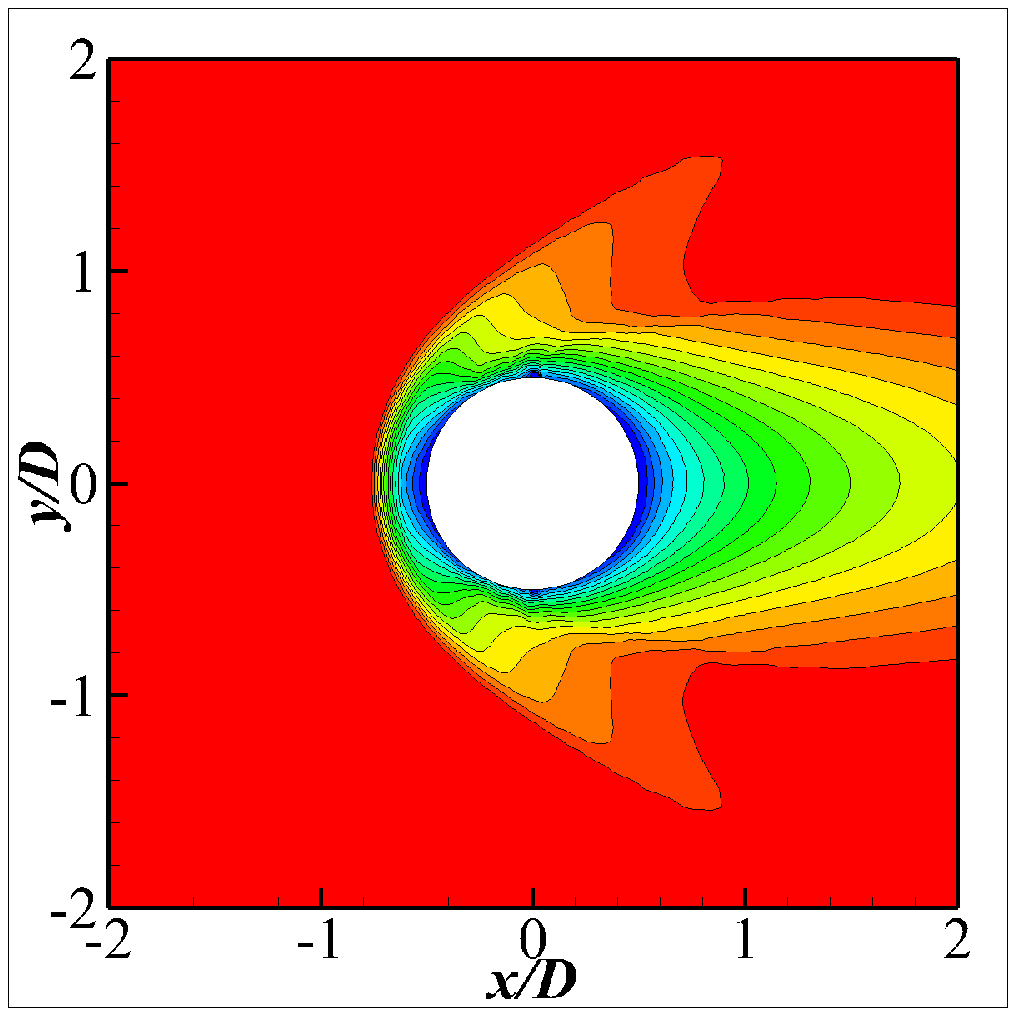}
  \includegraphics[width=2in]{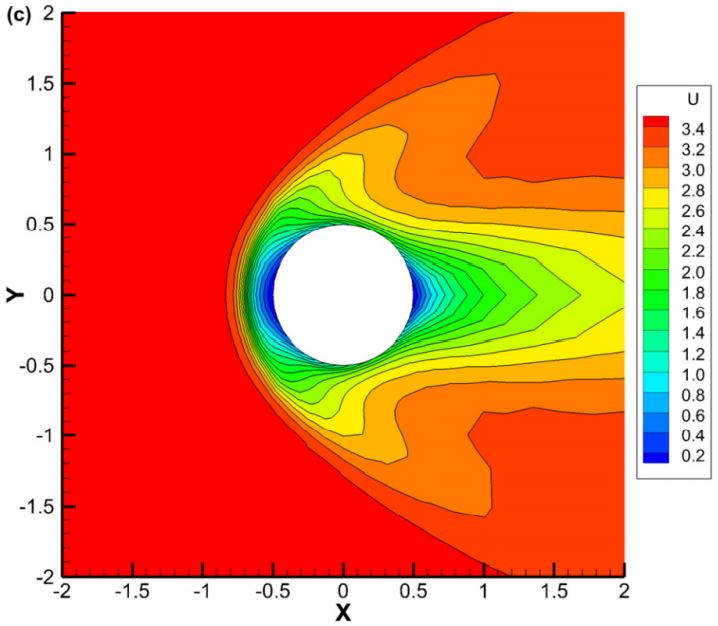}\\
  \includegraphics[width=2in]{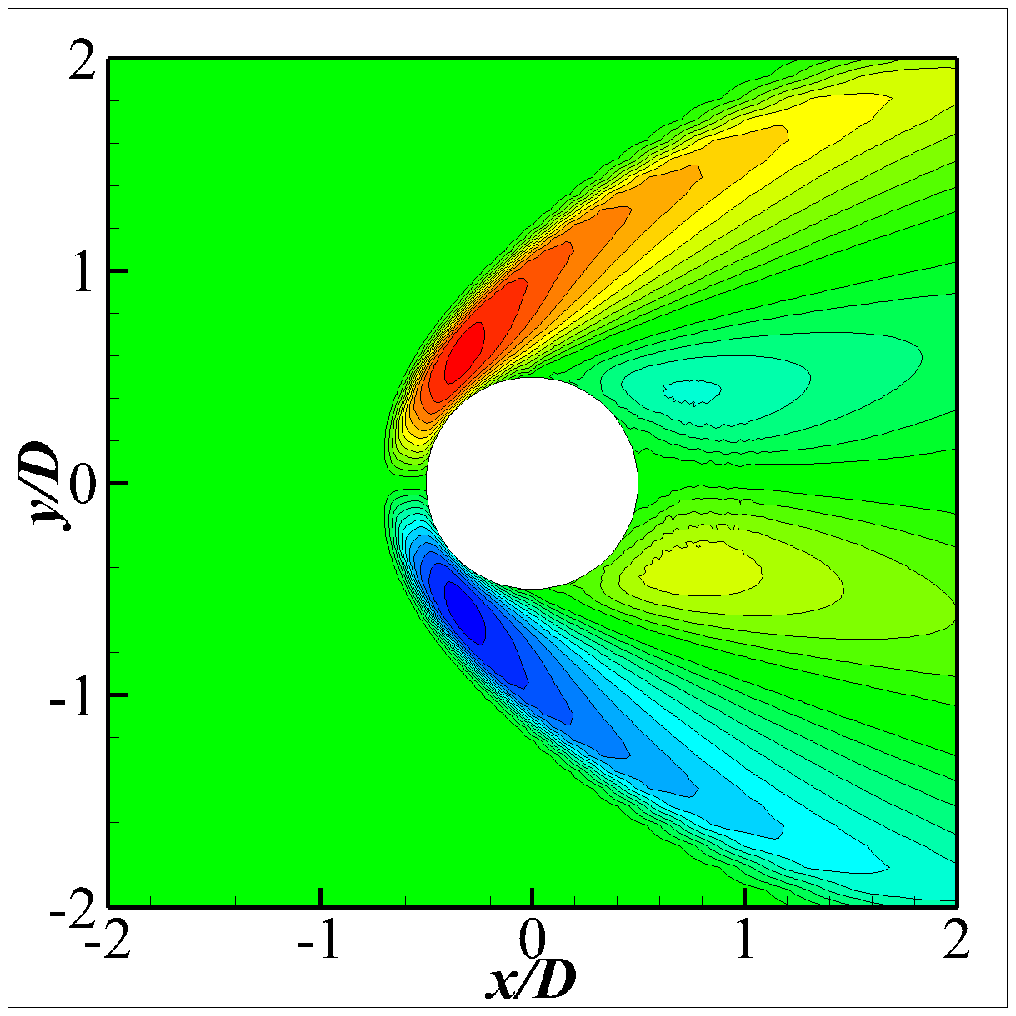}
  \includegraphics[width=2in]{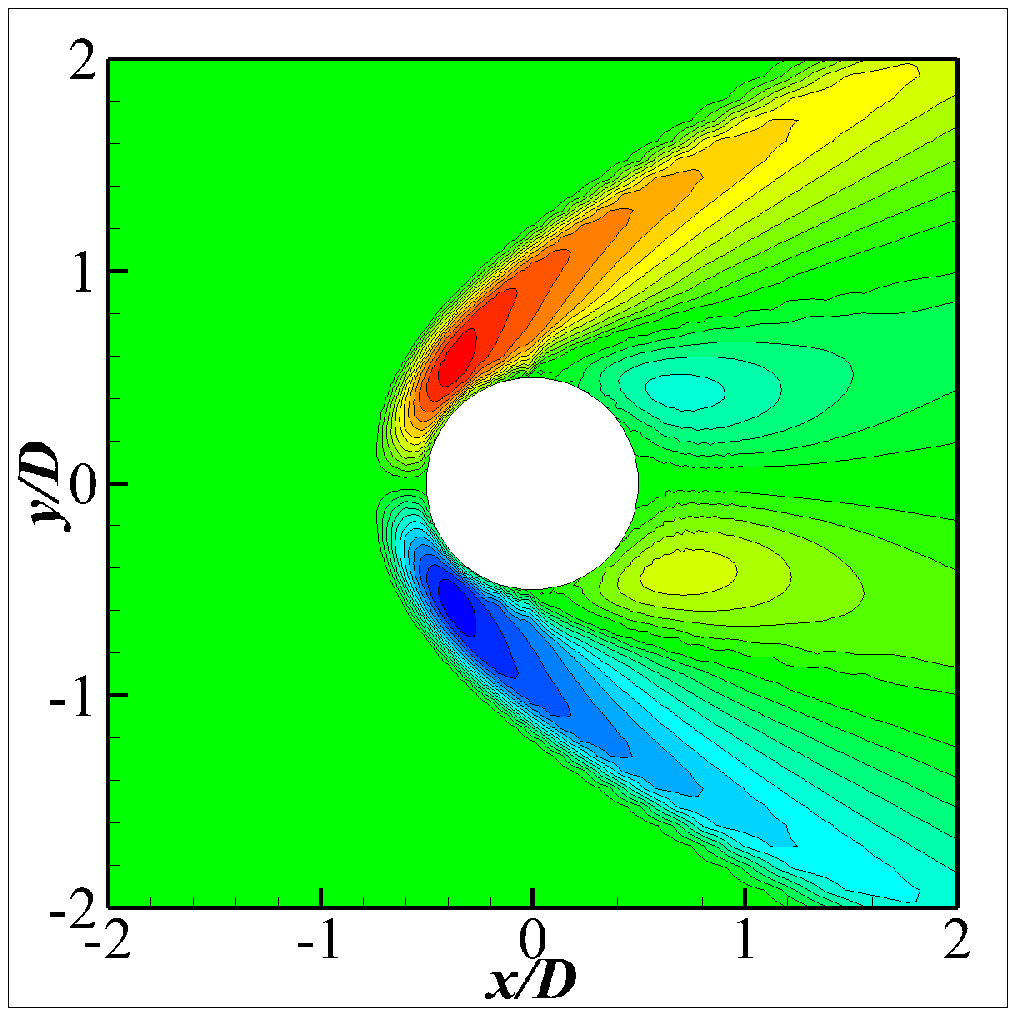}
  \includegraphics[width=2in]{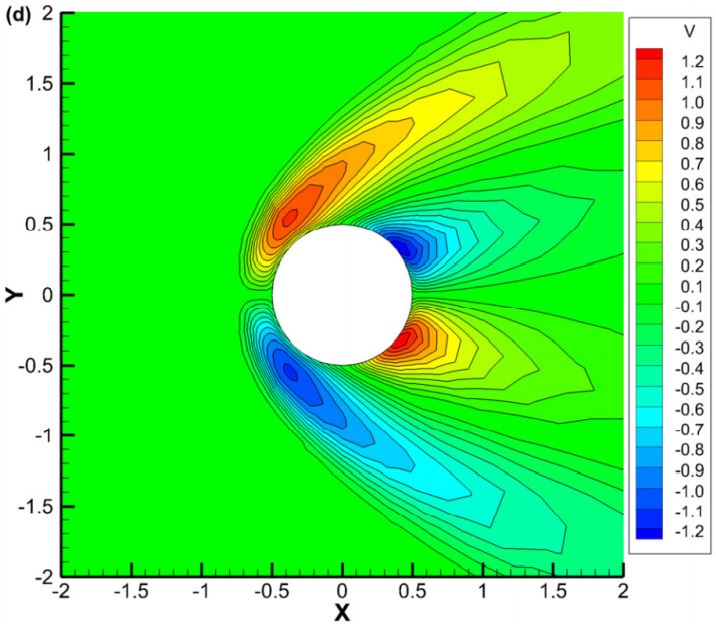}\\
  \end{center}
\caption{Density, pressure, stream-wise velocity and vertical velocity contours (from top to bottom) of flow around a sphere with $M = 3.834$ and $kn = 0.03$: present results(left column), present results with local Knudsen number modification (middle column) and results from Ref.~\cite{yang2019improved} (right column).}
\label{Fig:mspcontour}
\end{figure}

\begin{figure}
  \begin{center}
  \includegraphics[width=3in]{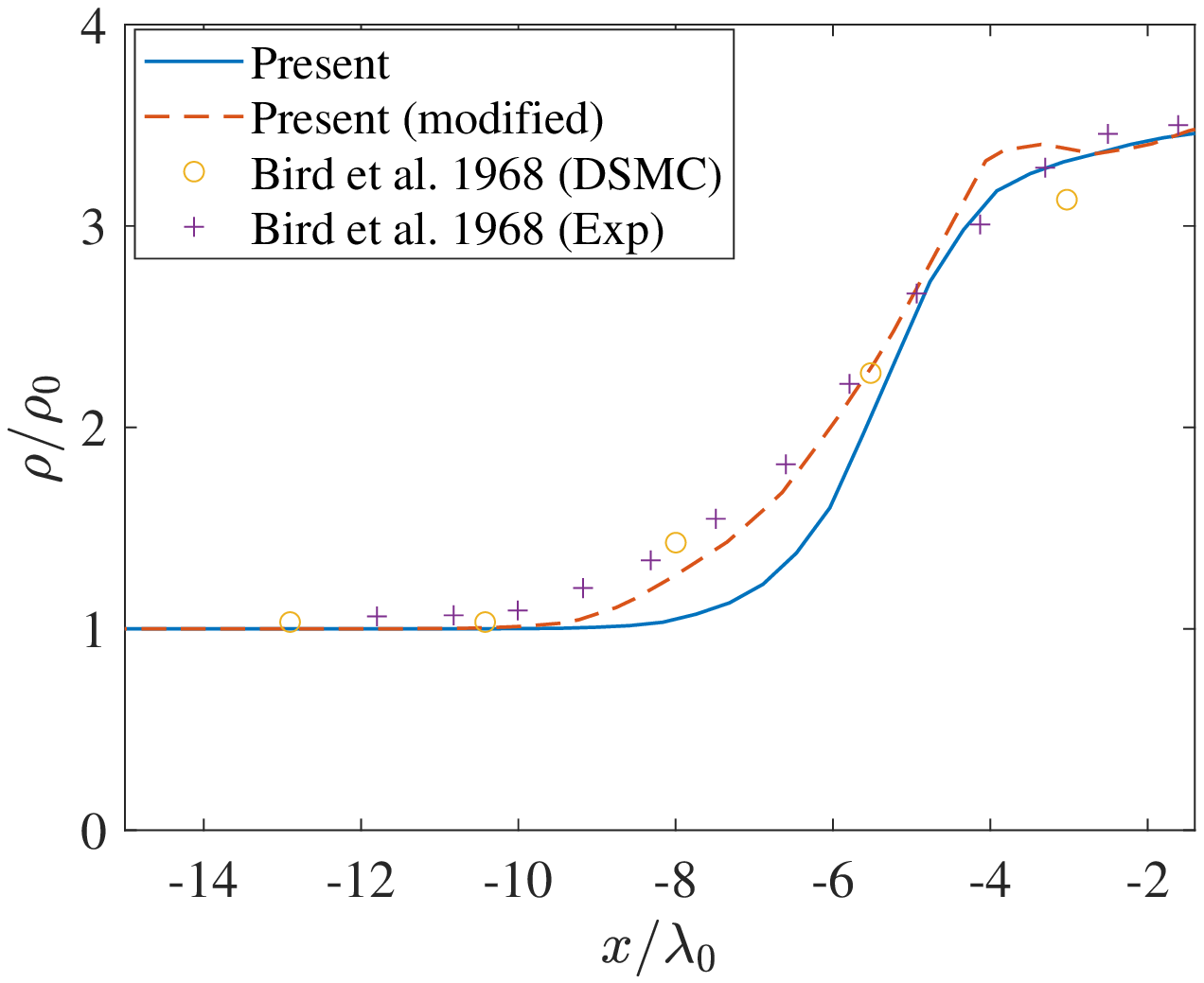}
  \includegraphics[width=3in]{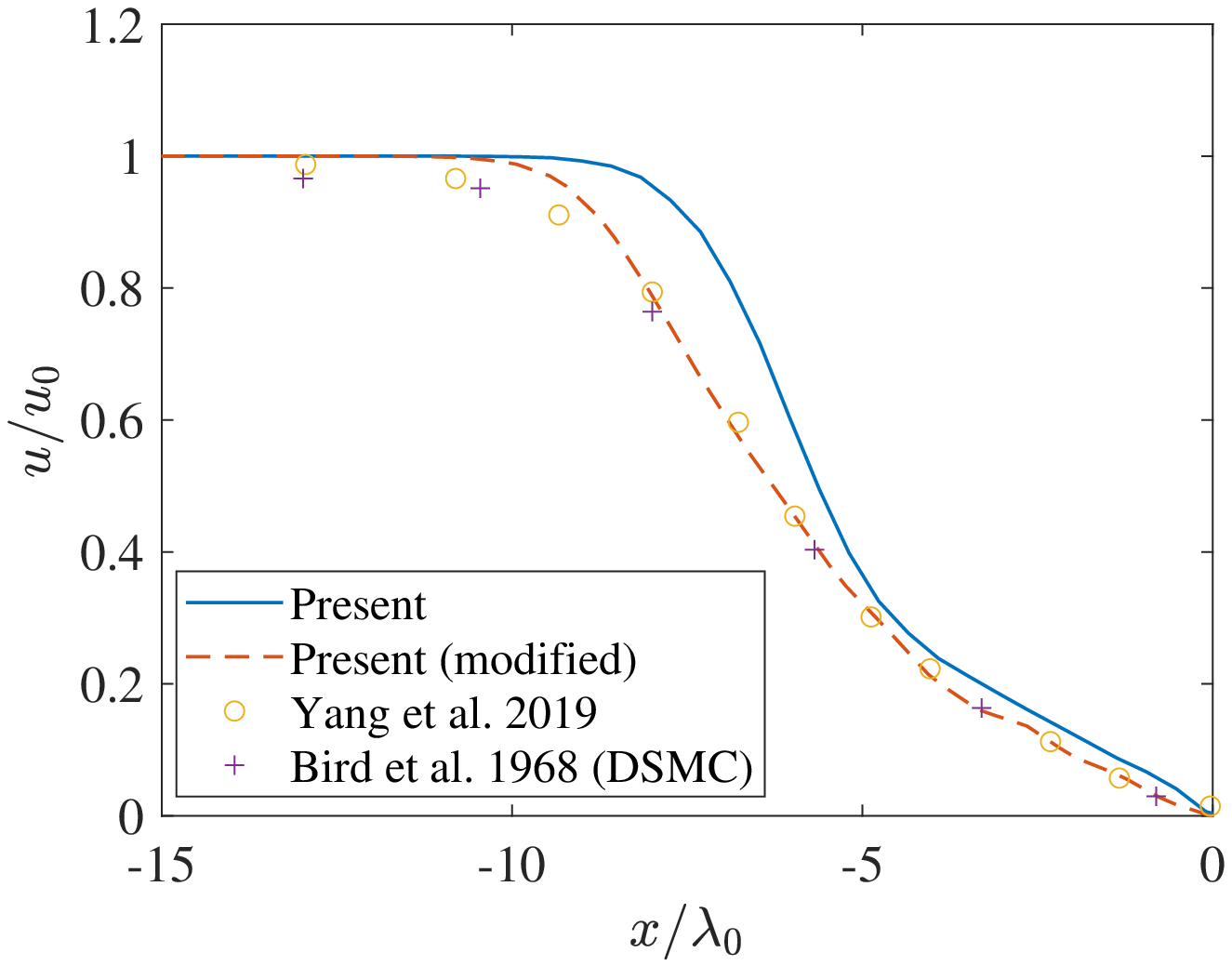}\\
  \includegraphics[width=3in]{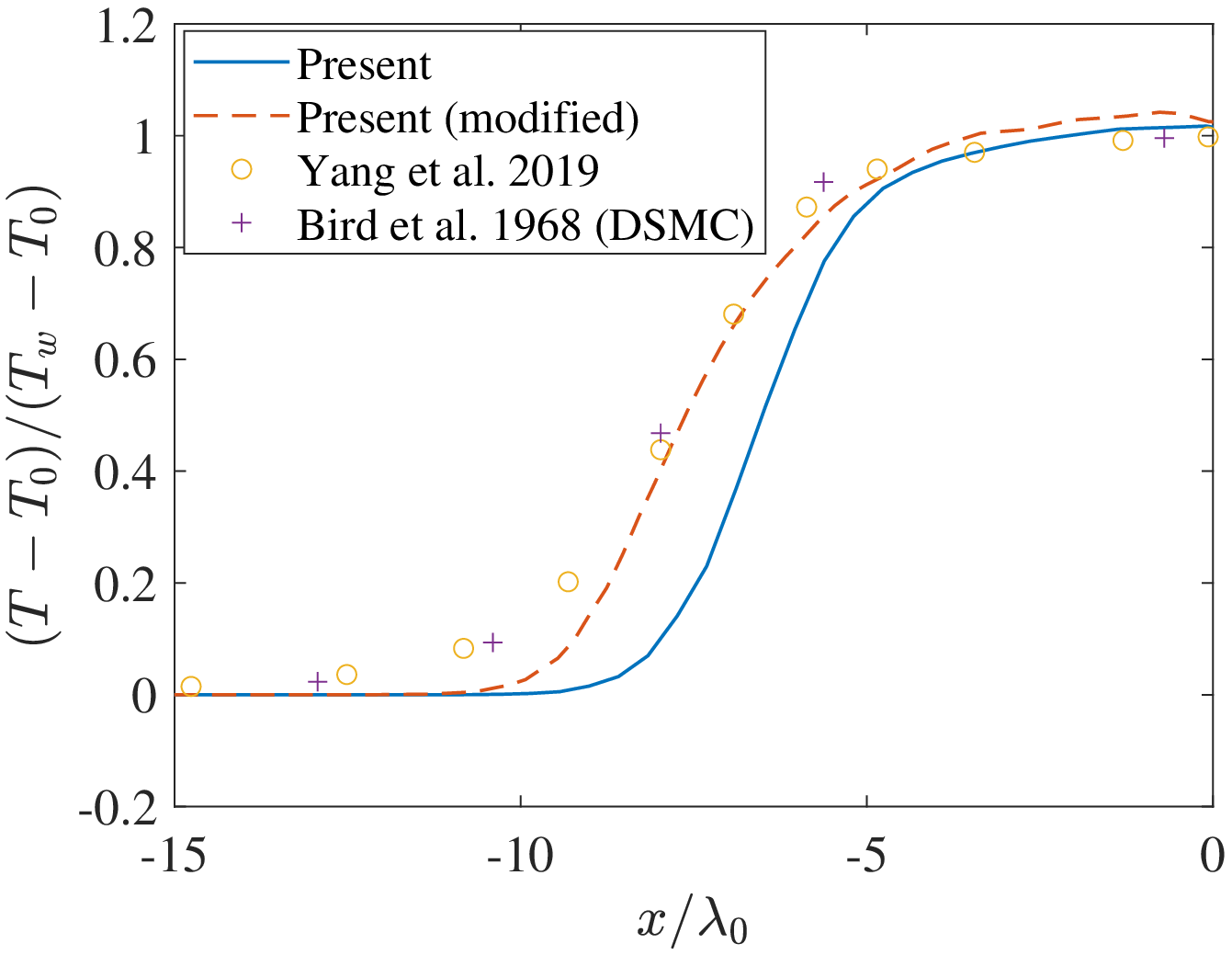}\\
  \end{center}
\caption{Density, stream-wise velocity and temperature profiles along the stagnation line.}
\label{Fig:mspzplane}
\end{figure}

It should be noted that this is a steady case, and Fig.~\ref{Fig:mspcontour} shows the contours of density, pressure, stream-wise velocity and vertical velocity when the steady condition is achieved, where the first column is directly calculated by the present method, the second column corresponds to the present method with local Knudsen number modification, and the results calculated by using DVM is also included in the third column for qualitative comparison. It is found that the detached bow-shape shock wave at the front of the sphere is well captured by the present method. However, the results from the present method show a sharper transition at the shock wave front compared with those obtained by using DVM. This phenomenon has been observed and explained as the failure of NSE at regions with high local Knudsen numbers (i.e., $Kn_{local} > 0.1$). The local Knudsen number can be calculated by using the local density gradient,
\begin{equation}
Kn_{local} = \frac{\Delta \rho}{\rho} Kn.
\end{equation}  

To overcome this local failure, a simple way is to add extra artificial viscosity to damp the sharp transition to be a diffused one. Here, we propose to add the artificial viscosity as follows
\begin{equation}
\mu = \mu_0 (T/T_0) ^ {\chi} + \frac{(Kn_{local} - 0.1)}{Kn} \mu_0.
\label{eq:extravis}
\end{equation}

When $kn_{local} \leq 0.1$, this model reduces to the original hard sphere model. The extra artificial viscosity is introduced for $kn_{local} > 0.1$ and it is linearly proportional to the local Knudsen number. By incorporating this artificial viscosity model into the present numerical method, a more diffused shock wave front is observed as show in the second column in Fig.~\ref{Fig:mspcontour}. A quantitative comparison of the flow field on a line in front of the sphere is presented in Fig.~\ref{Fig:mspzplane}, where the `modified' means the incorporation of artificial viscosity by using Eq.~\eqref{eq:extravis}. It is found that the proposed artificial viscosity criterion diffuses the shock wave front well to align the shock waves with the those obtained by using DVM, DSMC and experiment. The drag coefficient calculated by using the present method is about 1.32 which is very close to 1.36 in Ref.~\cite{yang2019improved}. It is also noted that the artificial viscosity model shows negligible effects on the drag coefficient even it is necessary to match the shock front. This also agrees with the previous observations in Refs.~\cite{lofthouse2008velocity} .

\begin{figure}
  \begin{center}
  \includegraphics[width=4in]{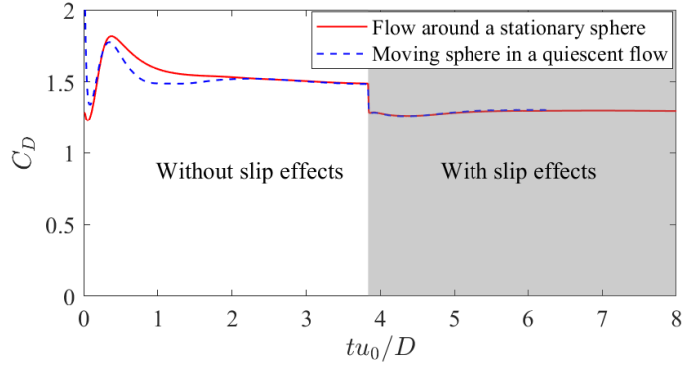}
  \end{center}
\caption{Time histories of $C_D$ for flow around a stationary sphere and moving sphere in a quiescent flow.}
\label{Fig:cdspheremv}
\end{figure}

To further validate the numerical method for handling moving geometries in 3D rarefied gas flow, a moving sphere in steady flow is considered instead of applying uniform flow around a stationary sphere. Non-uniform mesh in the spanwise and lateral directions are the same as in the uniform flow conditions, and a uniform mesh is used in the streamwise direction. The simulation is conducted with three stages, i.e., no-slip boundary condition stage, slip-velocity boundary condition stage and slip-velocity boundary condition with modified viscosity, to distinguish the effects of slip-velocity and modified viscosity. The slip velocity boundary condition and modification of viscosity based on Eq.~\eqref{eq:extravis} are activated after the previous quasi-steady states are achieved. The time histories of the drag coefficient are shown in Fig.~\ref{Fig:cdspheremv}. It is found that the results from moving sphere condition is consistent with its counterpart with uniform flow, indicating the validity of the numerical method for moving geometries. Specifically, the drag coefficient is around 1.60 with no-slip boundary condition, and drops to 1.30 when slip-velocity boundary condition is activated.

\begin{table}
 \caption{Parameters in the flapping wing simulation.}
 \label{table:flpcases}
\begin{center}
\renewcommand{\arraystretch}{1.5}
\setlength\tabcolsep{10pt}
\begin{tabular}{cccc}
  \hline
    No. & $Re$ &$Kn$  & Slip model\\
  \hline
    1/2 & 115.6 & 0.001 & On/Off\\
    3/4 & 23.1 & 0.005 & On/Off\\
    5/6 & 11.6 & 0.01 & On/Off\\
    7/8 & 5.8 & 0.02 & On/Off\\
    9/10 & 2.3 & 0.05 & On/Off\\
  \hline
\end{tabular}
\end{center}
\end{table}

\begin{figure}
  \begin{center}
  \includegraphics[width=3in]{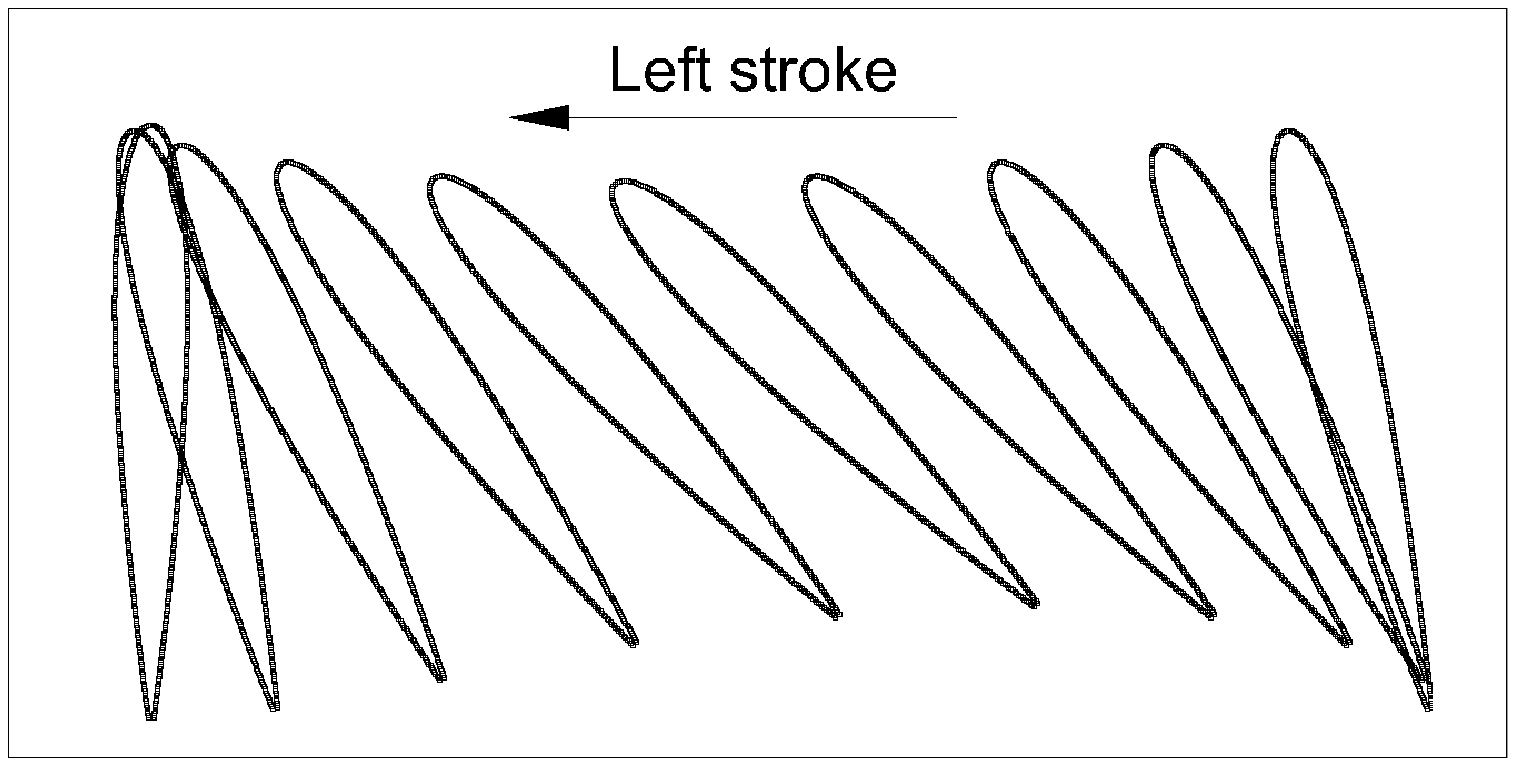}
  \includegraphics[width=3in]{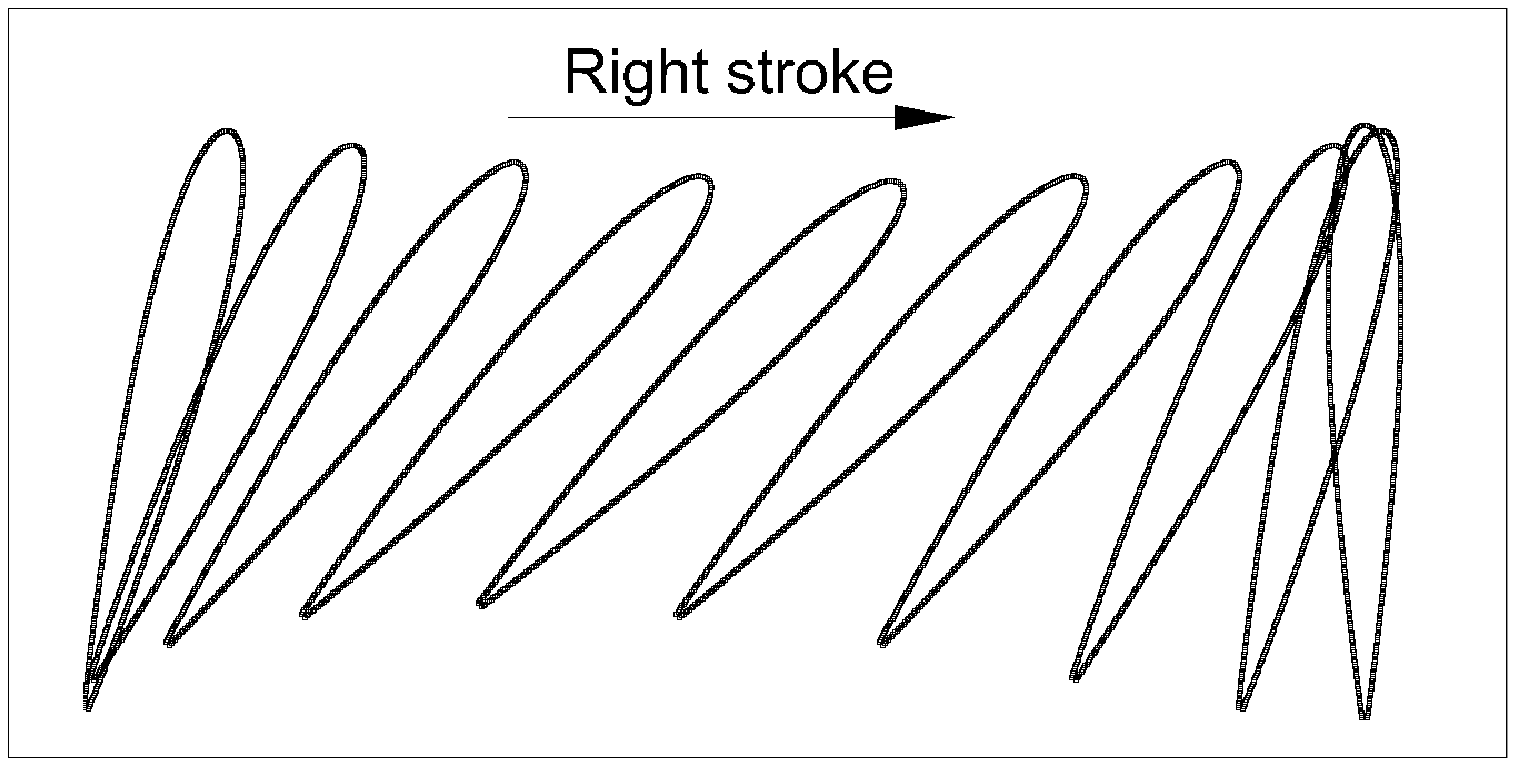}\\
  \end{center}
\caption{Schematics of the flapping airfoil.}
\label{Fig:flpmotion}
\end{figure}

\section{Aerodynamics of flapping wing hovering in rarefied gas flow}
\subsection{Flapping airfoil hovering in rarefied gas flow} \label{sec:naca2d}
With the development of planetary exploration, UAVs have drawn considerable attention for its significant maneuverability over the land rovers and the excellent ability in providing higher resolution images over satellites. The Intelligent with rotary-wings on Mars has made the flight out of Earth successful for the first time. This encourages the further study to explore the ubiquitous flapping-wing strategy for the flight on Mars. However, it is noted that the Mars has very rarefied atmosphere, and thus the micro aerial vehicles in such environment feature low Reynolds numbers and rarefied gas flow. Here, a flapping airfoil (NACA0015) hovering in rarefied gas flow is considered to understand the associated aerodynamic performance, as shown in Fig.~\ref{Fig:flpmotion}. The wing is governed by a combined stroke and pitching motion which is described as 
\begin{equation}
A = A_m sin(2 \pi f t + \frac{\pi}{2}), \quad \theta = \theta sin(2 \pi f t),
\label{eq:flpmotion}
\end{equation}
where $f$ is the flapping frequency, $A_m$ and $\theta$ are respectively the stroke and pitching amplitude. The parameters that govern this problem include Reynolds number and Knudsen number, which are defined as 
\begin{equation}
Re = \frac{\rho_0 U L}{\mu}, \quad Kn = \frac{\lambda}{L},
\label{eq:flpgvpara}
\end{equation} 
where $\rho_0$ is the initial fluid density, $\mu$ is the fluid viscosity, $L$ is the length of the wing, $\lambda$ is mean-free-path length of the fluid, and $U = 2 \pi f A_m$ which represents the maximum wing-tip velocity. Here, $A_m/L = 2.0$ and $\alpha_m = \pi/4$ are used according to Ref.~\cite{yin2010effect}. Based on the properties of the atmosphere on the surface of Mars, i.e., $p_0=600Pa$, $\rho_0 = 0.012 kg/m^3$ and $\gamma = 1.67$, $Kn$ ranging from 0.001 (close to continuum regime) to 0.05 (slip regime) are considered. The dynamic viscosity of the gas is then calculated by using Eq.~\eqref{eq:hsvis} according to the properties of the Martian atmosphere. Specifically, four Knudsen numbers, i.e, $Kn = 0.001$, 0.005, 0.01, 0.02 and 0.05, are examined and four cases (No. 2, 4, 6, 8 and 10 in Tab.~\ref{table:flpcases}) at the same Reynolds numbers without slip model are also examined to analysis the rarefied gas effects. It is noted that the Reynolds numbers considered here are normally lower than those having been previously studied~\cite{yin2010effect,tian2013force,shahzad2018effects}, as the ultra-low density atmosphere is considered. The parameters used in these simulations are shown in Tab.~\ref{table:flpcases}. It should be noted that a small Mach number ($M = U/c \approx 0.07$ with $c$ being the sound speed in the gas) is used in all simulations thus the gas can be considered as incompressible. To quantify the aerodynamic performance of the flapping wing, the lift coefficient, power coefficient and efficiency are defined as
\begin{equation}
C_x=\frac{2F_x}{\rho U^2 L},\quad C_L=\frac{2F_y}{\rho U^2 L},\quad C_P=\frac{-2 \int \boldsymbol{f}\cdot \boldsymbol{V} dl}{\rho U^3 L}, \quad \eta = \frac{C_{L,m}}{C_{P,m}},
\label{eq:ctcl}
\end{equation}
where $F_x$ and $F_y$ is the force acting on the wing by the ambient fluid in $x$ (horizontal) and $y$ (vertical) direction, respectively, $\boldsymbol{f}$ is the hydrodynamic traction on the wing, $\boldsymbol{V}$ is the velocity of the wing, and $C_{L,m}$ and $C_{P,m}$ are the time averaged lift and power coefficients, respectively. 

\begin{figure}
  \begin{center}
  \includegraphics[width=2in]{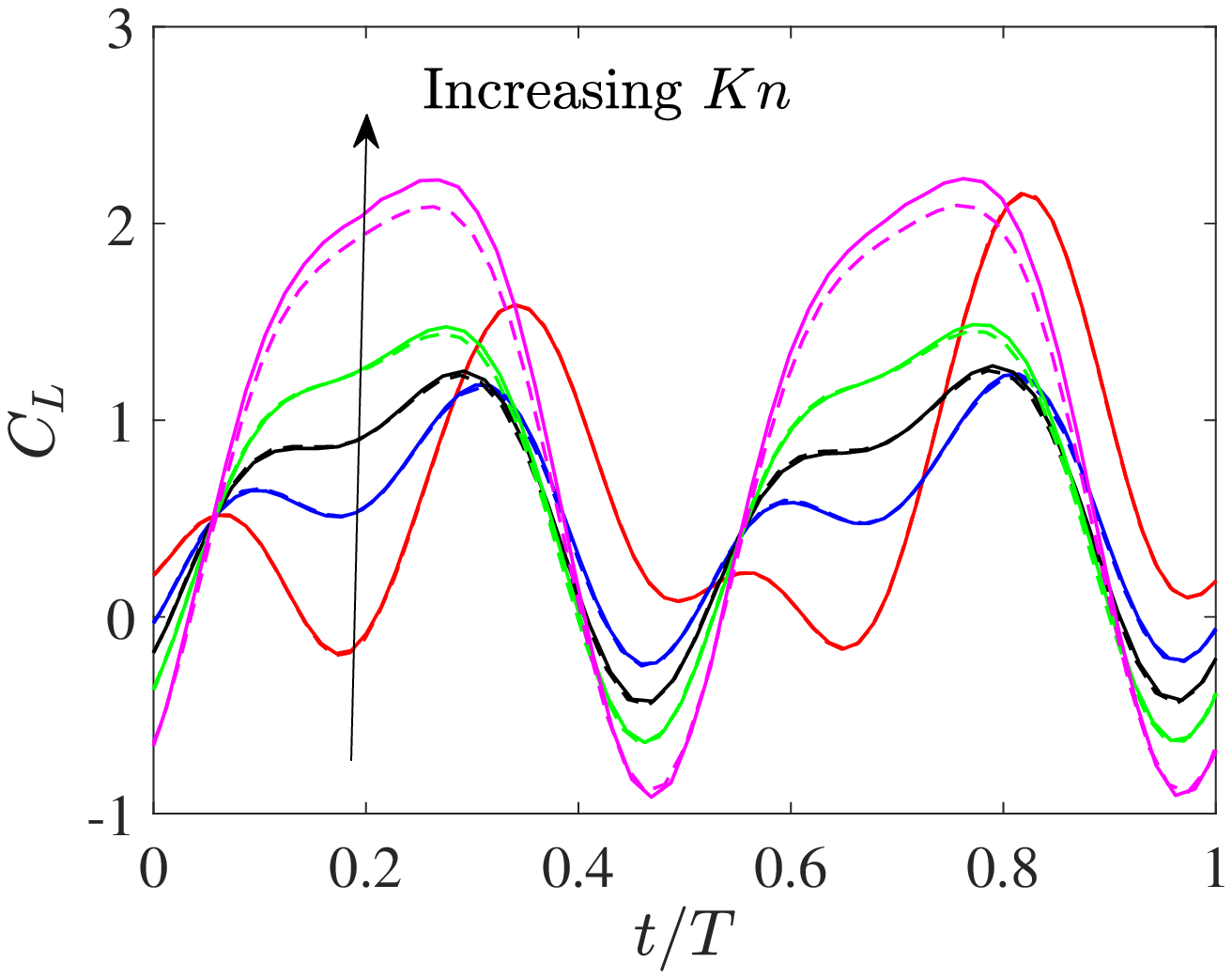}
  \includegraphics[width=2in]{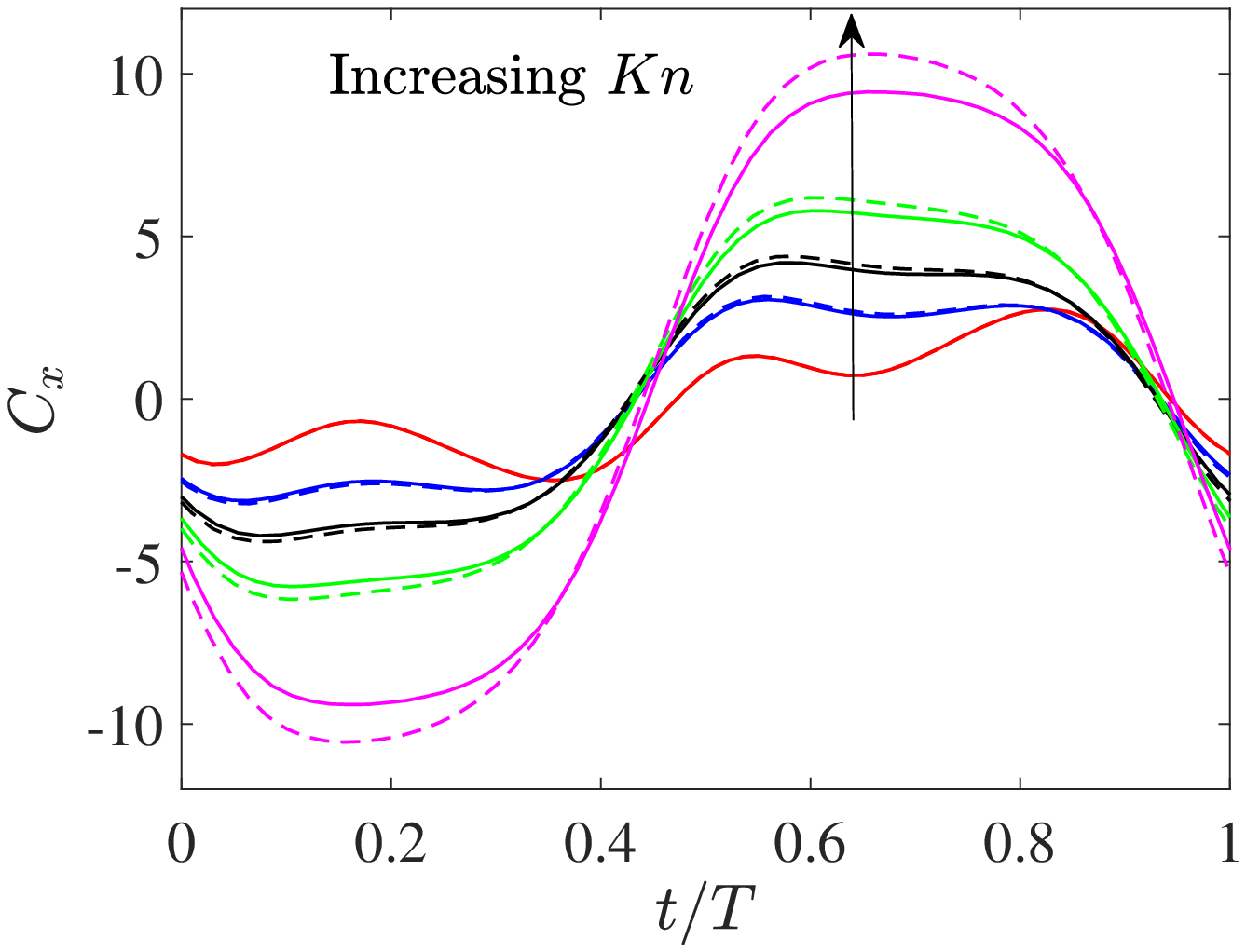}
  \includegraphics[width=2in]{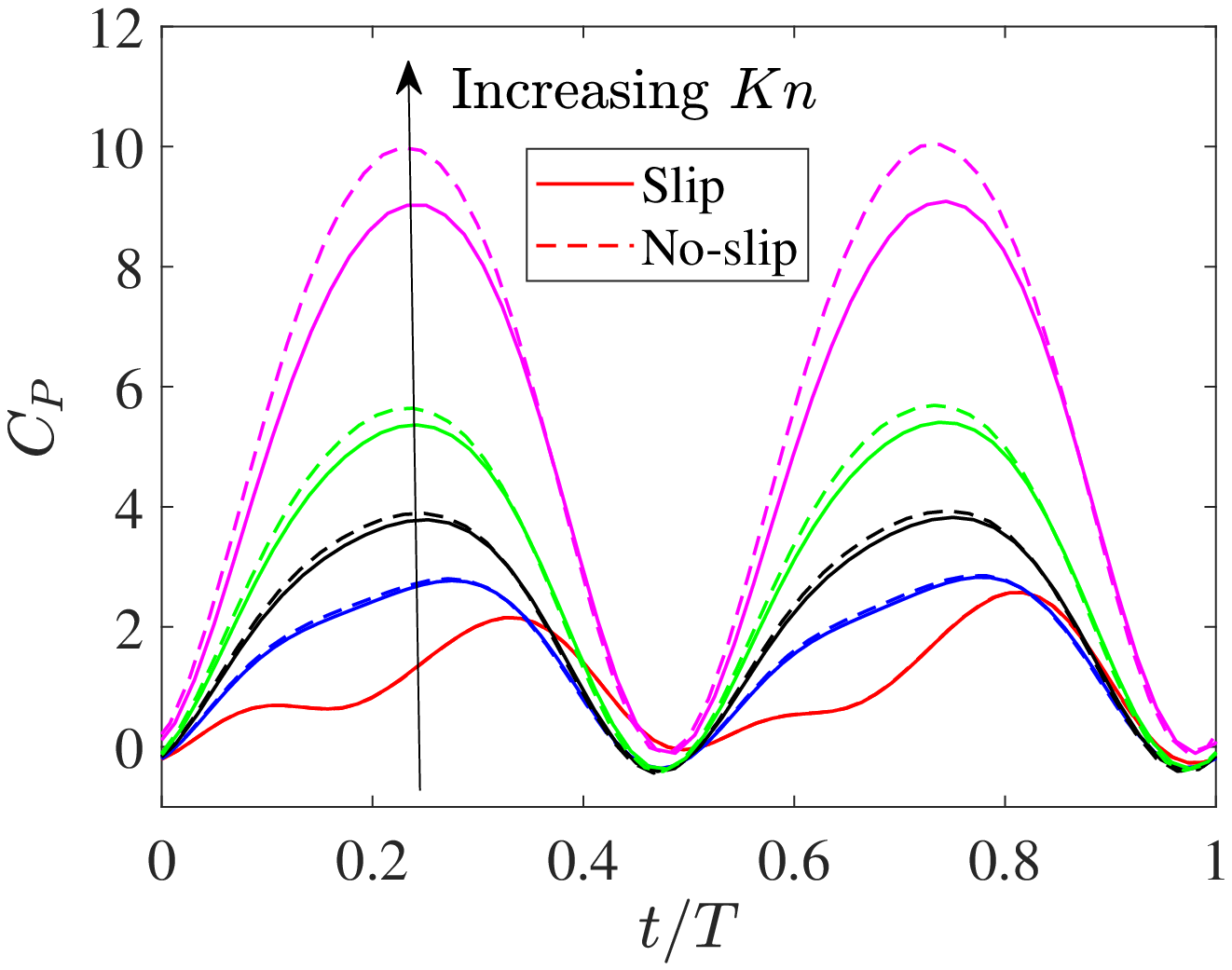}
  \end{center}
\caption{Time history of $C_L$, $C_x$ and $C_P$ at various $Kn$ from 0.001 to 0.05, where the solid line and dashed line represent the results with and without slip model, respectively.}
\label{Fig:flphis}
\end{figure}

\begin{figure}
  \begin{center}
  \includegraphics[width=1.5in]{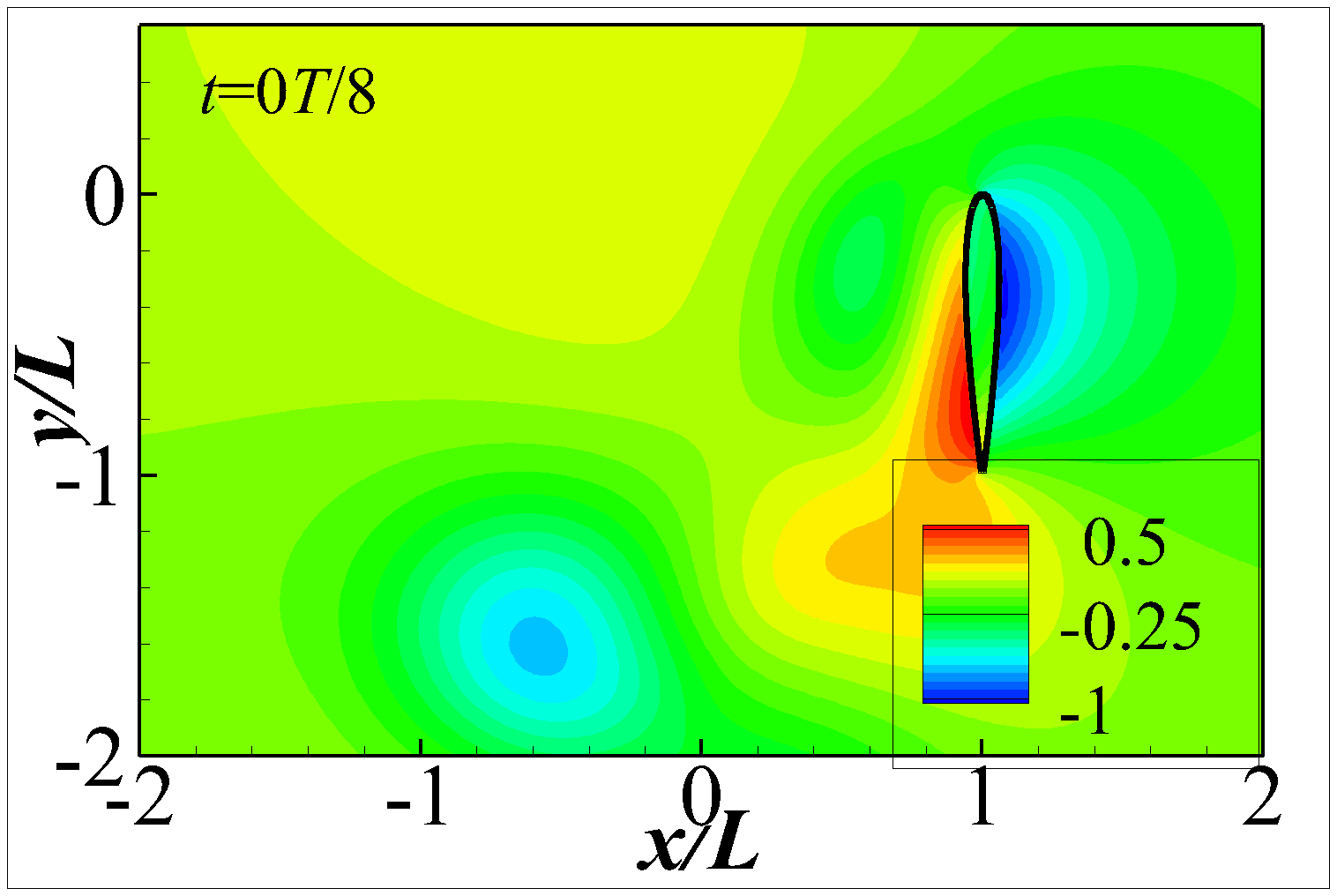}
  \includegraphics[width=1.5in]{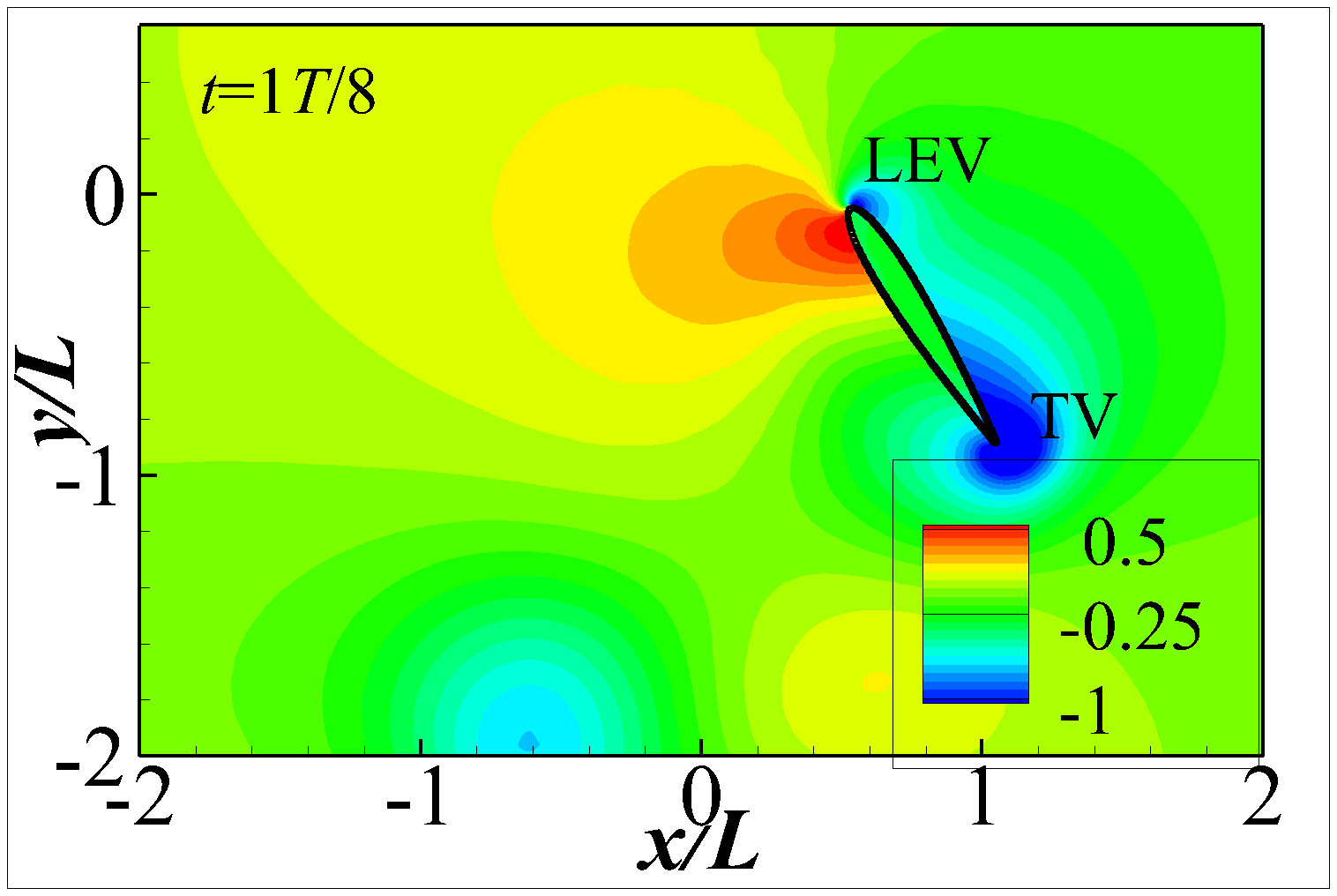}
  \includegraphics[width=1.5in]{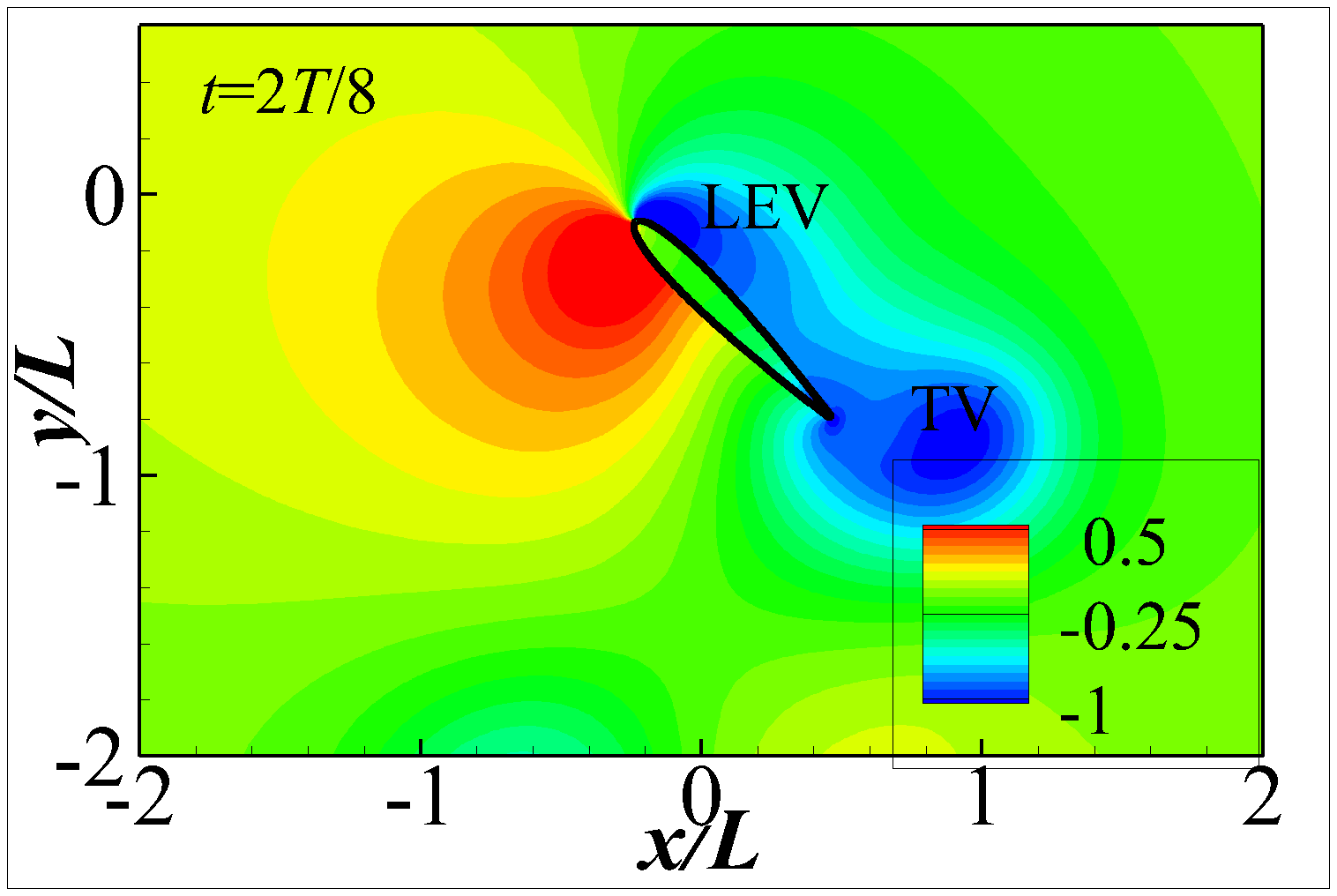}
  \includegraphics[width=1.5in]{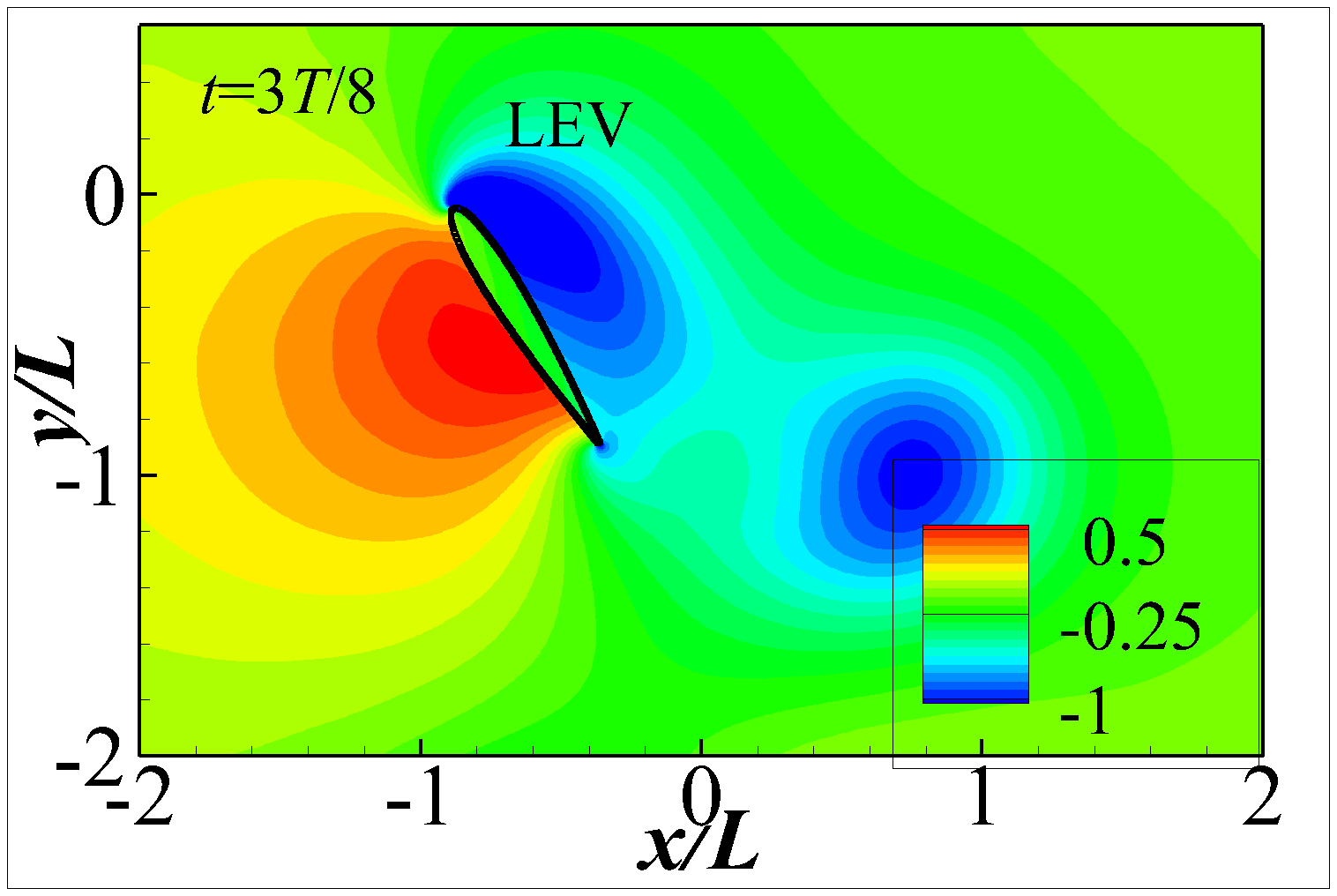}\\
  \includegraphics[width=1.5in]{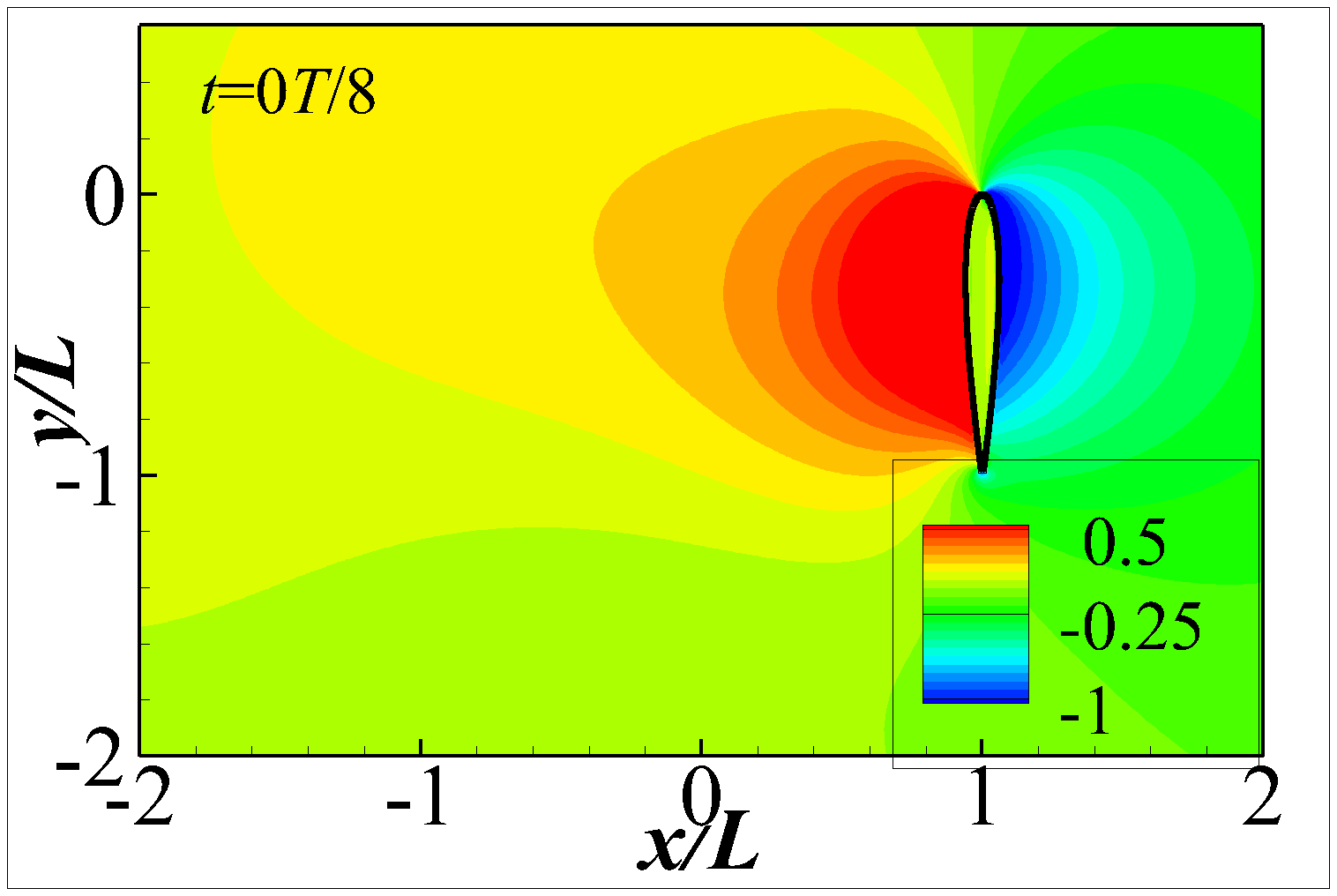}
  \includegraphics[width=1.5in]{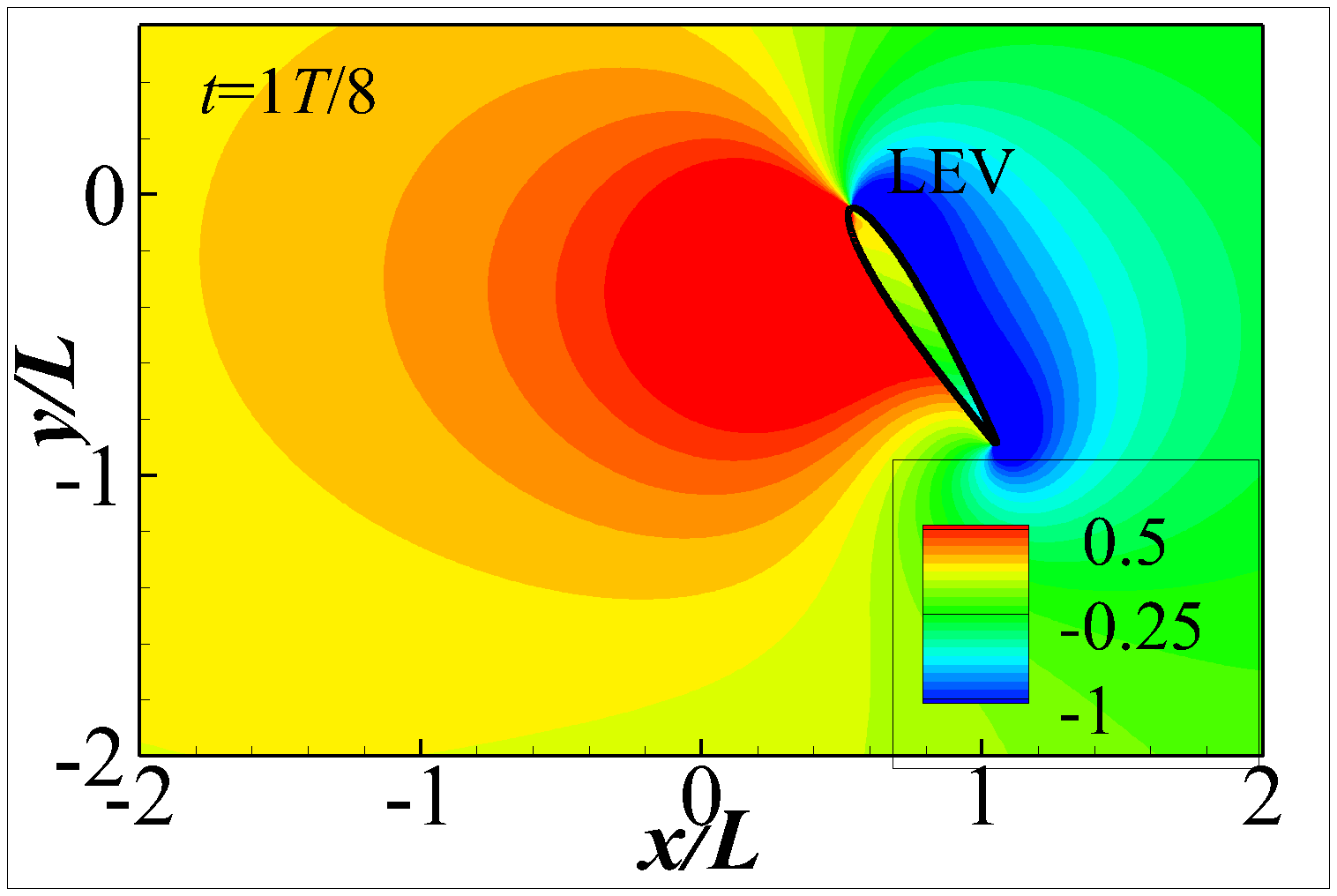}
  \includegraphics[width=1.5in]{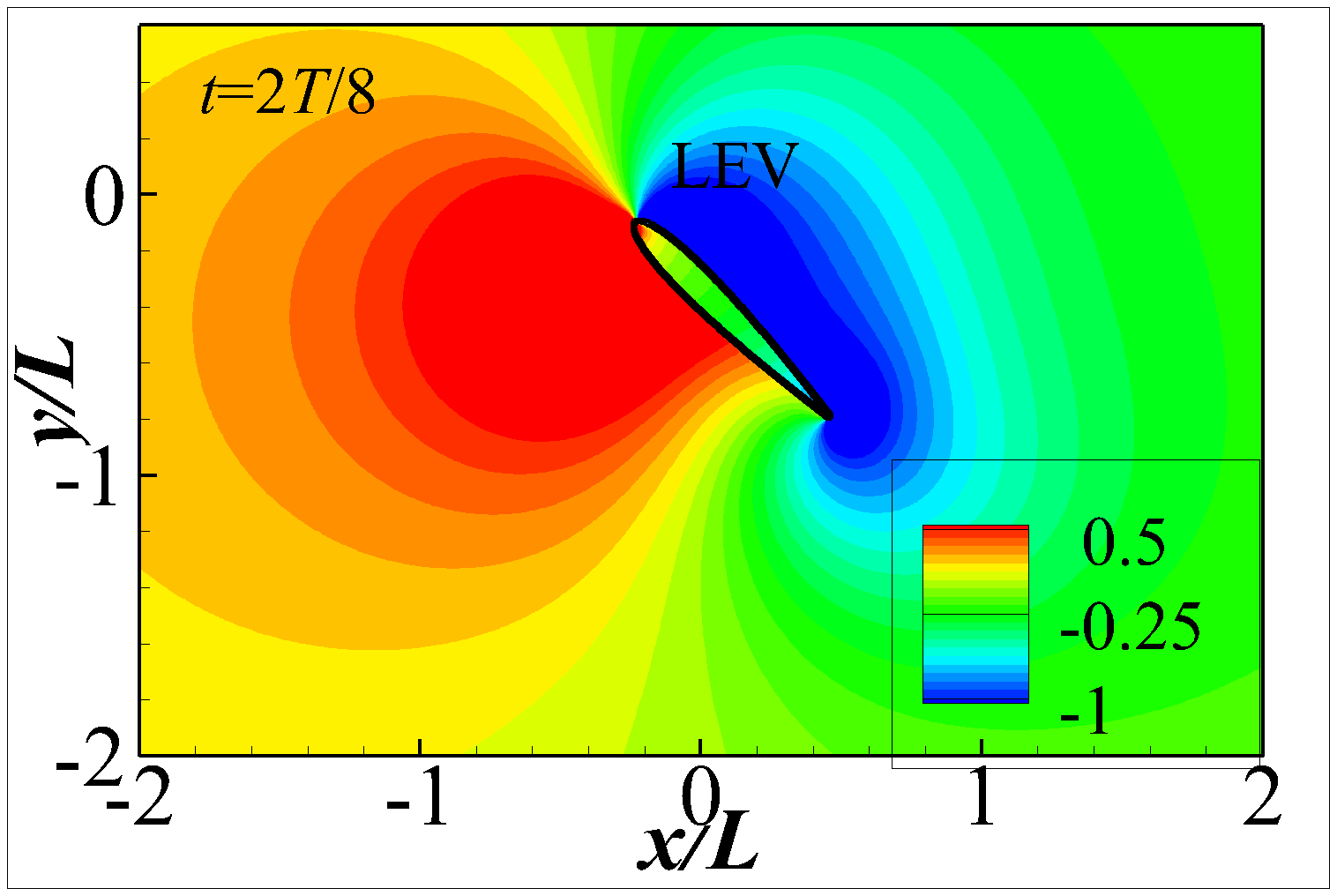}
  \includegraphics[width=1.5in]{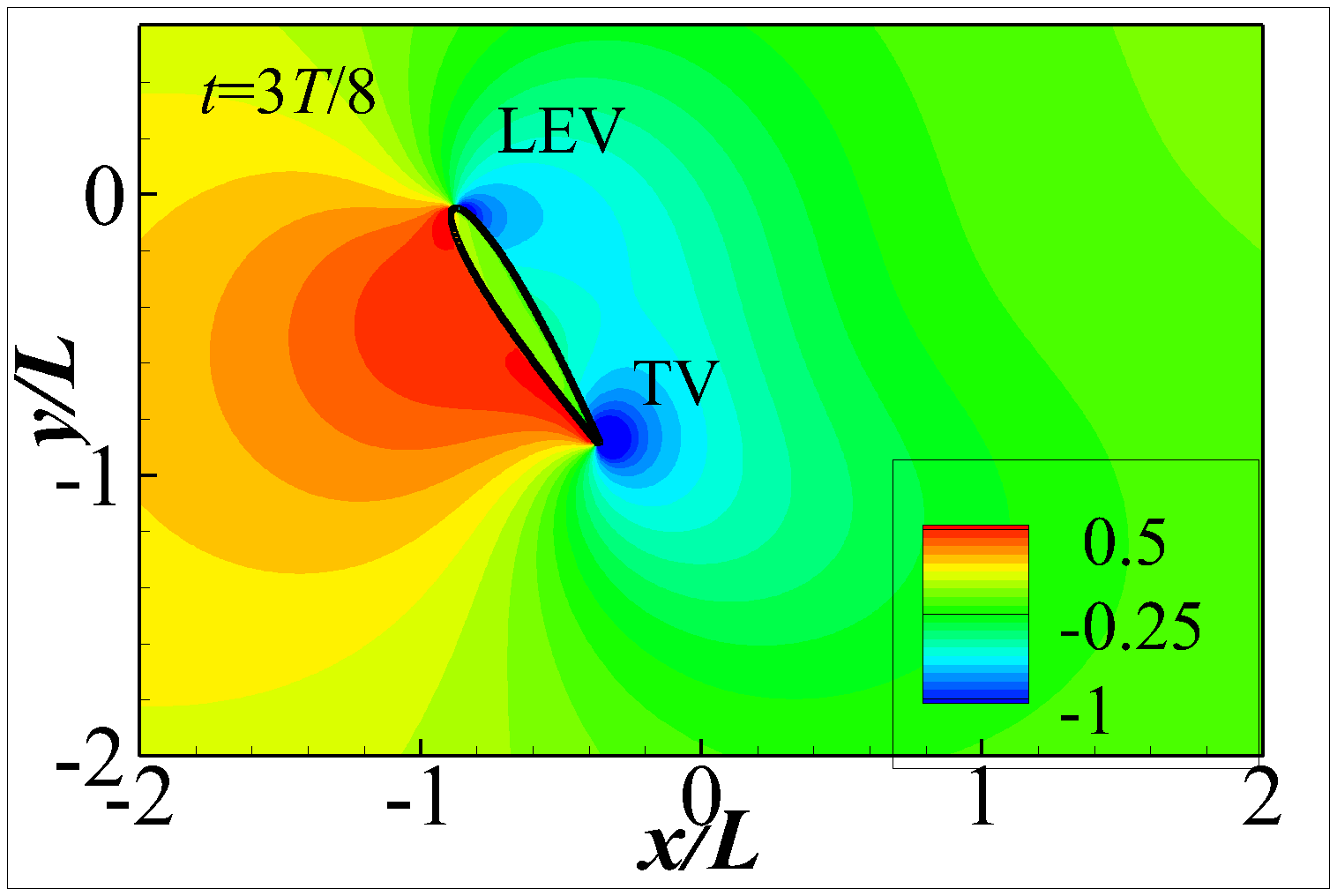}\\
  \end{center}
\caption{Instantaneous contours of $C_p$ in a half period at Re = 115.6 (top) and 23.1 (bottom).}
\label{Fig:flpcpcontour}
\end{figure}

\begin{figure}
  \begin{center}
  \includegraphics[width=2in]{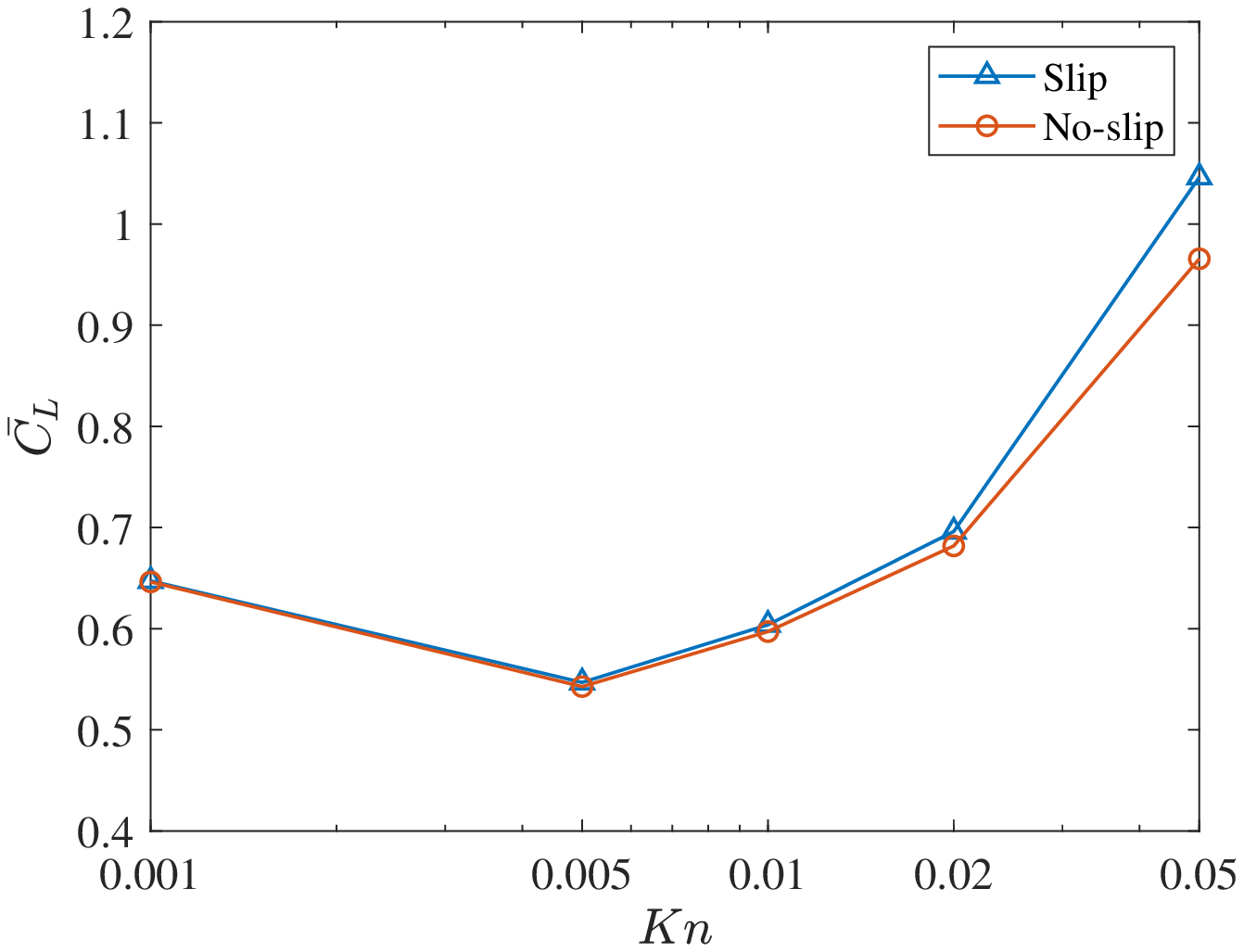}
  \includegraphics[width=2in]{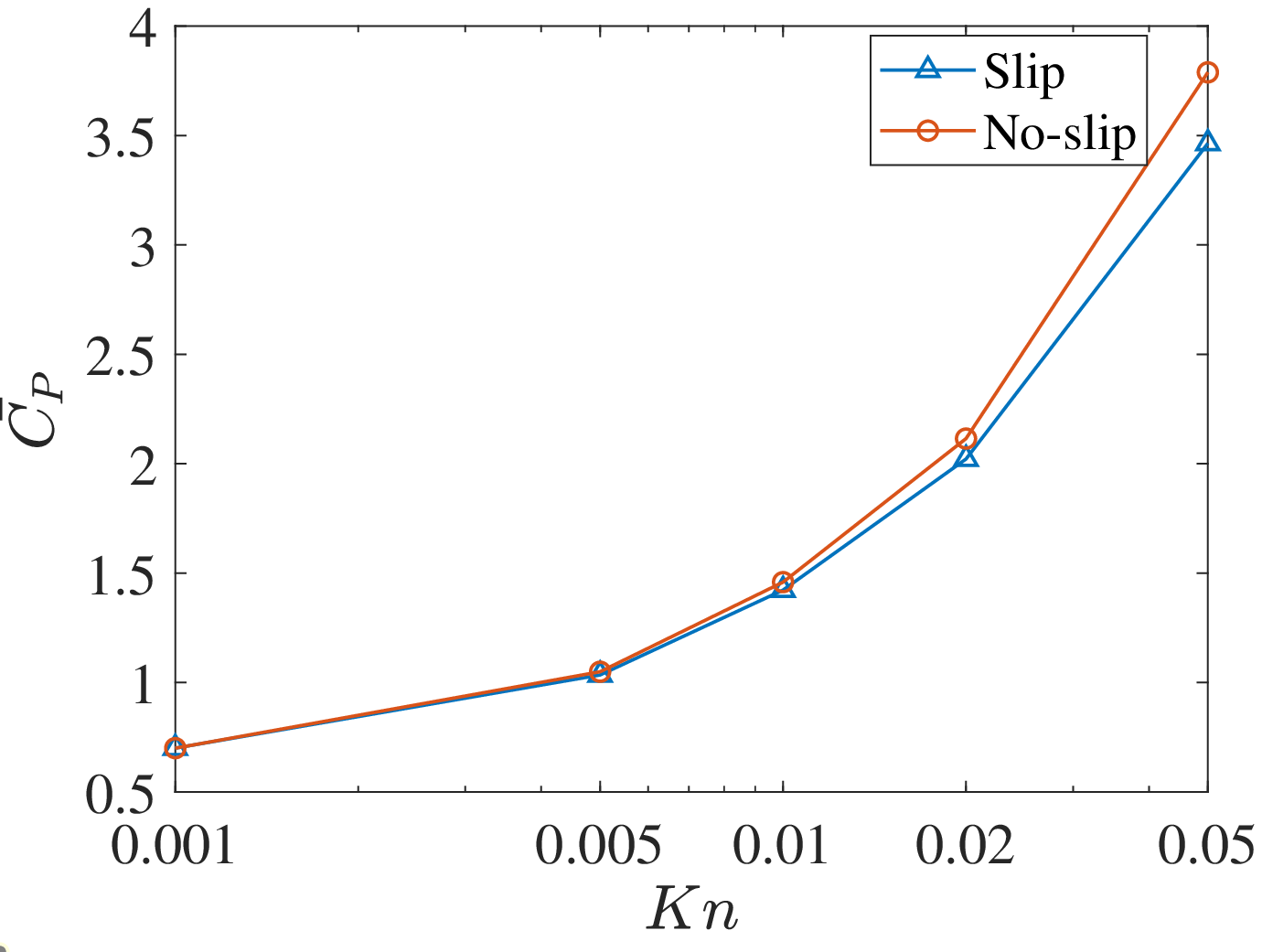}
  \includegraphics[width=2in]{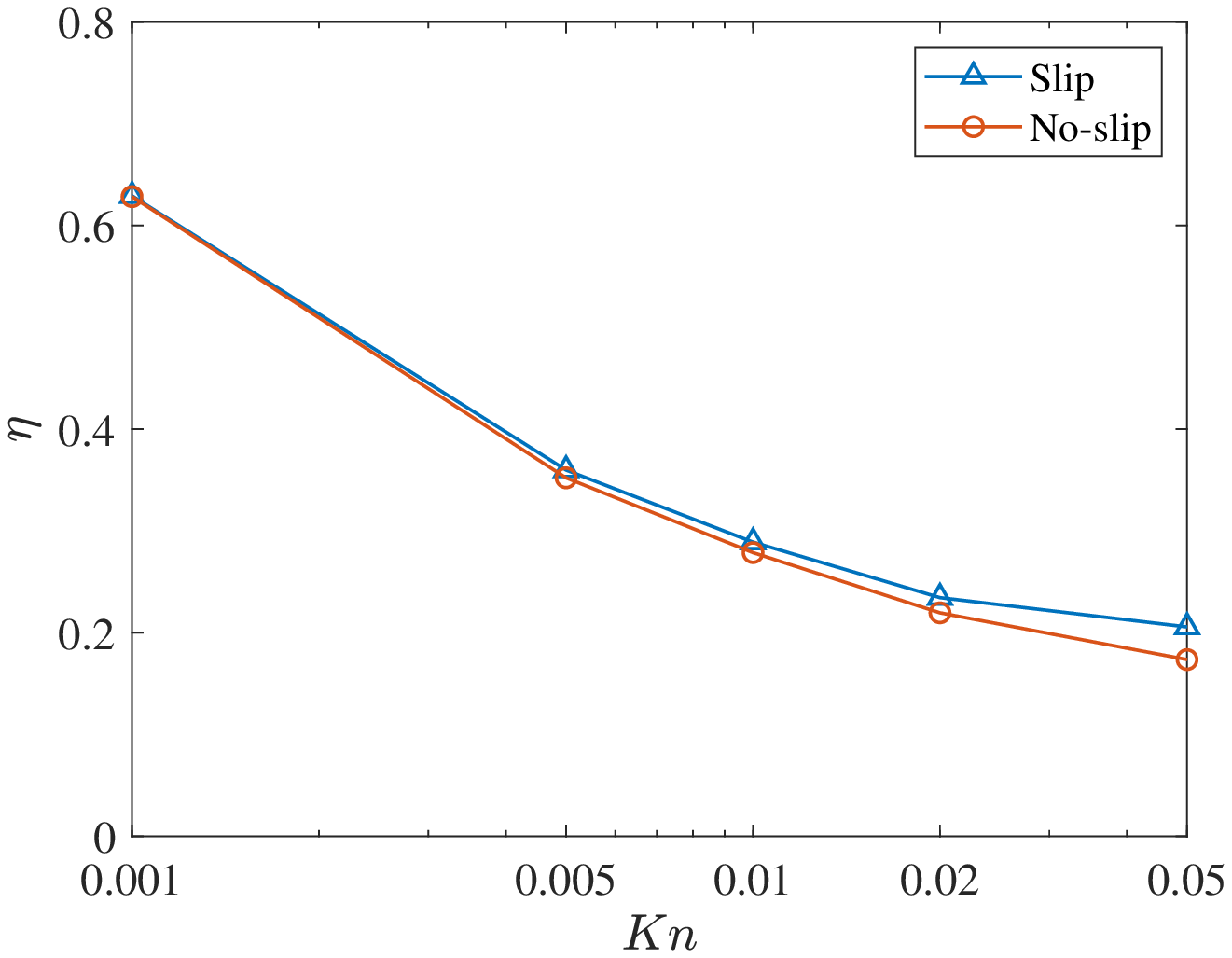}
  \end{center}
\caption{Time-averaged $C_L$ and $C_P$, and $\eta$ at various $Kn$ from 0.001 to 0.05.}
\label{Fig:flpmean}
\end{figure}

Fig.~\ref{Fig:flphis} shows the time histories of $C_L$, $C_x$ and $C_P$ in one period. First, similar time histories are observed for all conditions at $Kn \leq 0.005$ ($Re \leq 23.1$), i.e., two symmetric peaks at around $t = T/4$ and $3T/4$ of $C_L$ and $C_P$ in one period. The major differences are found at the amplitudes of these peaks. When $Re = 115.6$, another two peaks are observed at $t = 0.06T$ and $0.55T$. To explain the Reynolds number effects, the instantaneous contours of pressure coefficient $C_p$ ($=(p-p_0)/(0.5 \rho_0 U^2)$) in a half period are shown in Fig.~\ref{Fig:flpcpcontour} as the other half is approximately symmetric. It is found that a larger $C_p$ is generated at the pressure side at $Re = 23.1$ Compared with that at $Re = 115.6$, which thus makes the peaks of $C_L$ and $C_x$ much higher as shown in Fig.~\ref{Fig:flphis}. The complex time histories of $C_L$ at $Re = 115.6$ can be explained by the vortex shedding at the suction side. Specifically, the leading edge vortex (LEV) and trailing edge vortex (TEV) are close to each other at $Re = 23.1$ and thus only a single vortex can be observed as shown in Fig.~\ref{Fig:flpcpcontour} (bottom) at $t = T/8$ and $2T/$. This single vortex is attached to the suction side from $t = 0T/8$ to $2T/8$. While, the LEV and TEV are detached from each other at $Re = 115.6$ as shown in Fig.~\ref{Fig:flpcpcontour} (top) at $t = 2T/8$. After the TEV separates from the trailing edge, the negative pressure at the suction side increases and thus a low $C_L$ is observed at $t = 0.2T$. With the airfoil continues to stroke to the left and pitch anti-clock-wise, the TEV sheds to the downstream and a big LEV is formed as shown in Fig.~\ref{Fig:flpcpcontour} (top) at $t = 3T/8$, which corresponds to the delayed peak of $C_L$ as shown in Fig.~\ref{Fig:flphis}. 

\begin{figure}
  \begin{center}
  \includegraphics[width=3in]{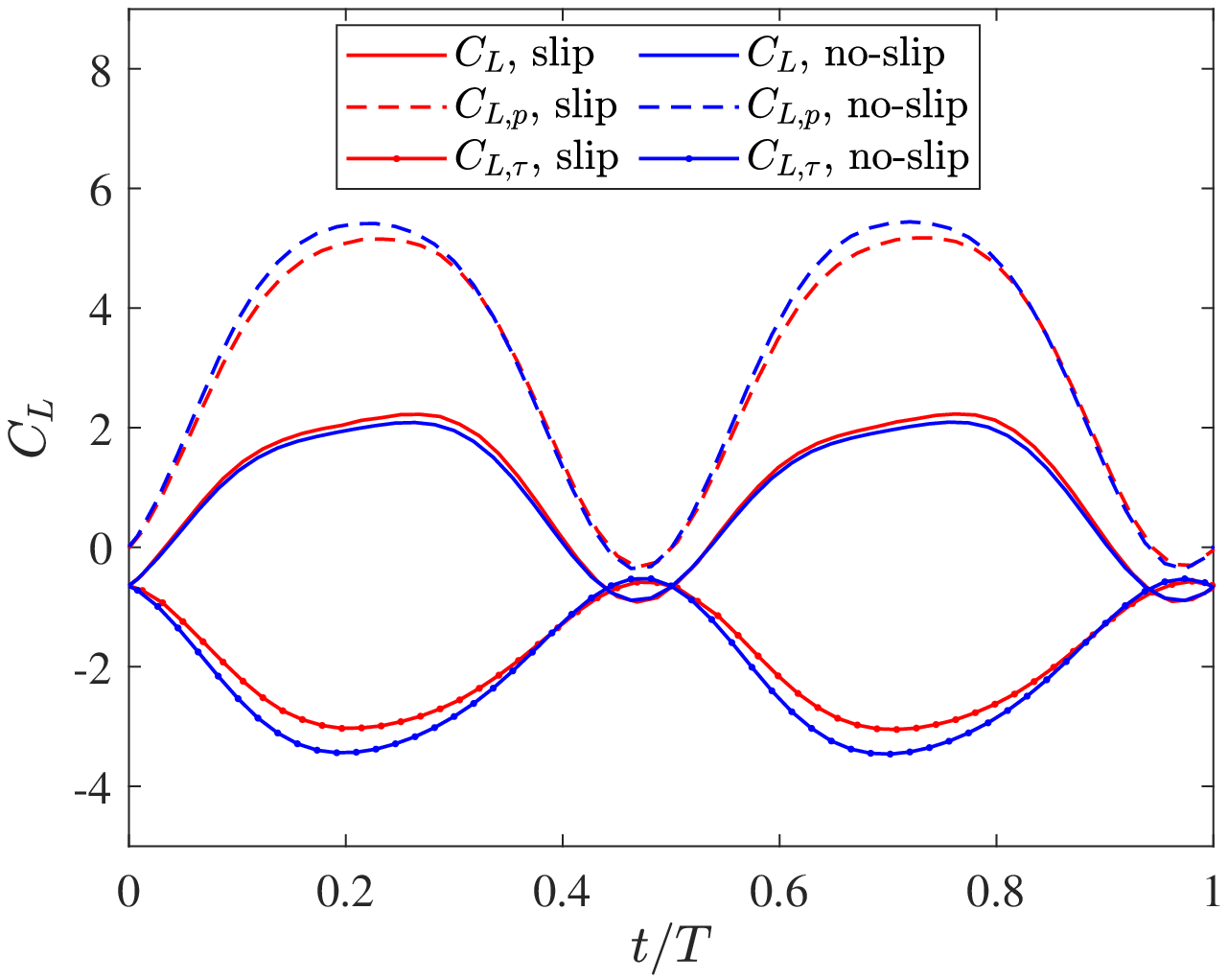}
  \includegraphics[width=3in]{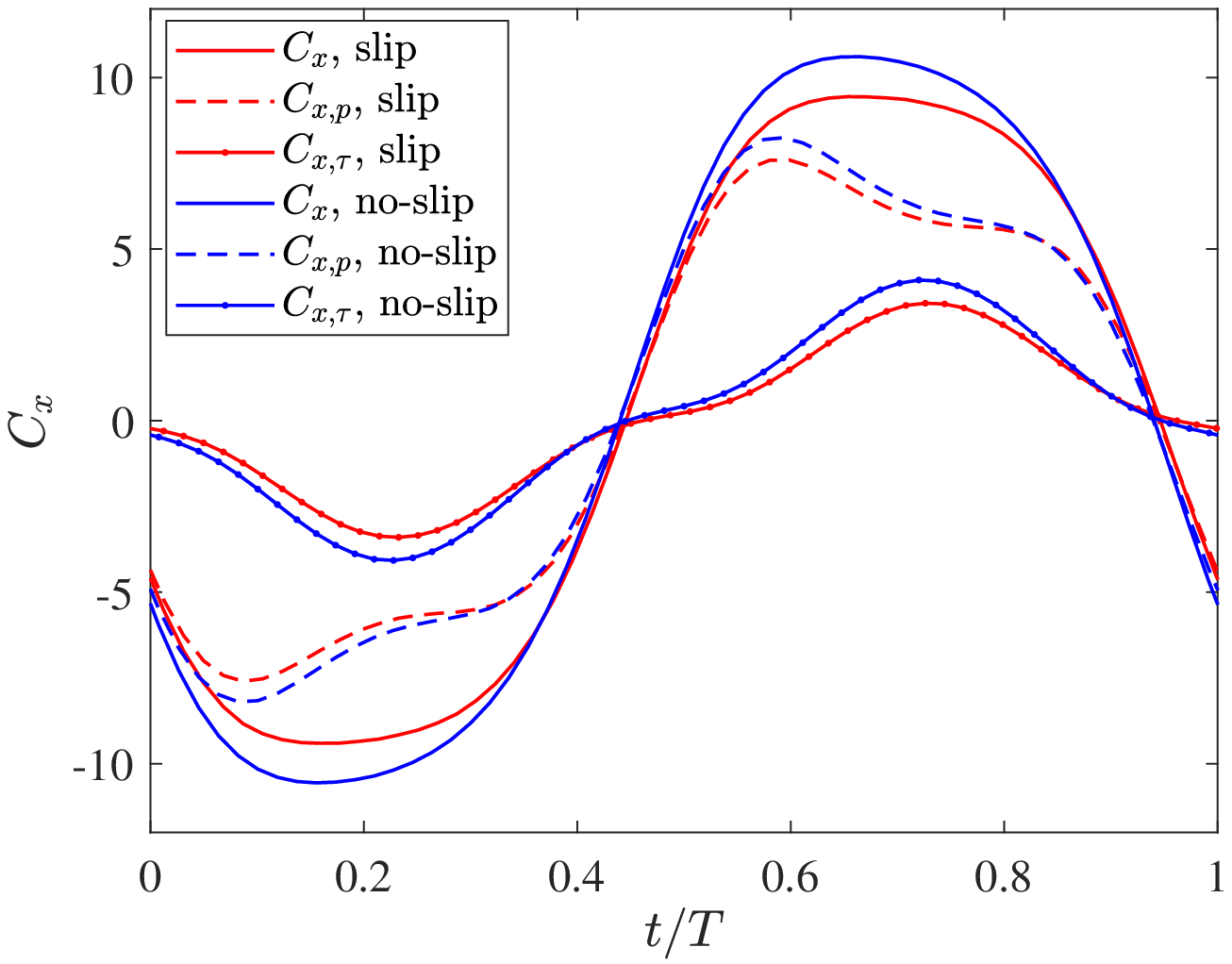}\\
  \end{center}
\caption{Time history of $C_L$ and $C_x$ at $Kn = 0.05$.}
\label{Fig:flphiskn}
\end{figure}

\begin{figure}
  \begin{center}
  \includegraphics[width=1.5in]{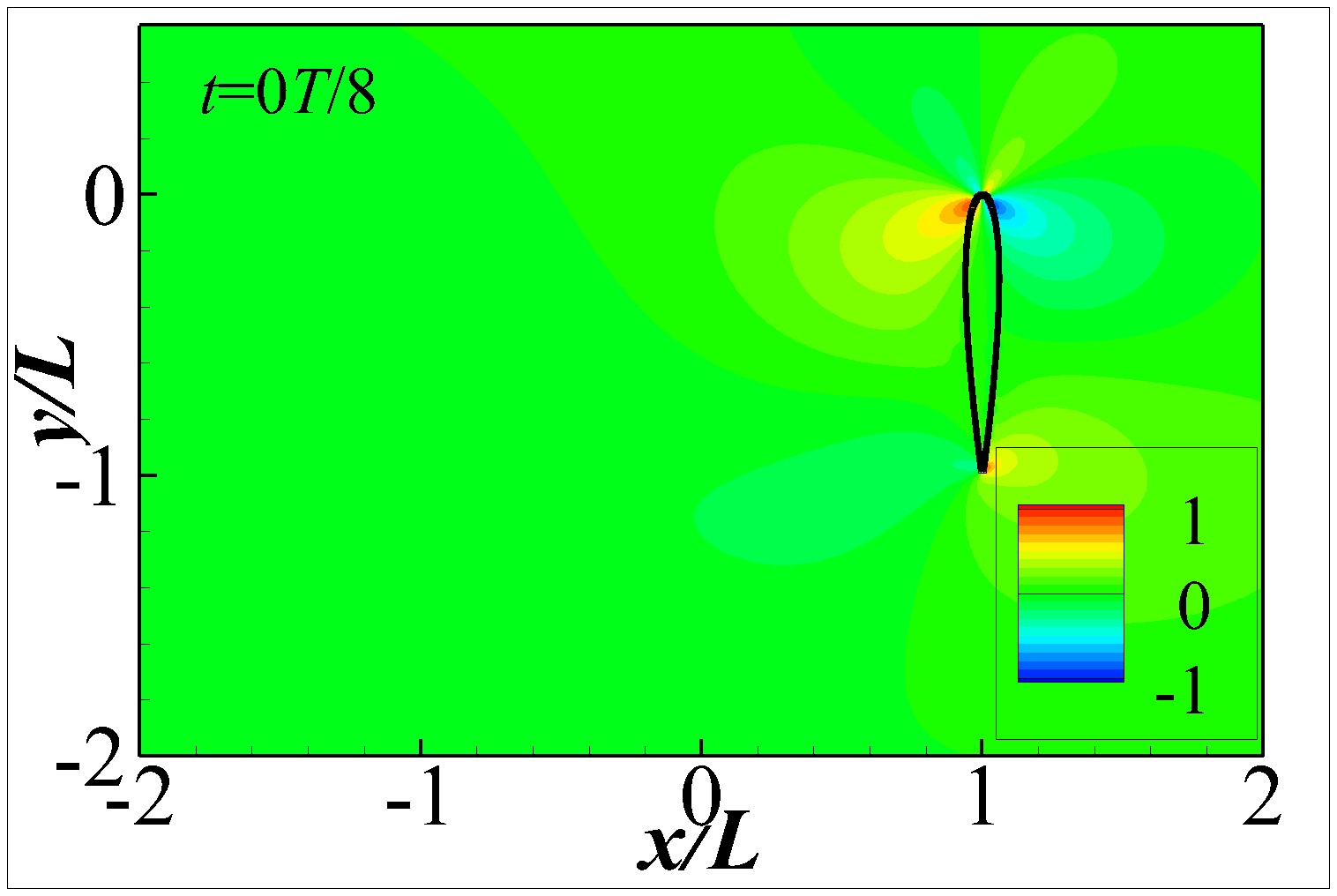}
  \includegraphics[width=1.5in]{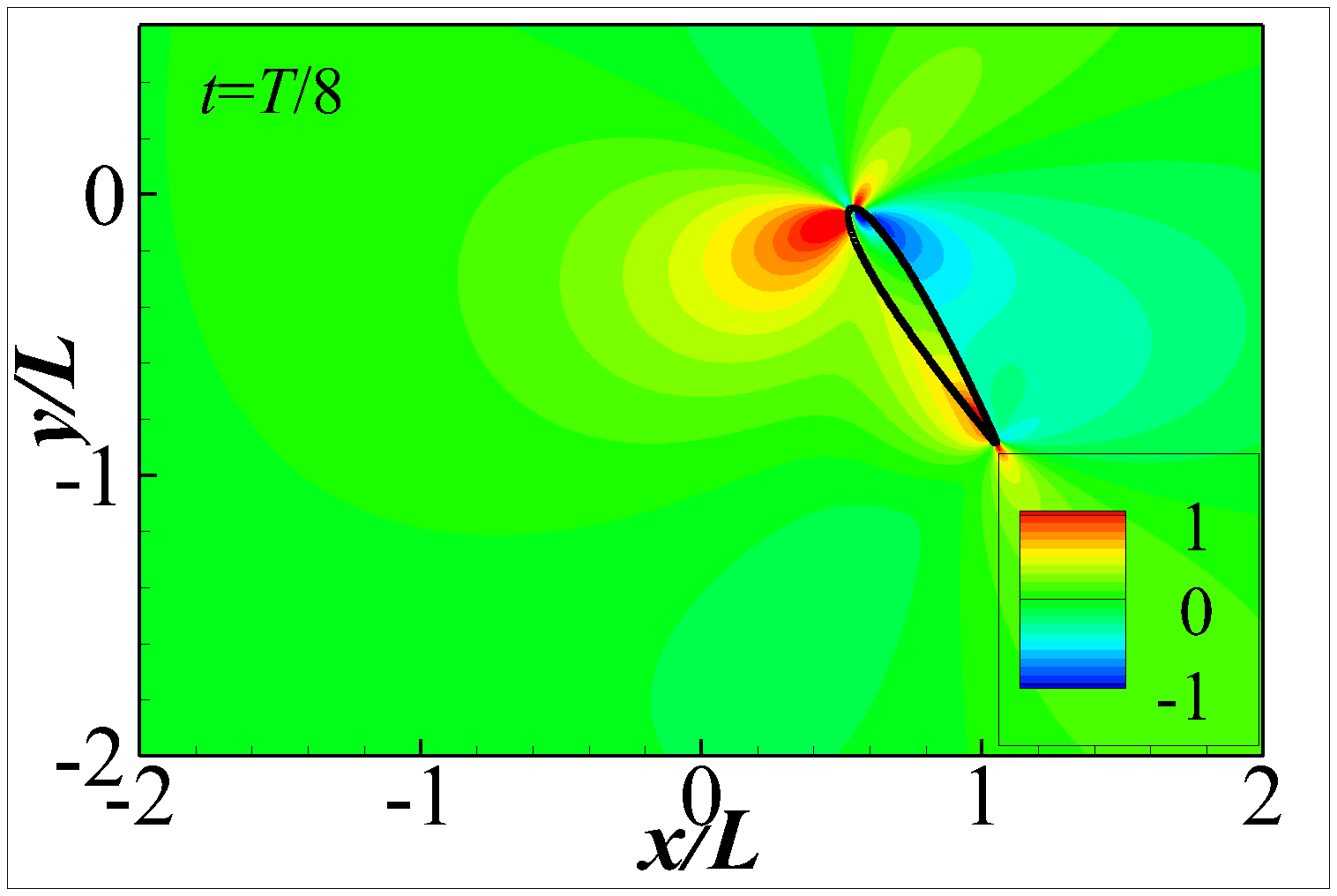}
  \includegraphics[width=1.5in]{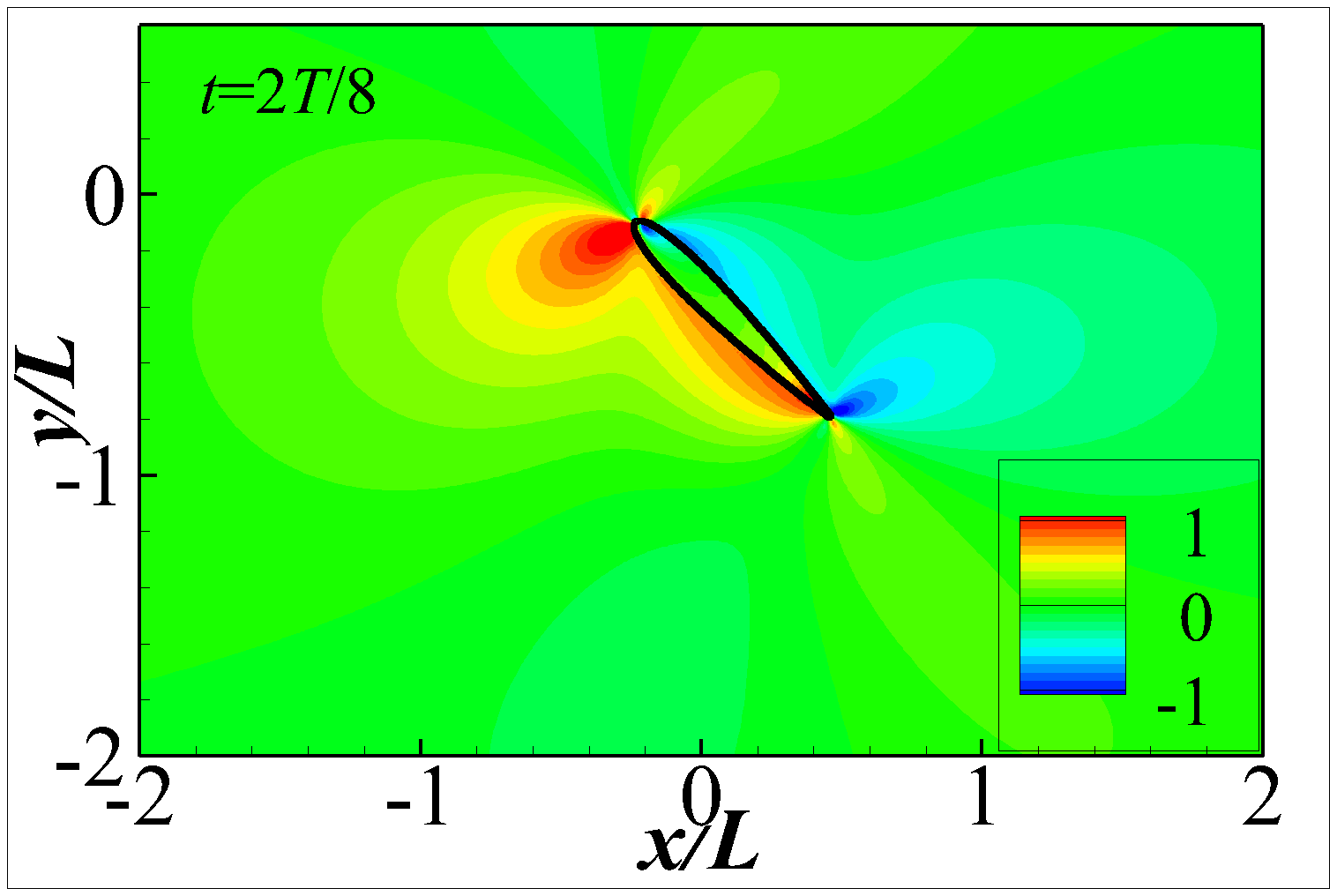}
  \includegraphics[width=1.5in]{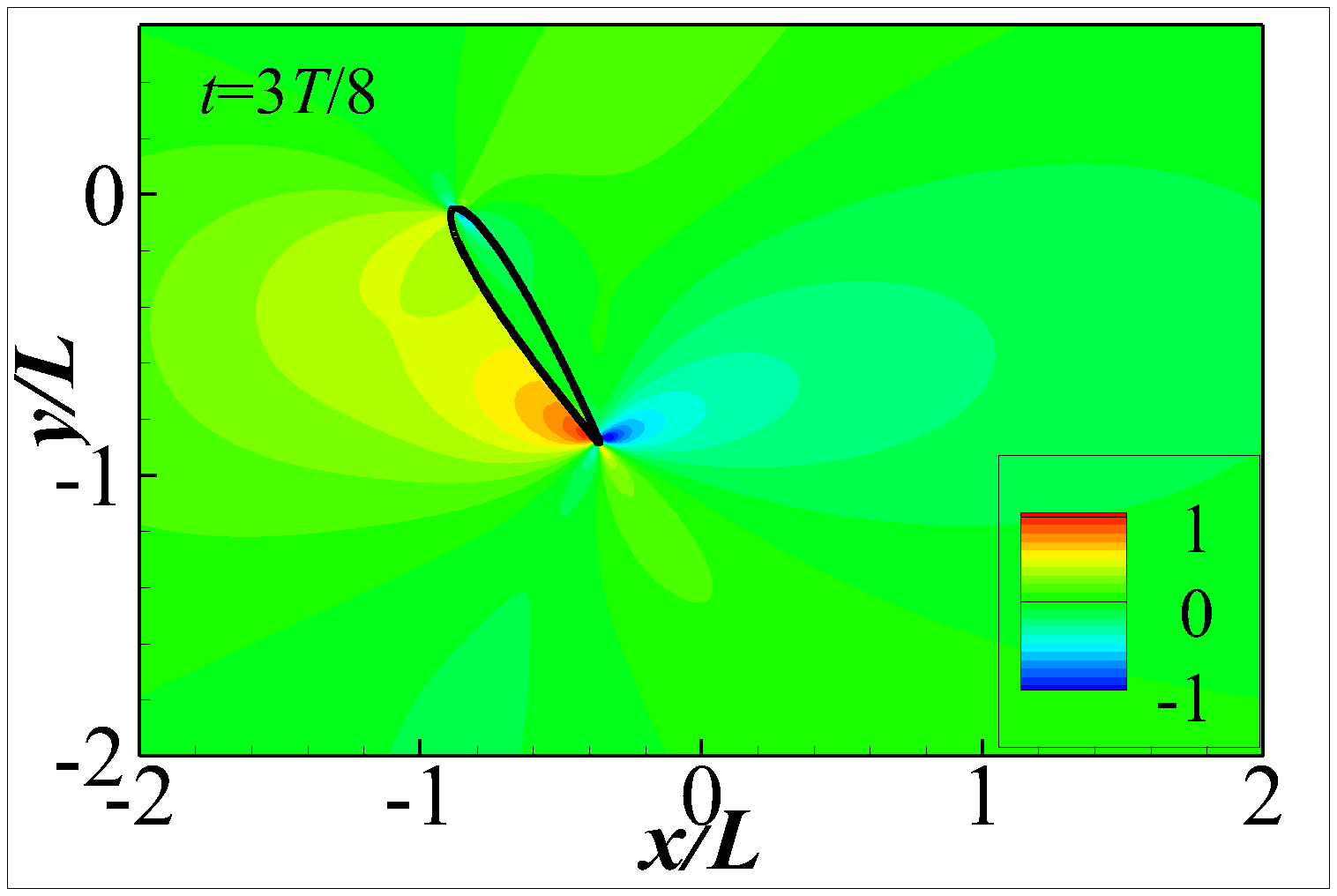}\\
  \includegraphics[width=1.5in]{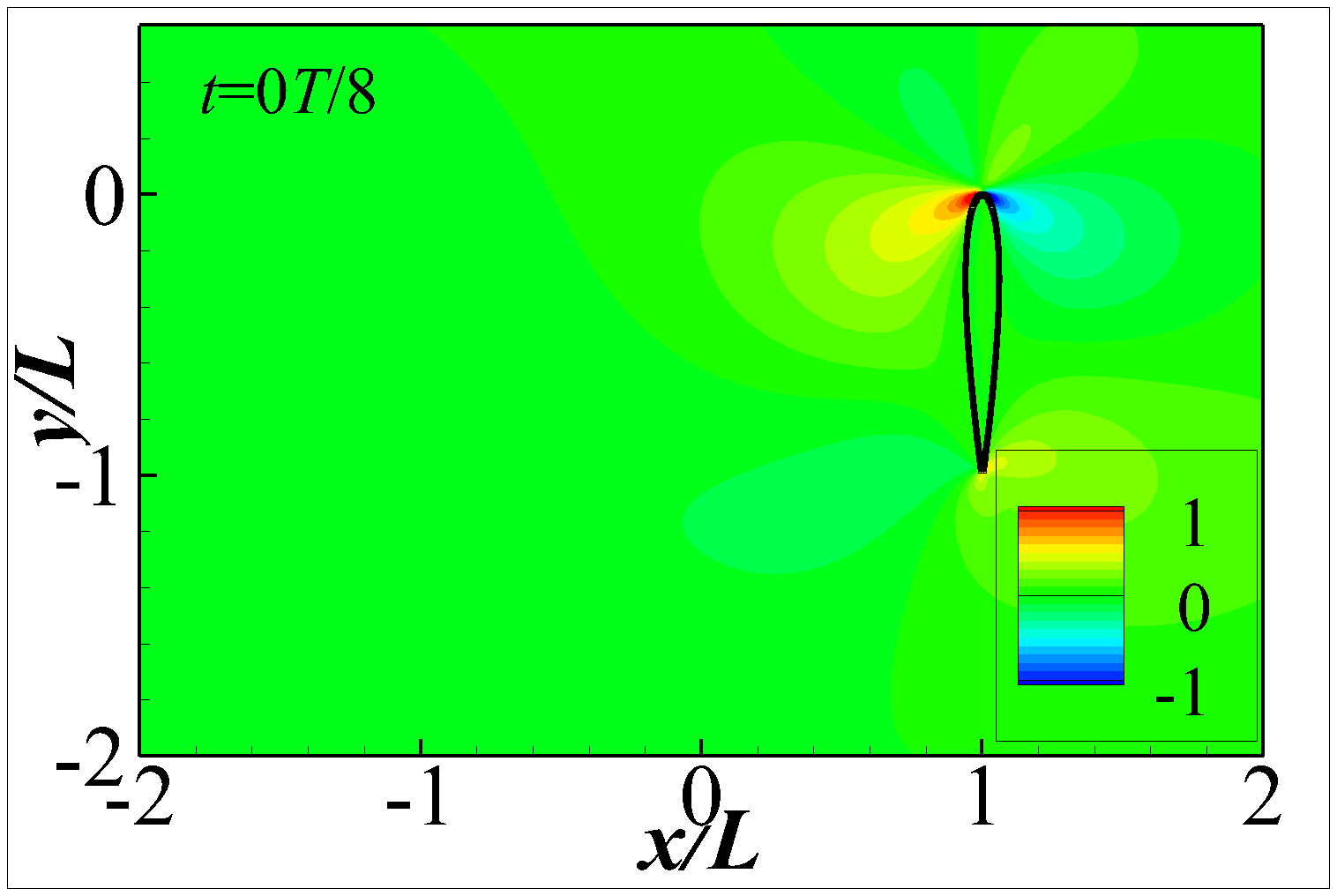}
  \includegraphics[width=1.5in]{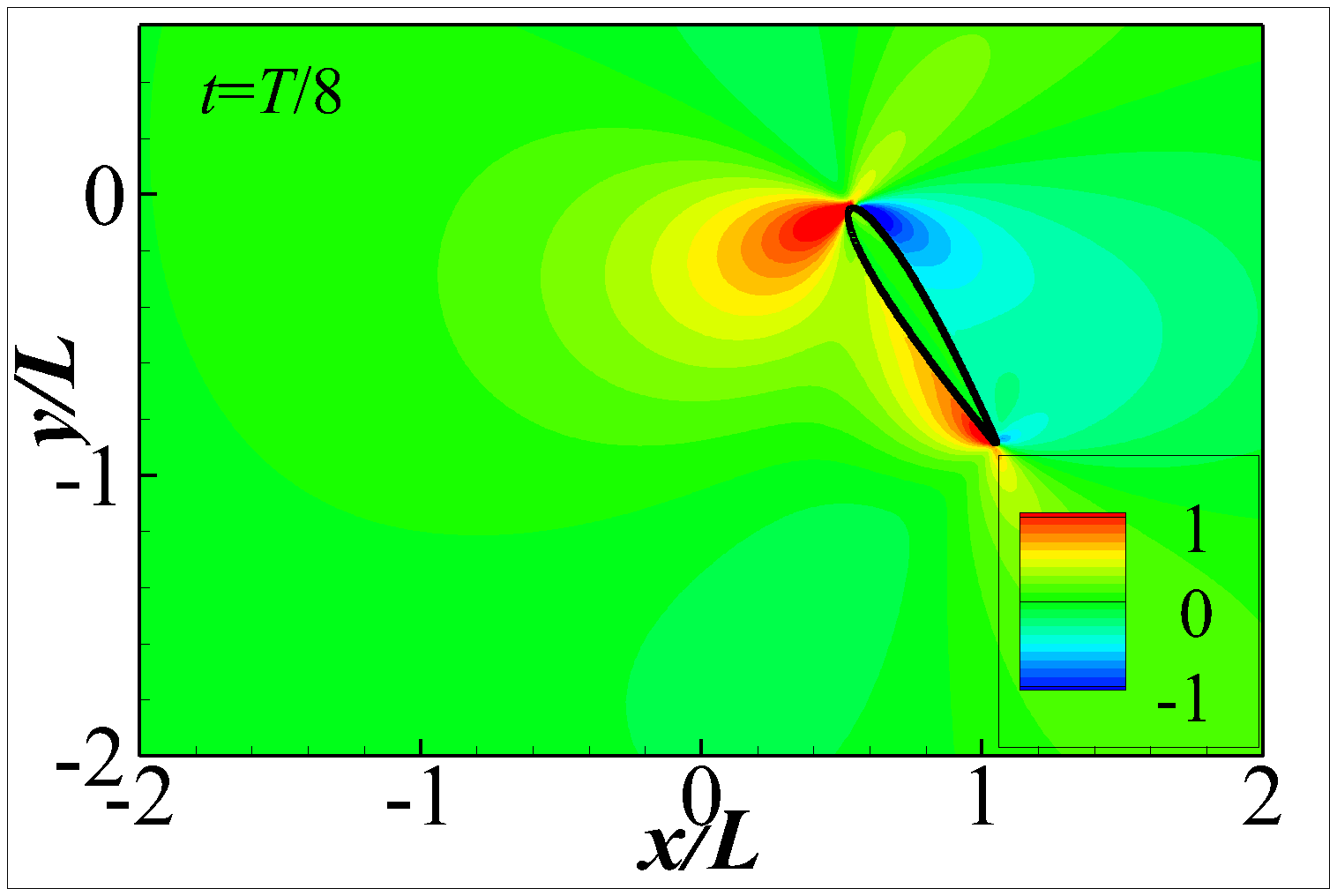}
  \includegraphics[width=1.5in]{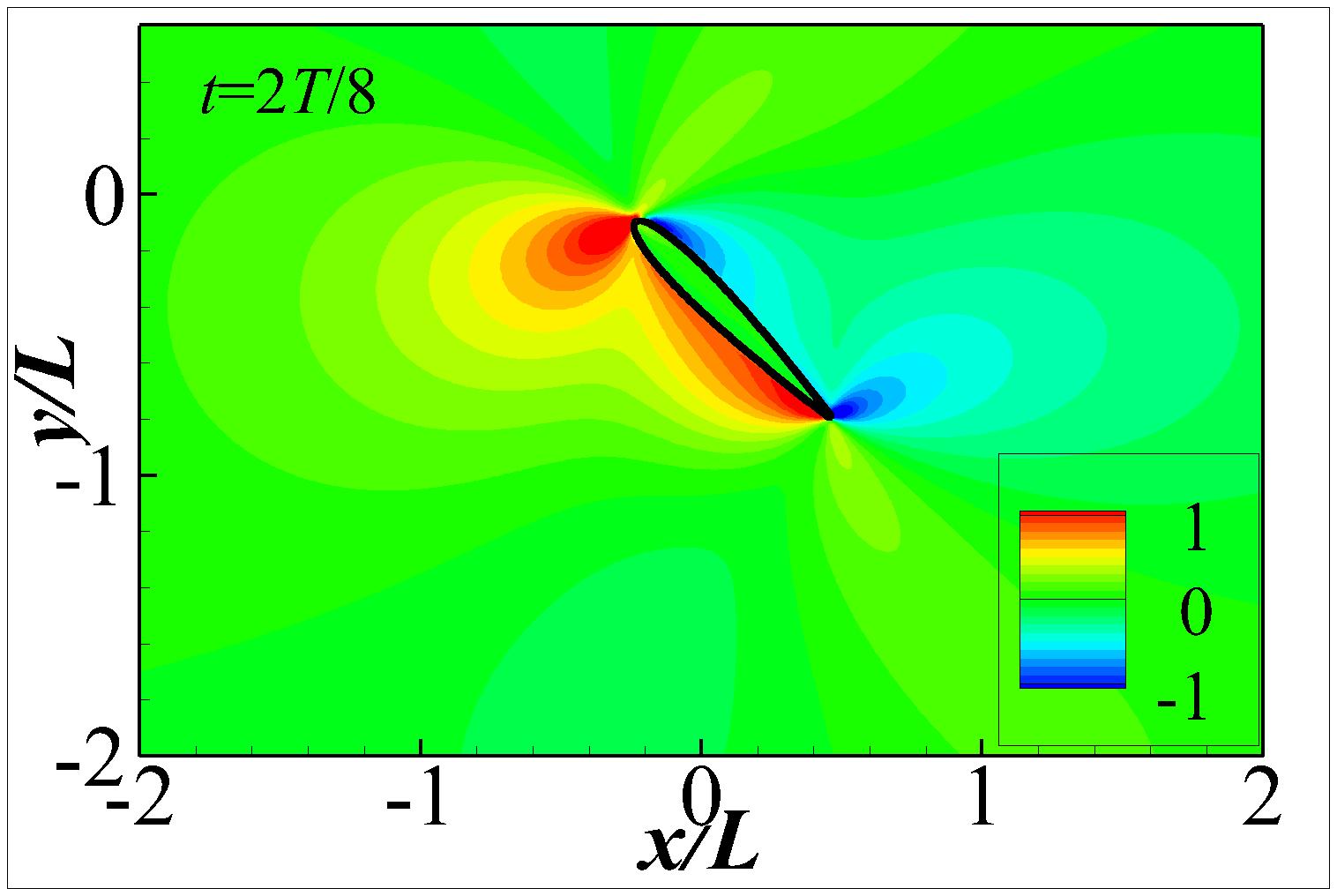}
  \includegraphics[width=1.5in]{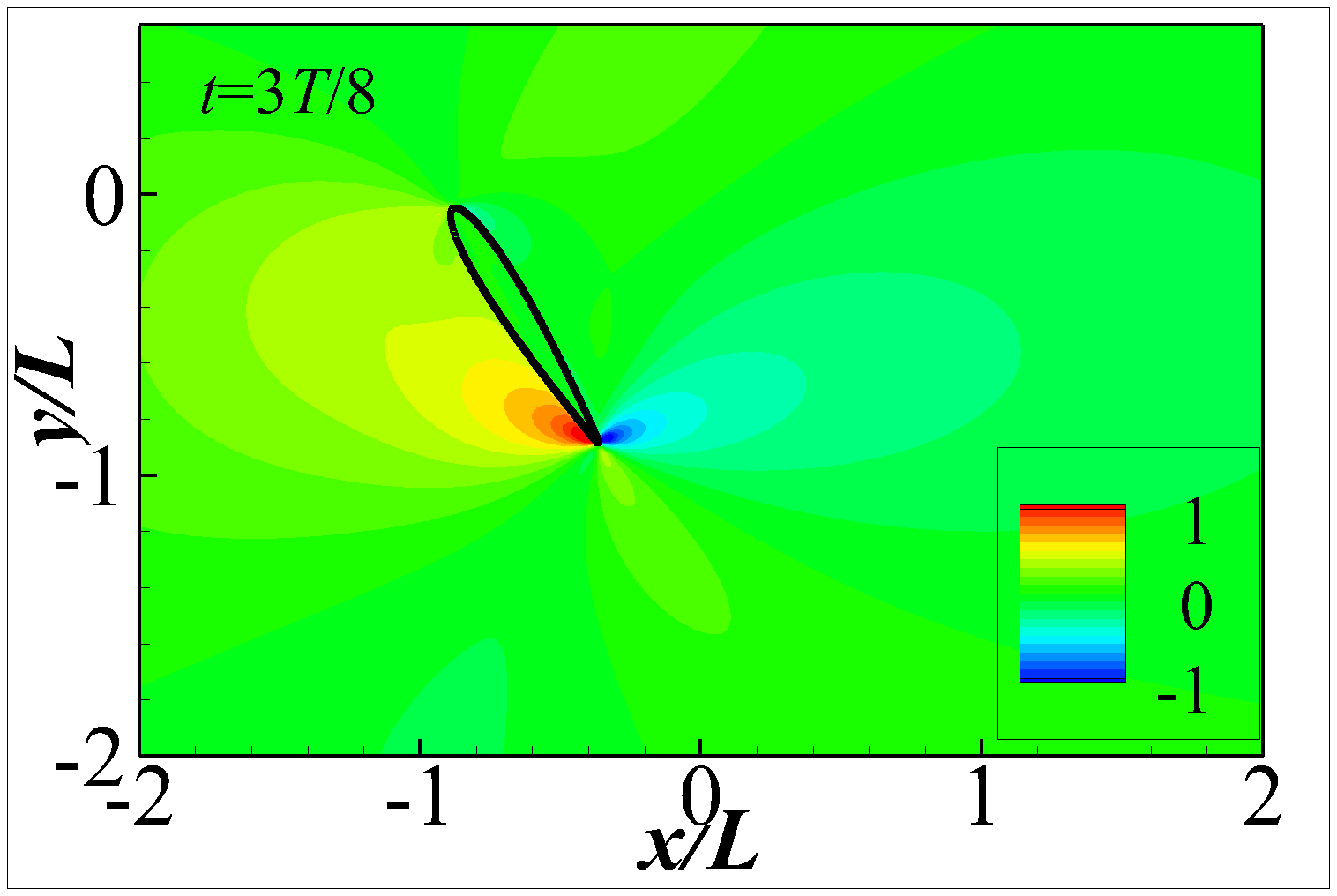}\\
  \end{center}
\caption{Instantaneous contours of vorticity in a half period at Re = 2.3, $Kn = 0.05$ with (top) and without (bottom) slip model.}
\label{Fig:flpknvortcontour}
\end{figure}

Second, the effect of rarefied gas is negligible (less than 1\%) at low Knudsen numbers, i.e., $Kn \leq 0.01$, but it increases with $Kn$. Specifically, the slip velocities enhance the lift generation, reduces the horizontal force and the power consumption. This is because the increasing $Kn$ relief the shear stress at the surface of airfoil, while the viscous stress contributes to the generation of negative lift and the positive horizontal force. Therefore, the rarefied gas effects enhance the aerodynamic performance of airfoil during hovering flight. To quantify these effects, a better comparison of the time-averaged lift coefficient ($\bar{C_L}$) and power coefficient ($\bar{C_P}$), and the efficiency is shown in Fig.~\ref{Fig:flpmean}. It confirms the enhancement of slip velocities on the aerodynamic performance of the hovering arifoil, and such enhancements becomes observable when $Kn \geq 0.01$. $\bar{C_L}$ is increased from 0.97 to 1.05 at $Kn = 0.05$, while $\bar{C_P}$ decreases from 5.56 to 5.08. Therefore, the efficiency $\eta$ increases from 0.197 to 0.234 with an enhancement of around 20\%. To explain the slip velocities effects, $C_L$ and $C_x$ are decomposed into the pressure part ($C_{L,p}$ and $C_{x,p}$) and viscous part ($C_{L,\tau}$ and $C_{x,\tau}$), and the time histories at $Kn = 0.05$ are shown in Fig.~\ref{Fig:flphiskn}. When the slip velocity is involved, the amplitudes of both pressure and viscous components are decreased. While, the decrease of viscous component is higher than that of the pressure component at $Kn = 0.05$ for the very low Reynolds number, i.e., $Re = 2.3$. As the pressure and viscous contribute to the positive and negative of lift generation, the overall effects of rarefied gas leads to the increase of $C_L$. On the other hand, the the contribution of pressure and viscous are opposite to the horizontal force, which thus leads to the decrease of $C_x$. As the amplitude of $C_x$ is much higher than $C_L$, $C_P$ as a combined effects of both vertical and horizontal forces decreases when the slip velocity is included.

\subsection{A 3D flapping wing hovering in rarefied gas flow}

\begin{figure}
 \begin{center}
  \includegraphics[width=3.0in]{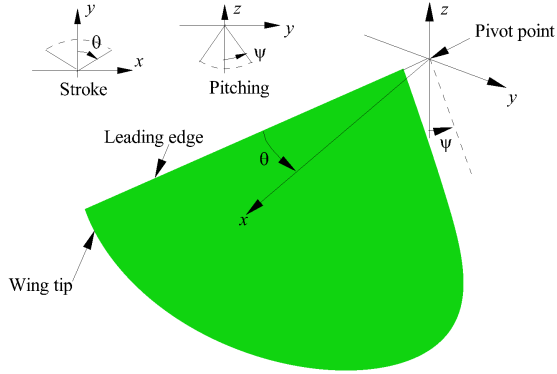}
  \end{center}
\caption{Schematic of a wing flapping in three-dimensional domain~\cite{wang2020numerical}.}
\label{Fig:wing3dsch}
\end{figure}

In this section, we further consider a 3D flapping wing hovering in rarefied gas flow. The wing geometry is generated based on the model proposed by Ellington~\cite{ellington1984aerodynamics} with $\bar{r_1} = 0.43$. The wing geometry and its motion are shown in Fig.~\ref{Fig:wing3dsch} and more details can be found in Refs.~\cite{wang2020numerical}. The wing undergoes stroke and pitching motion according to the following equations
\begin{equation}
\theta(t)=\frac{\theta_m}{2} {\rm sin}(2\pi f_0 t+ \frac{\pi}{2}),\quad \psi(t)=\frac{\psi_m}{2}{\rm sin}(2 \pi f_0 t+ \phi),
\label{eq:flapping_motion}
\end{equation}
where $\theta$ and $\psi$ are respectively the stroke and pitching angle, $\phi$ is the phase angle between wing stroke and pitching motion, $f_0$ is the flapping frequency, $\theta_m$ and $\psi_m$ are the amplitude of stroke and pitching, respectively. The governing parameters for this physical problem including Knudsen number, Mach number,  wing-to-fluid mass ratio, dimensionless flapping frequency (represents the wing flexibility with 0 means rigid) are given by
\begin{equation}
Kn = \frac{\lambda}{\bar{c}}, \quad M=\frac{U}{c}, \quad m^*=\frac{\rho_s h}{\rho \bar{c}},\quad \omega^*=\frac{2\pi f_0}{\omega_n}, 
\label{eq:flapping_para}
\end{equation}
where $U=2 f_0 (R+0.1\bar{c}) \theta_m$ is the mean stroke velocity of the leading edge, $c$ is the sound speed of the fluid, $\rho$ is the density of the fluid, $\rho_s$ is the density of the wing, $h$ is the thickness of the wing, $\omega_n=k_n^2/\bar{c}^2\sqrt{E_B/\rho_s}$ with $k_n=1.8751$ (the frequency of the first natural vibration mode of a sing beam with fixed leading edge~\cite{tian2013force}) and $E_B = E_y h^3/12$ being the bending rigidity ($E_y$ is the Young's modulus of the wing). Here, $\beta_m=2\pi/3$, $\psi = \pi/4$, $\phi=0$, $AR=\bar{c}/R=1.5$ are used for the wing geometry and motion~\cite{dai2012dynamic,shahzad2016effects}. Different with the nearly incompressible flow regime in Section \ref{sec:naca2d}, the compressible with a Mach number of 0.4 is considered as the high flapping velocity is capable to generate high lift force in the ultra-low density Martian atmosphere. A flexible wing with $m^*=50.0$ and $\omega^*=0.3$ (chosen based on Ref.~\cite{wang2022numerical} considering Martian environment) is examined at $Kn = 0.03$ within the slip flow regime, with its counter-part without slip flow effects being simulated for comparison. The lift coefficient which quantifies the aerodynamic performance of the wing is defined as $C_L=\frac{2F_z}{\rho U^2 \bar{c}}$, where $F_z$ are the force acting on the wing by the ambient fluid in $z$ direction. Fig.~ shows a direct comparison of the time histories of $C_L$ for the the flexible wing considered with no-slip and without slip boundary conditions. It is observed that the overall $C_L$ generated by the wing with slip boundary condition outperforms its counterpart working with no-slip boundary condition slightly, which is attributed to the balance of the reductions of the negative lift contribution from friction and the positive lift contribution from pressure as expected. This implies that the slip flow regime has the potential to benefit the aerodynamic performance of flapping wing by tuning its kinematics, which will be studied in the future.

\begin{figure}
 \begin{center}
  \includegraphics[width=3.5in]{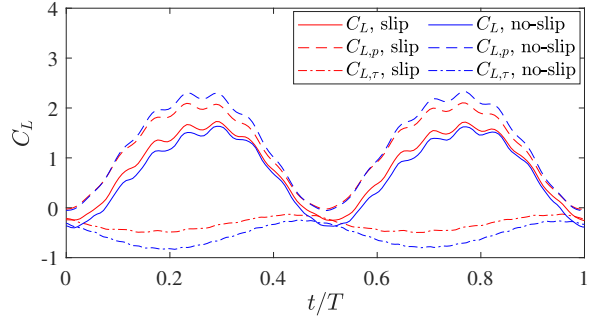}
  \end{center}
\caption{Time histories of $C_L$ generated by a flexible wing with $m^*=50$ and $\omega^*=0.3$ at $Kn=0.03$.}
\label{Fig:clwing3d}
\end{figure}

\section{Conclusions}
%%%%%%%%%%%%%%
In this paper, an immersed boundary method for the fluid--structure--thermal interaction in rarefied gas flow is presented. In this method, the slip model is incorporated with the penalty immersed boundary method to address the velocity and temperature jump conditions at the fluid--structure interface in rarefied gas flow within slip regime. 

Several validations are conducted including Poiseuille flow in a 2D pipe, flow around a 2D NACA airfoil, moving square cylinder in a 2D pipe, flow around a sphere and moving sphere in quiescent flow. The present results show good agreement with the previous published data obtained by other methods, and it confirms the the good ability of the proposed method in handling fluid--structure--thermal interaction for both incompressible and highly compressible rarefied gas flow. 

To overcome the incapability of Navier-Stokes equations at high local Knudsen numbers in supersonic flow, an artificial viscosity is introduced to ease the sharp transition at the shock wave front. The validation of flow around a sphere confirms the improved prediction of density, velocity and temperature profiles along the stagnation line.

Inspired by Martian exploration, the application of proposed method to study the aerodynamics of a flapping wing hovering in rarefied gas flow is conducted in both 2D and 3D domain considering incompressible and subsonic flow, respectively. The 2D simulation in compressible flow indicate that the rarefied gas effect benefits the lift generation due to the relief of shear stress which contributes to the negative lift generation. The 3D simulation in subsonic flow show that the slip boundary condition benefits the lift generation slightly and shed some lights for the further studies to optimize flapping wings with variations in geometry and kinetics for the flight in rarefied gas.

\section*{Acknowledgements}
%%%%%%%%%%%%%%
This work was partially supported by Australian Research Council Discovery Project (project number DP200101500) and conducted with the assistance of resources from the National Computational Infrastructure (NCI), which is supported by the Australian Government.
%%%%%%%%%%%%
% References
%%%%%%%%%%%%

\bibliographystyle{./class/model1-num-names}
\bibliography{paper}

\begin{thebibliography}{50}
\expandafter\ifx\csname natexlab\endcsname\relax\def\natexlab#1{#1}\fi
\providecommand{\bibinfo}[2]{#2}
\ifx\xfnm\relax \def\xfnm[#1]{\unskip,\space#1}\fi
%Type = Article
\bibitem[{Wang and Tian(2019)}]{wang2019numerical}
\bibinfo{author}{L.~Wang}, \bibinfo{author}{F.-B. Tian},
\newblock \bibinfo{title}{Numerical simulation of flow over a parallel
  cantilevered flag in the vicinity of a rigid wall},
\newblock \bibinfo{journal}{Physical Review E} \bibinfo{volume}{99}
  (\bibinfo{year}{2019}) \bibinfo{pages}{053111}.
%Type = Article
\bibitem[{Tian et~al.(2010)Tian, Luo, Zhu, and Lu}]{tian2010interaction}
\bibinfo{author}{F.-B. Tian}, \bibinfo{author}{H.~Luo},
  \bibinfo{author}{L.~Zhu}, \bibinfo{author}{X.-Y. Lu},
\newblock \bibinfo{title}{Interaction between a flexible filament and a
  downstream rigid body},
\newblock \bibinfo{journal}{Physical Review E} \bibinfo{volume}{82}
  (\bibinfo{year}{2010}) \bibinfo{pages}{026301}.
%Type = Article
\bibitem[{Toja-Silva et~al.(2018)Toja-Silva, Kono, Peralta, Lopez-Garcia, and
  Chen}]{toja2018review}
\bibinfo{author}{F.~Toja-Silva}, \bibinfo{author}{T.~Kono},
  \bibinfo{author}{C.~Peralta}, \bibinfo{author}{O.~Lopez-Garcia},
  \bibinfo{author}{J.~Chen},
\newblock \bibinfo{title}{A review of computational fluid dynamics ({CFD})
  simulations of the wind flow around buildings for urban wind energy
  exploitation},
\newblock \bibinfo{journal}{Journal of Wind Engineering and Industrial
  Aerodynamics} \bibinfo{volume}{180} (\bibinfo{year}{2018})
  \bibinfo{pages}{66--87}.
%Type = Article
\bibitem[{Kamakoti and Shyy(2004)}]{kamakoti2004fluid}
\bibinfo{author}{R.~Kamakoti}, \bibinfo{author}{W.~Shyy},
\newblock \bibinfo{title}{Fluid--structure interaction for aeroelastic
  applications},
\newblock \bibinfo{journal}{Progress in Aerospace Sciences}
  \bibinfo{volume}{40} (\bibinfo{year}{2004}) \bibinfo{pages}{535--558}.
%Type = Article
\bibitem[{Huang et~al.(2021)Huang, Tian, Young, and Lai}]{huang2021transition}
\bibinfo{author}{Q.~Huang}, \bibinfo{author}{F.-B. Tian},
  \bibinfo{author}{J.~Young}, \bibinfo{author}{J.~C. Lai},
\newblock \bibinfo{title}{Transition to chaos in a two-sided collapsible
  channel flow},
\newblock \bibinfo{journal}{Journal of Fluid Mechanics} \bibinfo{volume}{926}
  (\bibinfo{year}{2021}).
%Type = Article
\bibitem[{Dowell and Hall(2001)}]{dowell2001modeling}
\bibinfo{author}{E.~H. Dowell}, \bibinfo{author}{K.~C. Hall},
\newblock \bibinfo{title}{Modeling of fluid-structure interaction},
\newblock \bibinfo{journal}{Annual Review of Fluid Mechanics}
  \bibinfo{volume}{33} (\bibinfo{year}{2001}) \bibinfo{pages}{445--490}.
%Type = Article
\bibitem[{Yang et~al.(2019)Yang, Shu, Yang, and Wu}]{yang2019improved}
\bibinfo{author}{L.~Yang}, \bibinfo{author}{C.~Shu}, \bibinfo{author}{W.~Yang},
  \bibinfo{author}{J.~Wu},
\newblock \bibinfo{title}{An improved three-dimensional implicit discrete
  velocity method on unstructured meshes for all knudsen number flows},
\newblock \bibinfo{journal}{Journal of Computational Physics}
  \bibinfo{volume}{396} (\bibinfo{year}{2019}) \bibinfo{pages}{738--760}.
%Type = Article
\bibitem[{Tang et~al.(2005)Tang, Tao, and He}]{tang2005lattice}
\bibinfo{author}{G.~Tang}, \bibinfo{author}{W.~Tao}, \bibinfo{author}{Y.~He},
\newblock \bibinfo{title}{Lattice {B}oltzmann method for gaseous microflows
  using kinetic theory boundary conditions},
\newblock \bibinfo{journal}{Physics of Fluids} \bibinfo{volume}{17}
  (\bibinfo{year}{2005}) \bibinfo{pages}{058101}.
%Type = Article
\bibitem[{Ahangar et~al.(2019)Ahangar, Ayani, and
  Esfahani}]{ahangar2019simulation}
\bibinfo{author}{E.~K. Ahangar}, \bibinfo{author}{M.~B. Ayani},
  \bibinfo{author}{J.~A. Esfahani},
\newblock \bibinfo{title}{Simulation of rarefied gas flow in a microchannel
  with backward facing step by two relaxation times using {L}attice {B}oltzmann
  method--slip and transient flow regimes},
\newblock \bibinfo{journal}{International Journal of Mechanical Sciences}
  \bibinfo{volume}{157} (\bibinfo{year}{2019}) \bibinfo{pages}{802--815}.
%Type = Article
\bibitem[{Bhagat et~al.(2019)Bhagat, Gijare, and
  Dongari}]{bhagat2019implementation}
\bibinfo{author}{A.~Bhagat}, \bibinfo{author}{H.~Gijare},
  \bibinfo{author}{N.~Dongari},
\newblock \bibinfo{title}{Implementation of knudsen layer phenomena in rarefied
  high-speed gas flows},
\newblock \bibinfo{journal}{Journal of Aerospace Engineering}
  \bibinfo{volume}{32} (\bibinfo{year}{2019}) \bibinfo{pages}{04019100}.
%Type = Article
\bibitem[{Lockerby et~al.(2004)Lockerby, Reese, Emerson, and
  Barber}]{lockerby2004velocity}
\bibinfo{author}{D.~A. Lockerby}, \bibinfo{author}{J.~M. Reese},
  \bibinfo{author}{D.~R. Emerson}, \bibinfo{author}{R.~W. Barber},
\newblock \bibinfo{title}{Velocity boundary condition at solid walls in
  rarefied gas calculations},
\newblock \bibinfo{journal}{Physical Review E} \bibinfo{volume}{70}
  (\bibinfo{year}{2004}) \bibinfo{pages}{017303}.
%Type = Article
\bibitem[{Maxwell(1879)}]{maxwell1879vii}
\bibinfo{author}{J.~C. Maxwell},
\newblock \bibinfo{title}{Vii. on stresses in rarified gases arising from
  inequalities of temperature},
\newblock \bibinfo{journal}{Philosophical Transactions of the Royal Society of
  London}  (\bibinfo{year}{1879}) \bibinfo{pages}{231--256}.
%Type = Article
\bibitem[{Mitsuya(1993)}]{mitsuya1993modified}
\bibinfo{author}{Y.~Mitsuya},
\newblock \bibinfo{title}{Modified reynolds equation for ultra-thin film gas
  lubrication using 1.5-order slip-flow model and considering surface
  accommodation coefficient},
\newblock \bibinfo{journal}{Journal of Tribology} \bibinfo{volume}{115}
  (\bibinfo{year}{1993}) \bibinfo{pages}{289--294}.
%Type = Article
\bibitem[{Loyalka et~al.(1975)Loyalka, Petrellis, and
  Storvick}]{loyalka1975some}
\bibinfo{author}{S.~Loyalka}, \bibinfo{author}{N.~Petrellis},
  \bibinfo{author}{T.~Storvick},
\newblock \bibinfo{title}{Some numerical results for the {BGK} model: {T}hermal
  creep and viscous slip problems with arbitrary accomodation at the surface},
\newblock \bibinfo{journal}{The Physics of Fluids} \bibinfo{volume}{18}
  (\bibinfo{year}{1975}) \bibinfo{pages}{1094--1099}.
%Type = Inproceedings
\bibitem[{Maccormack(1989)}]{maccormack1989nonequilibrium}
\bibinfo{author}{R.~Maccormack},
\newblock \bibinfo{title}{Nonequilibrium effects for hypersonic transitional
  flows using continuum approach},
\newblock in: \bibinfo{booktitle}{27th Aerospace Sciences Meeting}, p.
  \bibinfo{pages}{461}.
%Type = Article
\bibitem[{Wu(2008)}]{wu2008slip}
\bibinfo{author}{L.~Wu},
\newblock \bibinfo{title}{A slip model for rarefied gas flows at arbitrary
  {K}nudsen number},
\newblock \bibinfo{journal}{Applied Physics Letters} \bibinfo{volume}{93}
  (\bibinfo{year}{2008}) \bibinfo{pages}{253103}.
%Type = Book
\bibitem[{Kennard et~al.(1938)}]{kennard1938kinetic}
\bibinfo{author}{E.~H. Kennard}, et~al., \bibinfo{title}{Kinetic theory of
  gases}, volume \bibinfo{volume}{483}, \bibinfo{publisher}{McGraw-hill New
  York}, \bibinfo{year}{1938}.
%Type = Article
\bibitem[{Tucny et~al.(2020)Tucny, Vidal, Leclaire, and
  Bertrand}]{tucny2020comparison}
\bibinfo{author}{J.-M. Tucny}, \bibinfo{author}{D.~Vidal},
  \bibinfo{author}{S.~Leclaire}, \bibinfo{author}{F.~Bertrand},
\newblock \bibinfo{title}{Comparison of existing and extended boundary
  conditions for the simulation of rarefied gas flows using the lattice
  {B}oltzmann method},
\newblock \bibinfo{journal}{International Journal of Modern Physics C}
  \bibinfo{volume}{31} (\bibinfo{year}{2020}) \bibinfo{pages}{2050070}.
%Type = Article
\bibitem[{Fan et~al.(2001)Fan, Boyd, Cai, Hennighausen, and
  Candler}]{fan2001computation}
\bibinfo{author}{J.~Fan}, \bibinfo{author}{I.~D. Boyd}, \bibinfo{author}{C.-P.
  Cai}, \bibinfo{author}{K.~Hennighausen}, \bibinfo{author}{G.~V. Candler},
\newblock \bibinfo{title}{Computation of rarefied gas flows around a {NACA}
  0012 airfoil},
\newblock \bibinfo{journal}{AIAA Journal} \bibinfo{volume}{39}
  (\bibinfo{year}{2001}) \bibinfo{pages}{618--625}.
%Type = Article
\bibitem[{Le et~al.(2015)Le, Shoja-Sani, and Roohi}]{le2015rarefied}
\bibinfo{author}{N.~T. Le}, \bibinfo{author}{A.~Shoja-Sani},
  \bibinfo{author}{E.~Roohi},
\newblock \bibinfo{title}{Rarefied gas flow simulations of {NACA} 0012 airfoil
  and sharp 25--55-deg biconic subject to high order nonequilibrium boundary
  conditions in {CFD}},
\newblock \bibinfo{journal}{Aerospace Science and Technology}
  \bibinfo{volume}{41} (\bibinfo{year}{2015}) \bibinfo{pages}{274--288}.
%Type = Article
\bibitem[{Lofthouse et~al.(2008)Lofthouse, Scalabrin, and
  Boyd}]{lofthouse2008velocity}
\bibinfo{author}{A.~J. Lofthouse}, \bibinfo{author}{L.~C. Scalabrin},
  \bibinfo{author}{I.~D. Boyd},
\newblock \bibinfo{title}{Velocity slip and temperature jump in hypersonic
  aerothermodynamics},
\newblock \bibinfo{journal}{Journal of Thermophysics and Heat Transfer}
  \bibinfo{volume}{22} (\bibinfo{year}{2008}) \bibinfo{pages}{38--49}.
%Type = Article
\bibitem[{Boyd et~al.(1995)Boyd, Chen, and Candler}]{boyd1995predicting}
\bibinfo{author}{I.~D. Boyd}, \bibinfo{author}{G.~Chen}, \bibinfo{author}{G.~V.
  Candler},
\newblock \bibinfo{title}{Predicting failure of the continuum fluid equations
  in transitional hypersonic flows},
\newblock \bibinfo{journal}{Physics of Fluids} \bibinfo{volume}{7}
  (\bibinfo{year}{1995}) \bibinfo{pages}{210--219}.
%Type = Article
\bibitem[{Shi et~al.(2019)Shi, Yang, Jin, He, and Wang}]{shi2019wall}
\bibinfo{author}{B.~Shi}, \bibinfo{author}{X.~Yang}, \bibinfo{author}{G.~Jin},
  \bibinfo{author}{G.~He}, \bibinfo{author}{S.~Wang},
\newblock \bibinfo{title}{Wall-modeling for large-eddy simulation of flows
  around an axisymmetric body using the diffuse-interface immersed boundary
  method},
\newblock \bibinfo{journal}{Applied Mathematics and Mechanics}
  \bibinfo{volume}{40} (\bibinfo{year}{2019}) \bibinfo{pages}{305--320}.
%Type = Article
\bibitem[{Capizzano(2011)}]{capizzano2011turbulent}
\bibinfo{author}{F.~Capizzano},
\newblock \bibinfo{title}{Turbulent wall model for immersed boundary methods},
\newblock \bibinfo{journal}{AIAA Journal} \bibinfo{volume}{49}
  (\bibinfo{year}{2011}) \bibinfo{pages}{2367--2381}.
%Type = Article
\bibitem[{Wang and Tian(2018)}]{wang2018heat}
\bibinfo{author}{L.~Wang}, \bibinfo{author}{F.-B. Tian},
\newblock \bibinfo{title}{Heat transfer in non-newtonian flows by a hybrid
  immersed boundary--lattice {B}oltzmann and finite difference method},
\newblock \bibinfo{journal}{Applied Sciences} \bibinfo{volume}{8}
  (\bibinfo{year}{2018}) \bibinfo{pages}{559}.
%Type = Article
\bibitem[{Huang and Tian(2019)}]{huang2019recent}
\bibinfo{author}{W.-X. Huang}, \bibinfo{author}{F.-B. Tian},
\newblock \bibinfo{title}{Recent trends and progress in the immersed boundary
  method},
\newblock \bibinfo{journal}{Proceedings of the Institution of Mechanical
  Engineers, Part C: Journal of Mechanical Engineering Science}
  \bibinfo{volume}{233} (\bibinfo{year}{2019}) \bibinfo{pages}{7617--7636}.
%Type = Article
\bibitem[{Ghias et~al.(2007)Ghias, Mittal, and Dong}]{ghias2007sharp}
\bibinfo{author}{R.~Ghias}, \bibinfo{author}{R.~Mittal},
  \bibinfo{author}{H.~Dong},
\newblock \bibinfo{title}{A sharp interface immersed boundary method for
  compressible viscous flows},
\newblock \bibinfo{journal}{Journal of Computational Physics}
  \bibinfo{volume}{225} (\bibinfo{year}{2007}) \bibinfo{pages}{528--553}.
%Type = Article
\bibitem[{Mittal et~al.(2008)Mittal, Dong, Bozkurttas, Najjar, Vargas, and
  Von~Loebbecke}]{mittal2008versatile}
\bibinfo{author}{R.~Mittal}, \bibinfo{author}{H.~Dong},
  \bibinfo{author}{M.~Bozkurttas}, \bibinfo{author}{F.~Najjar},
  \bibinfo{author}{A.~Vargas}, \bibinfo{author}{A.~Von~Loebbecke},
\newblock \bibinfo{title}{A versatile sharp interface immersed boundary method
  for incompressible flows with complex boundaries},
\newblock \bibinfo{journal}{Journal of Computational Physics}
  \bibinfo{volume}{227} (\bibinfo{year}{2008}) \bibinfo{pages}{4825--4852}.
%Type = Inproceedings
\bibitem[{Luo et~al.(2009)Luo, Dai, and Ferreira~de Sousa}]{luo2009hybrid}
\bibinfo{author}{H.~Luo}, \bibinfo{author}{H.~Dai},
  \bibinfo{author}{P.~Ferreira~de Sousa},
\newblock \bibinfo{title}{A hybrid formulation to suppress the numerical
  oscillations caused by immersed moving boundaries},
\newblock in: \bibinfo{booktitle}{APS Division of Fluid Dynamics Meeting
  Abstracts}, volume~\bibinfo{volume}{62}, pp. \bibinfo{pages}{EL--006}.
%Type = Article
\bibitem[{Seo and Mittal(2011)}]{seo2011sharp}
\bibinfo{author}{J.~H. Seo}, \bibinfo{author}{R.~Mittal},
\newblock \bibinfo{title}{A sharp-interface immersed boundary method with
  improved mass conservation and reduced spurious pressure oscillations},
\newblock \bibinfo{journal}{Journal of Computational Physics}
  \bibinfo{volume}{230} (\bibinfo{year}{2011}) \bibinfo{pages}{7347--7363}.
%Type = Article
\bibitem[{Wang et~al.(2017)Wang, Currao, Han, Neely, Young, and
  Tian}]{wang2017immersed}
\bibinfo{author}{L.~Wang}, \bibinfo{author}{G.~M. Currao},
  \bibinfo{author}{F.~Han}, \bibinfo{author}{A.~J. Neely},
  \bibinfo{author}{J.~Young}, \bibinfo{author}{F.-B. Tian},
\newblock \bibinfo{title}{An immersed boundary method for fluid--structure
  interaction with compressible multiphase flows},
\newblock \bibinfo{journal}{Journal of Computational Physics}
  \bibinfo{volume}{346} (\bibinfo{year}{2017}) \bibinfo{pages}{131--151}.
%Type = Article
\bibitem[{Meng and Zhang(2011)}]{meng2011accuracy}
\bibinfo{author}{J.~Meng}, \bibinfo{author}{Y.~Zhang},
\newblock \bibinfo{title}{Accuracy analysis of high-order lattice {B}oltzmann
  models for rarefied gas flows},
\newblock \bibinfo{journal}{Journal of Computational Physics}
  \bibinfo{volume}{230} (\bibinfo{year}{2011}) \bibinfo{pages}{835--849}.
%Type = Article
\bibitem[{Liu et~al.(1994)Liu, Osher, and Chan}]{liu1994weighted}
\bibinfo{author}{X.-D. Liu}, \bibinfo{author}{S.~Osher},
  \bibinfo{author}{T.~Chan},
\newblock \bibinfo{title}{Weighted essentially non-oscillatory schemes},
\newblock \bibinfo{journal}{Journal of Computational Physics}
  \bibinfo{volume}{115} (\bibinfo{year}{1994}) \bibinfo{pages}{200--212}.
%Type = Misc
\bibitem[{Zienkiewicz and Taylor(2000)}]{zienkiewicz2000finite}
\bibinfo{author}{O.~Zienkiewicz}, \bibinfo{author}{R.~Taylor},
  \bibinfo{title}{Finite element method, 5th edn. {V}olume 1-the basis},
  \bibinfo{year}{2000}.
%Type = Article
\bibitem[{Wang et~al.(2017)Wang, Han, and Zhou}]{wang2017projection}
\bibinfo{author}{L.~Wang}, \bibinfo{author}{F.~Han}, \bibinfo{author}{Q.~Zhou},
\newblock \bibinfo{title}{The projection angles of fragments from a cylindrical
  casing filled with charge initiated at one end},
\newblock \bibinfo{journal}{International Journal of Impact Engineering}
  \bibinfo{volume}{103} (\bibinfo{year}{2017}) \bibinfo{pages}{138--148}.
%Type = Article
\bibitem[{Tian et~al.(2014)Tian, Dai, Luo, Doyle, and Rousseau}]{tian2014fluid}
\bibinfo{author}{F.-B. Tian}, \bibinfo{author}{H.~Dai},
  \bibinfo{author}{H.~Luo}, \bibinfo{author}{J.~F. Doyle},
  \bibinfo{author}{B.~Rousseau},
\newblock \bibinfo{title}{Fluid--structure interaction involving large
  deformations: 3{D} simulations and applications to biological systems},
\newblock \bibinfo{journal}{Journal of Computational Physics}
  \bibinfo{volume}{258} (\bibinfo{year}{2014}) \bibinfo{pages}{451--469}.
%Type = Article
\bibitem[{Goldstein et~al.(1993)Goldstein, Handler, and
  Sirovich}]{goldstein1993modeling}
\bibinfo{author}{D.~Goldstein}, \bibinfo{author}{R.~Handler},
  \bibinfo{author}{L.~Sirovich},
\newblock \bibinfo{title}{Modeling a no-slip flow boundary with an external
  force field},
\newblock \bibinfo{journal}{Journal of Computational Physics}
  \bibinfo{volume}{105} (\bibinfo{year}{1993}) \bibinfo{pages}{354--366}.
%Type = Article
\bibitem[{Peskin(2002)}]{peskin2002immersed}
\bibinfo{author}{C.~S. Peskin},
\newblock \bibinfo{title}{The immersed boundary method},
\newblock \bibinfo{journal}{Acta Numerica} \bibinfo{volume}{11}
  (\bibinfo{year}{2002}) \bibinfo{pages}{479--517}.
%Type = Article
\bibitem[{Hadjiconstantinou(2003)}]{hadjiconstantinou2003comment}
\bibinfo{author}{N.~G. Hadjiconstantinou},
\newblock \bibinfo{title}{Comment on {C}ercignani’s second-order slip
  coefficient},
\newblock \bibinfo{journal}{Physics of Fluids} \bibinfo{volume}{15}
  (\bibinfo{year}{2003}) \bibinfo{pages}{2352--2354}.
%Type = Article
\bibitem[{Kim et~al.(2008)Kim, Pitsch, and Boyd}]{kim2008slip}
\bibinfo{author}{S.~H. Kim}, \bibinfo{author}{H.~Pitsch},
  \bibinfo{author}{I.~D. Boyd},
\newblock \bibinfo{title}{Slip velocity and {K}nudsen layer in the lattice
  {B}oltzmann method for microscale flows},
\newblock \bibinfo{journal}{Physical Review E} \bibinfo{volume}{77}
  (\bibinfo{year}{2008}) \bibinfo{pages}{026704}.
%Type = Article
\bibitem[{Shterev et~al.(2019)Shterev, Manoach, and
  Stefanov}]{shterev2019hybrid}
\bibinfo{author}{K.~Shterev}, \bibinfo{author}{E.~Manoach},
  \bibinfo{author}{S.~Stefanov},
\newblock \bibinfo{title}{Hybrid numerical approach to study the interaction of
  the rarefied gas flow in a microchannel with a cantilever},
\newblock \bibinfo{journal}{International Journal of Non-Linear Mechanics}
  \bibinfo{volume}{117} (\bibinfo{year}{2019}) \bibinfo{pages}{103239}.
%Type = Article
\bibitem[{Vogenitz et~al.(1968)Vogenitz, Bird, Broadwell, and
  Rungaldier}]{vogenitz1968theoretical}
\bibinfo{author}{F.~Vogenitz}, \bibinfo{author}{G.~Bird},
  \bibinfo{author}{J.~Broadwell}, \bibinfo{author}{H.~Rungaldier},
\newblock \bibinfo{title}{Theoretical and experimental study of rarefied
  supersonic flows about several simple shapes.},
\newblock \bibinfo{journal}{AIAA Journal} \bibinfo{volume}{6}
  (\bibinfo{year}{1968}) \bibinfo{pages}{2388--2394}.
%Type = Article
\bibitem[{Yin and Luo(2010)}]{yin2010effect}
\bibinfo{author}{B.~Yin}, \bibinfo{author}{H.~Luo},
\newblock \bibinfo{title}{Effect of wing inertia on hovering performance of
  flexible flapping wings},
\newblock \bibinfo{journal}{Physics of Fluids} \bibinfo{volume}{22}
  (\bibinfo{year}{2010}) \bibinfo{pages}{111902}.
%Type = Article
\bibitem[{Tian et~al.(2013)Tian, Luo, Song, and Lu}]{tian2013force}
\bibinfo{author}{F.-B. Tian}, \bibinfo{author}{H.~Luo},
  \bibinfo{author}{J.~Song}, \bibinfo{author}{X.-Y. Lu},
\newblock \bibinfo{title}{Force production and asymmetric deformation of a
  flexible flapping wing in forward flight},
\newblock \bibinfo{journal}{Journal of Fluids and Structures}
  \bibinfo{volume}{36} (\bibinfo{year}{2013}) \bibinfo{pages}{149--161}.
%Type = Article
\bibitem[{Shahzad et~al.(2018)Shahzad, Tian, Young, and
  Lai}]{shahzad2018effects}
\bibinfo{author}{A.~Shahzad}, \bibinfo{author}{F.-B. Tian},
  \bibinfo{author}{J.~Young}, \bibinfo{author}{J.~C. Lai},
\newblock \bibinfo{title}{Effects of hawkmoth-like flexibility on the
  aerodynamic performance of flapping wings with different shapes and aspect
  ratios},
\newblock \bibinfo{journal}{Physics of Fluids} \bibinfo{volume}{30}
  (\bibinfo{year}{2018}) \bibinfo{pages}{091902}.
%Type = Article
\bibitem[{Wang and Tian(2020)}]{wang2020numerical}
\bibinfo{author}{L.~Wang}, \bibinfo{author}{F.-B. Tian},
\newblock \bibinfo{title}{Numerical study of sound generation by
  three-dimensional flexible flapping wings during hovering flight},
\newblock \bibinfo{journal}{Journal of Fluids and Structures}
  \bibinfo{volume}{99} (\bibinfo{year}{2020}) \bibinfo{pages}{103165}.
%Type = Article
\bibitem[{Ellington(1984)}]{ellington1984aerodynamics}
\bibinfo{author}{C.~P. Ellington},
\newblock \bibinfo{title}{The aerodynamics of hovering insect flight. i. the
  quasi-steady analysis},
\newblock \bibinfo{journal}{Philosophical Transactions of the Royal Society of
  London. B, Biological Sciences} \bibinfo{volume}{305} (\bibinfo{year}{1984})
  \bibinfo{pages}{1--15}.
%Type = Article
\bibitem[{Dai et~al.(2012)Dai, Luo, and Doyle}]{dai2012dynamic}
\bibinfo{author}{H.~Dai}, \bibinfo{author}{H.~Luo}, \bibinfo{author}{J.~F.
  Doyle},
\newblock \bibinfo{title}{Dynamic pitching of an elastic rectangular wing in
  hovering motion},
\newblock \bibinfo{journal}{Journal of Fluid Mechanics} \bibinfo{volume}{693}
  (\bibinfo{year}{2012}) \bibinfo{pages}{473--499}.
%Type = Article
\bibitem[{Shahzad et~al.(2016)Shahzad, Tian, Young, and
  Lai}]{shahzad2016effects}
\bibinfo{author}{A.~Shahzad}, \bibinfo{author}{F.-B. Tian},
  \bibinfo{author}{J.~Young}, \bibinfo{author}{J.~C. Lai},
\newblock \bibinfo{title}{Effects of wing shape, aspect ratio and deviation
  angle on aerodynamic performance of flapping wings in hover},
\newblock \bibinfo{journal}{Physics of Fluids} \bibinfo{volume}{28}
  (\bibinfo{year}{2016}) \bibinfo{pages}{111901}.
%Type = Article
\bibitem[{Wang et~al.(2022)Wang, Tian, and Liu}]{wang2022numerical}
\bibinfo{author}{L.~Wang}, \bibinfo{author}{F.-B. Tian},
  \bibinfo{author}{H.~Liu},
\newblock \bibinfo{title}{Numerical study of three-dimensional flapping wings
  hovering in ultra-low-density atmosphere},
\newblock \bibinfo{journal}{Physics of Fluids} \bibinfo{volume}{34}
  (\bibinfo{year}{2022}) \bibinfo{pages}{041903}.

\end{thebibliography}

%%%%%%%%%%%%%%%%%%%%%%%%%%%%%%%%%%%%%%%%%%
\end{document}